\documentclass[final,3p,times]{elsarticle}

\usepackage{mathpazo}
\usepackage{latexsym}
\usepackage{amssymb}
\usepackage{amsthm}
\usepackage[cmex10]{amsmath}

\usepackage{mathtools}

\usepackage{graphicx}
\usepackage{float}
\usepackage{subfig}
\usepackage[]{placeins}
\usepackage[]{algorithm}
\usepackage{algpseudocode}
\usepackage{multirow}
\usepackage[subnum]{cases}
\usepackage{tabularx}

\usepackage{subfig}

\DeclarePairedDelimiter\ceil{\lceil}{\rceil}

\DeclarePairedDelimiter\abs{\lvert}{\rvert}   % |absolute value|
\newtheorem{theorem}{Theorem}
\newtheorem{proposition}[theorem]{Proposition}% 
\newtheorem{lemma}{Lemma}
\newtheorem{definition}{Definition}%

\makeatletter
\renewcommand\appendix{\par
  \setcounter{section}{0}%
  \setcounter{subsection}{0}%
  \setcounter{equation}{0}%
  \setcounter{table}{0}%------------ << add
  \setcounter{figure}{0}%----------- << add
  \gdef\theequation{\@Alph\c@section.\arabic{equation}}%
  \gdef\thefigure{\@Alph\c@section.\arabic{figure}}%
  \gdef\thetable{\@Alph\c@section.\arabic{table}}%
  \gdef\thesection{\Alph{section}}%
  \@addtoreset{equation}{section}%
  \@addtoreset{table}{section}%----- << add
  \@addtoreset{figure}{section}%---- << add
}
\makeatother

\begin{document}

\begin{frontmatter}

%% \cortex[cor1]{}

\title{Distributed P2P quantile tracking with relative value error}

\author [unile] {Marco Pulimeno}
\ead{marco.pulimeno@unisalento.it}
\author [unile] {Italo Epicoco}
\ead{italo.epicoco@unisalento.it}
\author [unile] {Massimo~Cafaro \corref{cor1}}
\ead{massimo.cafaro@unisalento.it}
\cortext[cor1]{Corresponding author}
\affiliation[unile]{organization={University of Salento, Dept. of Engineering for Innovation},
             addressline={Via per Monteroni},
             city={Lecce},
             postcode={73100},
             %state={Puglia},
             country={Italy}}

\begin{abstract}
	In this paper we present \textsc{DUDDSketch}, a distributed version of the \textsc{UDDSketch} algorithm for accurate tracking of quantiles. The algorithm is a fully decentralized, gossip-based distributed protocol working in the context of unstructured P2P networks. We discuss the algorithm's design and formally prove its correctness. We also show, through extensive experimental results, that the algorithm converges to the results provided by the sequential algorithm, which is a fundamental and highly desirable property.

\end{abstract}


\end{frontmatter}

\section{Introduction}
\label{introduction}
In this paper we design and analyze a novel distributed algorithm for precise quantile tracking with relative value error in the context of P2P systems. Here, we focus on large univariate datasets or streams. Denoting by $D = \{x_i\}_i^n$ a dataset of real values $x_i \in \mathbb{R}$, let $n=\abs{D}$ be its cardinality.  Each item $x_i$ in $D$ may appear in $D$ with a multiplicity greater than 1, making $D$ a multiset (or \emph{bag}).

The computation of quantiles is one of the pre-processing steps that can be performed on such a dataset. Other pre-processing steps that can be performed on such a dataset include the computation of \emph{count}, \emph{mean} and \emph{standard deviation}. These pre-processing steps help characterize the fundamental statistical properties of the dataset that is being used. Quantiles are a type of order statistic \cite{buragohain2009quantiles}, and they offer a description of the statistical properties of the underlying dataset that is more robust than using the mean and variance in the presence of outliers. As a result, they offer a better input characterization and a more accurate depiction of the data. 

The estimation of the cumulative distribution function (CDF) of an input can be done using quantiles, which is the primary reason why computing quantiles is so important \cite{buragohain2009quantiles}. A parametric approach can provide a statistical description of the data distribution (knowing the distribution and its parameters) when certain a priori information about the data distribution are known or assumptions can be made; however, in many real-world scenarios, no information is available and no assumptions can be made on the input; hence, computing the quantiles is necessary to determine how the data is distributed over its domain. Therefore, the activity of understanding data distribution is a necessary part of the data profiling process in data analysis. 

In order to do an accurate computation of quantiles one must have explicit knowledge of all of the values included in $D$ and must use a quantile algorithm. Since quantiles must be computed with regard to the full dataset, we may say that they are holistic statistics. Nevertheless, the computation for quantiles is not as straightforward as the computation for aggregation statistics such as, for instance, the mean of the values. 

When working with large datasets, these requirements cannot be met. For example, in the streaming model we have to process the input stream using a very limited amount of memory; moreover, data points arrive one at a time, and an algorithm has a bounded amount of time to process the current item. Ideally, each incoming item should be processed in  worst-case constant time. Therefore, the resulting algorithms' worst-case complexity is $O(n)$ where $n$ is the size of the input (it is worth recalling here that a stream may be either bounded, in which case $n$ is the fixed length of the stream, or it may be infinite: in this case, $n$ refers to the length of the stream up to and including the latest item to be processed).

The design of ad-hoc summaries, which reside in main memory and whose size is independent of the length of the stream, is the only possible solution to support data mining algorithms in these circumstances \cite{gibbons98}, because I/O operations on disks take too much time (due to the latency), which implies that an algorithm cannot rely on disk accesses in these situations. The difficulty is maintaining quantiles over the stream whilst summaries are maintained incrementally (i.e., as the input stream changes over time). 

To this purpose, algorithms can only provide an \emph{approximate}  solution; the aim is to deliver the best possible estimate (with the smallest feasible error) within the limits of processing time and memory. Tracking quantiles on streams has garnered considerable attention \cite{chen2020survey, luo2016quantiles, buragohain2009quantiles}, and Section \ref{related} provides some of the most pertinent research findings. Concerning the distributed setting, a classical distributed application involving quantiles is related to the necessity to precisely assess the latency of a website, i.e., the delay experienced by users when submitting query requests \cite{Vitter}. In order to meet customer demand and deliver good QoS (Quality of Service) values, highly requested websites (e.g., a search engine) distribute the incoming traffic load (i.e., the users' queries) among multiple web server hosts. Due to the frequently skewed distribution of latency values, it is common practice to monitor specific quantiles, such as the 95th, 98th, and 99th percentiles \cite{latency}. To compute the overall latency of a website (across all of the associated web server hosts), quantiles must be precisely maintained for each host, and a distributed algorithm is necessary to aggregate individual host responses. According to the \emph{internet live stats} website (data acquired on December 14, 2021), Google now processes an average of over 40,000 search queries per second, which corresponds to over 3.5 billion searches per day and 1.2 trillion searches per year globally. The search requests are served by a number of servers estimated to exceed 1 million (the guess is based on Google energy consumption). 

Quantiles are also used in the Gigascope streaming database  \cite{gigascope} for the purpose of network applications and systems monitoring, and in database query optimizers \cite{approximate-quantiles} since they allow estimating the size of intermediate results, which are required in order to determine an optimal execution plan. Moreover, Internet service providers  heavily rely on the use of quantiles both for network health monitoring \cite{UDAFs} and to fullfill the requirement of a concise summary of value distribution across sensor networks \cite{shrivastava2004medians}. 

This work continues a line of research that we started by designing, initially, a sequential algorithm for accurately tracking quantiles in a streaming setting \cite{uddsketch}. Next, we showed that our algorithm is mergeable \cite{Agarwal} and designed a parallel version \cite{CAFARO2023155}. Here, we present a distributed protocol.

Our gossip-based distributed protocol is a fully decentralized  algorithm that operates in the setting of unstructured P2P networks. The accuracy of the quantile estimates carried out in a distributed setting should be comparable to the accuracy of those obtained in a setting that is centralized, in which all of the data are kept in a single location or are part of a single data stream. This is an essential and much sought after feature. We demonstrate that our distributed protocol maintains the same features as the sequential version and outputs results that are either identical or comparable to those provided by the sequential version with regard to the accuracy, thus satisfying the quality criteria discussed above.

It is worth recalling here that our algorithm provides a \textit{relative value error} guarantee (see Section \ref{notation} for a discussion of strengths and weaknesses with regard to r\textit{elative rank error}). The algorithm is based on a sketch data structure which maintains an exponential histogram of the input data, and works in a peer to peer (P2P) setting. In particular, we make use of a gossip--based protocol \cite{Demers:1987}, namely \textit{distributed averaging}, which is a consensus algorithm serving as a foundational tool for information dissemination of our P2P algorithm. 

The task of distributed averaging is to drive the states of the nodes of a network towards an agreed common value with regard to the initial values held by the agents; this value is the average of the agents' initial values. In such a protocol, the agents share their status information only with a few other connected agents called \textit{neighbours}.

A synchronous gossip--based protocol consists of a sequence of rounds in which each peer randomly selects one or more peers, exchanges its local state information with the selected peers and updates its local state by using the received information. Therefore, gossip can be thought as a simple way of self-organising the peers' interactions in consensus seeking.

This paper is organized as follows. In Section \ref{notation} we provide preliminary required definitions and the notation used. Related work is discussed in Section \ref{related}. Next, we introduce the distributed setting in Section \ref{distributed} and present our distributed algorithm in Section \ref{duddsketch}. Its correctness is formally proved in Section \ref{correctness}, whilst extensive experimental results are shown and discussed in Section \ref{results}. Finally, we draw our conclusions in Section \ref{conclusions}.

\section{Preliminary definitions and notation}
\label{notation}

In this Section, we will review the preliminary definitions as well as the notation that will be utilized throughout the rest of the work. We start off by providing the definitions of \textit{rank} and \textit{$q$-quantile}.

\begin{definition}(\textit{Rank})\\
 Given a set $S$ with $n$ elements drawn from a totally ordered universe set $\mathcal{U}$, the \textit{rank} of the element $x$, denoted by $R(x)$, is the number of elements in $S$ less than or equal to $x$, i.e.
 
\begin{equation}
R(x):= \Big| \{ z\in S \mid z\leq x \}\Big|.
\end{equation}
\end{definition}
%\vspace{1mm}

\begin{definition}(\textit{$q$-quantile})\\
Given a set $S$ with $n$ elements drawn from a totally ordered universe set $\mathcal{U}$ and a real number $0\leq q \leq 1$, the \textit{inferior $q$-quantile} (respectively \textit{superior $q$-quantile}) is the element $x_q$ whose rank in $S$  is such that

\begin{equation}
 x_q\in S \, \, :\, \, R(x_q)=\lfloor 1+q\cdot(n-1)\rfloor
\end{equation}

(respectively $R(x_q)=\lceil 1+q\cdot(n-1)\rceil$).
\end{definition}
%\vspace{1mm}

For instance, $x_0$ and $x_1$ represent, respectively, the \textit{minimum} and the \textit{maximum} element of the set $S$, whilst $x_{0.5}$ represents the \textit{median}.

Since accuracy is the most important property of an algorithm designed for tracking quantiles, we recall here how accuracy can be defined: either as \textit{rank accuracy} or \textit{relative accuracy}. 
%We shall show that relative accuracy is intrinsically a better choice, even though it is more difficult to attain. 

\begin{definition}(\textit{Rank accuracy})\\
\label{rank:accuracy}
Given an item $v$ and a tolerance $\alpha$, an estimate of the rank $\tilde{R}(v)$ is returned such that
 
\begin{equation}
\mid\tilde{R}(v)-R(v) \mid \leq \alpha \cdot n.
\end{equation}
\end{definition}

\begin{definition}(\textit{Relative accuracy})\\
\label{rank:relative-accuracy}
Given the item related to the $q$-quantile $x_q\in S$, the \textit{$\alpha$-accurate $q$-quantile} is defined as the item $\tilde{x}_q$ such that

\begin{equation}
\mid\tilde{x}_q-x_q \mid \leq \alpha \cdot x_q. 
\end{equation}

An algorithm is \textit{$(q_0,q_1) \alpha$-accurate} if it returns $\alpha$-accurate $q$-quantiles, for $q_0 \leq q \leq q_1$.
\end{definition}

Before discussing related work, we briefly recall the concepts of additive and multiplicative approximation, since the algorithms designed so far strive to provide either the former or the latter. Additive $(\epsilon n)$ approximation is defined as follows. 

\begin{definition}(\textit{Additive $(\epsilon n)$ approximation})\\
Given a stream $\sigma = {x_1, x_2, \cdots, x_n}$ (or a dataset) of length $n$ and an error parameter $0 < \epsilon < 1$, the $\epsilon$-approximate $\phi$-quantile is any element $x \in \sigma$ with rank $R(x) = |\{y_i \in \sigma: y_i \leq x \}|$ (i.e., the number of elements less than or equal to $x$) such that $(\phi - \epsilon) n \leq R(x) \leq (\phi + \epsilon) n$. An approximation $\hat R(x)$ is additive if $|R(x) - \hat R(x) | \leq \epsilon n$.
\end{definition}

\begin{definition}(\textit{Multiplicative approximation})\\
A multiplicative approximation requires, given $0 < \epsilon < 1$ and an approximation $\hat R(x)$, that $|R(x) - \hat R(x)| \leq \epsilon R(x)$.
\end{definition}

It is a well-known fact that, for an algorithm tracking quantiles, achieving a multiplicative guarantee is strictly more difficult than achieving an additive one \cite{REQ, REQ-JOURNAL}. That said, \cite{REQ, REQ-JOURNAL} also remark that the relative value error guarantee, which corresponds to Definition \ref{rank:relative-accuracy}, only makes sense for data universes with a notion of magnitude and distance (e.g., numerical data), and that the definition is not invariant to natural data transformations, such as incrementing every data item by a large constant. In contrast, the standard notion of relative error, which corresponds to Definition \ref{rank:accuracy}, does not refer to the data items themselves, only to their ranks, and is arguably of more general applicability. Despite these limitations, relative value error can be used in many applications, notably for the problem of monitoring network latencies which is a quintessential application that demands relative error \cite{Masson}. In spite of the fact that a considerable amount of time has been spent concentrating the research attention on the design of sketches and other data structures supporting rank accuracy, many datasets include heavy tails causing algorithms that provide rank accuracy to actually return values with arbitrary large relative errors.

Finally, we introduce the notion of \textit{mergeability} and \textit{one-way mergeability} of stream summaries \cite{Agarwal, Agarwal-journal}. Let $S()$ be a summarization algorithm, $D$ a dataset and $\varepsilon$ an error parameter. It is worth noting here that $S()$ may be a one-to-many mapping, i.e. there can be multiple valid outputs since this may depend on the order used to process $D$: different permutations of $D$ lead to different valid summaries. Let $S(D, \varepsilon)$ be a valid summary with error $\varepsilon$ for the dataset $D$ produced by $S()$, and let $k(n, \varepsilon)$ be the maximum size of any $S(D, \varepsilon)$ for any dataset $D$ consisting of $n$ items. We are now ready to state the following definitions.

\begin{definition}(\textit{Mergeability of summaries})\\
A summarization method $S()$ is said to be mergeable if there exists an algorithm $\mathcal{A}$ that produces a summary $S\left(D_1 \uplus D_2, \varepsilon\right)$ from any two input summaries $S\left(D_1, \varepsilon\right)$ and $S\left(D_2, \varepsilon\right)$. Here, $\uplus$ denotes multiset addition. Note that, by definition, the size of the merged summary produced by A is at most $k\left(\left|D_1\right|+\left|D_2\right|, \varepsilon\right)$. If $k(n, \varepsilon)$ is independent of $n$, which we can denote by $k(\varepsilon)$, then the size of each $S\left(D_1, \varepsilon\right)$, $S\left(D_2, \varepsilon\right)$ and the summary produced by $\mathcal{A}$ is at most $k(\varepsilon)$.
\end{definition}

The previous definition requires that, given two summaries on two datasets, there is an algorithm that can merge the two input summaries into a single output summary on the two datasets combined together, whilst preserving both the error and size guarantees. Mergeability is an important property since it allows designing parallel and distributed algorithms in which the input data is partitioned among the processors and processed locally by each processor. The intermediate outputs produced by the processors need to be aggregated to output the final result related to the union of the local inputs. The aggregation step can be done in an arbitrary pairwise fashion, and the aggregation function corresponds to the notion of mergeability: each processor produces a summary of its share of the input and the aggregation function is the merge operation. A weaker, restricted form of mergeability, called one-way mergeability, is defined as follows. 

\begin{definition}(\textit{One-Way Mergeability of summaries})\\
A summary $S(D, \varepsilon)$ is one-way mergeable if there exist two algorithms $\mathcal{A}_1$ and $\mathcal{A}_2$ such that: (1) given any dataset $D$, $\mathcal{A}_2$ creates a summary of $D$ as $S(D, \varepsilon)$; (2) given any $S\left(D_2, \varepsilon\right)$ produced by $\mathcal{A}_2$ and any $S\left(D_1, \varepsilon\right)$ produced by $\mathcal{A}_1$ or $\mathcal{A}_2$, $\mathcal{A}_1$ builds a merged summary $S\left(D_1 \uplus D_2, \varepsilon\right)$.
\end{definition}

\section{Related work}
\label{related}

The vast majority of the algorithms proposed in the literature for approximate quantile computing are not \textit{mergeable} \cite{Agarwal, Agarwal-journal} i.e., they cannot be utilized in a distributed or parallel system, and others are only \textit{one-way mergeable} which prevents anyway their use in a distributed or parallel setting. Notable examples of one-way mergeable algorithms include Greenwald-Khanna \cite{greenwald2001space} and the \textsf{t-digest} \cite{tdunnings, DUNNING2021100049} algorithms. For details related to the one-way mergeability of Greenwald-Khanna and \textsf{t-digest}, see respectively \cite{Agarwal, Agarwal-journal} and \cite{Masson}.

Among the mergeable algorithms for quantile approximation it is worth recalling here \textsf{q-digest} \cite{shrivastava2004medians}, \textsf{M-Sketch} \cite{gan2018moment},  \textsf{KLL} \cite{KLL}, \textsf{DCS} \cite{DCS}, \textsc{REQ} \cite{REQ}, \textsc{DDSketch} \cite{Masson} and \textsc{UDDSketch} \cite{uddsketch}. The Quantile Digest, known as \textsf{q-digest}, uses a sketch data structure for approximating quantiles with a guaranteed error bound on the accuracy, where the accuracy that can be obtained improves with the amount of space allowed. Like an equi-depth histogram, but with the added flexibility of allowing buckets to overlap, the deterministic \textsf{q-digest} stores its data in \emph{variable-sized} buckets.

The integer numbers in the range $[0, U-1]$ define the universe $\mathbb{U}$ from which the input data are drawn. The literature refers to this model as the \emph{fixed-size universe} model. Hence, the assumption used in \cite{shrivastava2004medians} is that $\mathbb{U}$ is a finite, countable, positive-range universe set of integer values. This is on the one hand the approach's greatest strength, but also on the other hand its greatest weakness, as it drastically restricts its applicability to other situations (e.g., neither negative nor real values are allowed in this model). Another drawback relates to the inability to delete  values from the sketch: the algorithm is only designed for the \emph{cash-register} model. Lastly, \textsf{q-digest} consumes $O(\frac{1}{\epsilon} \lg U)$ memory words.

Moment Sketch, often referred to as \textsf{M-Sketch}, is a summary designed for effective mergeability. The summary uses the statistical moments to approximate quantiles and has a very small memory footprint and update overhead. The algorithm has been designed for the cash register model and is deterministic. An \emph{order}-$k$ moment sketch is a summary made up of an array of $2k + 3$ floating point values. The value $k$ is the highest power used to compute the moments. The values stored into the array correspond to information extracted from the processed dataset $D \subseteq \mathbb{R}$. These are $x_{min}$ and $x_{max}$, respectively the minimum and maximum values observed thus far; $n$, the number of items processed thus far; the sample moments $\mu_i = \frac{1}{n} \sum_{x \in D} x^i$ and the sample logarithmic moments $\nu_i = \frac{1}{n} \sum_{x \in D} \log^i (x)$ for $1 \le i \le k$. The space required is $O(k)$, but owing to the fact that $k$ is a small constant it can be considered as being $O(1)$. Rather than the underlying distribution, the moment sketch is targeted at monitoring the moments of an empirical continuous interval to approximate quantiles. The \emph{method of moments} and the \emph{principle of maximum entropy} are used to estimate quantiles. The algorithm was tested against current solutions \textsc{GK} and $\mathit{t}\mathsf{-digest}$ and found to be faster, with a reduced memory footprint for the merging operation whilst reaching the same $\epsilon$ error. Moment Sketch accuracy, on the other hand, is worse than \textsc{DDSketch} \cite{Masson} due to a higher relative error.

The \textsc{KLL} randomized algorithm  \cite{KLL}, has been designed to guarantee an additive approximation with at least $1-\delta$ probability, with $0 <  \delta < 1$. Actually, the authors devised two variants of \textsc{KLL}. One of them is known to be space optimal among randomized algorithms, requiring only $O(\frac{1}{\epsilon} \lg \lg (\frac{1}{\delta}))$ space, but with the serious drawback of being not mergeable. The other variant instead is mergeable but looses the space optimality, requiring $O(\frac{1}{\epsilon} \lg^2 \lg (\frac{1}{\delta}))$ space. \textsc{KLL} has been designed to work in the cash-register model. KLL$^\pm$ \cite{kllpm}, is a recently proposed updated version, which works in the \emph{bounded deletion} model (which only allows a fraction $\alpha$ of the inserted items to be removed).

The \textsc{KLL} algorithm uses a hierarchy of \emph{compactors} with different capacities in each level of the hierarchy. A compactor is a buffer of size $k$ storing items of the same weight $w$ as they are ingested from a stream. The height of each compactor is denoted by the symbol $h$; more specifically, the height of the first compactor is equal to $h = 1$, and the height of the final compactor in the chain is equal to $h = H$. In the event that a compactor reaches its capacity, the items stored within are sorted and then compacted, resulting in a sequence of $k/2$ items each of weight $2w$ in which only the even or odd items are retained (with equal probability), whilst the other items are simply thrown away. The items that were picked are then fed into another compactor, which has a height of $h+1$. There could be at most $H \le \lceil \log (n/k) \rceil$ compactors in the chain, $n$ being the length of the stream. 

The randomized algorithm known as Dyadic Count Sketch (\textsc{DCS}) \cite{DCS} has been designed for the fixed-size universe model, in which the data are drawn from an integer domain $[u] = [0, 1, \cdots, u-1]$. \textsc{DCS} applies a dyadic structure to the universe $[u]$ and makes use of a \textsc{Count-Sketch} data structure in order to estimate the frequency of items at each level of the dyadic structure. \textsc{DCS} exhibits the same drawback of \textsf{q-digest}, namely a bounded integer range. Due to the fact that deletions are allowed, the algorithm also works in the context of the more broad \emph{turnstile} model. Regarding the space required, \textsc{DCS} needs $O(\frac{1}{\epsilon} \lg^{1.5} u \lg^{1.5} (\frac{\lg u}{\epsilon}))$ space.

Relative-Error Quantiles Sketch  (\textsc{REQ}) \cite{REQ, REQ-JOURNAL} is one randomized algorithm designed to provide a multiplicative approximation, i.e., a relative error approximation. The space required is $O(\log^{1.5} (\epsilon n)/\epsilon)$. The algorithm does not allow deletions, so it works in the cash register model. As in \textsc{KLL}, the algorithm is based on a framework of relative-compactor objects. \textsc{REQ} is mergeable, however the proof of correctness of the merge procedure is challenging and complex since relative-error guarantees must also be satisfied by the combined sketch resulting from merging two sketches.

\subsection{The \textsc{DDSketch} algorithm}

The data summary of \textsc{DDSketch} \cite{Masson} consists of a group of buckets. The algorithm can process elements $x \in \mathbb R_{> 0}$ and requires two parameters as input in order to initialize its sketch. The first of these parameters, $\alpha$, is related to the accuracy desired by the user, and the second parameter, $m$, represents the maximum number of buckets that are permitted. The algorithm computes $\gamma = \frac{1+\alpha}{1-\alpha}$, a quantity which is then used to establish the boundaries of the $i$th bucket $B_i$. All of the values $x$ that satisfy the condition $\gamma^{i-1} < x \leq \gamma^i$ are handled by the bucket $B_i$, where $i = \ceil{\log_\gamma{x}}$. A bucket is only a counter variable that is initially initialized to zero. It is important to keep in mind that  \textsc{DDSketch} may also deal with negative values. This is done by exploiting another sketch in which an item $x \in \mathbb R_{< 0}$ is handled by inserting the value $-x$. 

To insert an item the algorithm determines in which bucket the item falls, then the corresponding bucket's counter is incremented by one; to remove an item the counter is instead decremented by one (if a counter goes to zero, the bucket is simply discarded). The summary is initially empty, with buckets dynamically added as required when processing input items. Bucket indexes are therefore dynamic in nature, and strictly related to the item $x$ being processed and the value of $gamma$. In order to prevent the summary from growing indefinitely, a collapsing mechanism is performed when the number of buckets in the summary becomes greater than $m$, the maximum number of buckets allowed. With \textsc{DDSketch}, the first two buckets with counts larger than zero are collapsed (however, if desired, the last two buckets may be equivalently collapsed). Denoting with $B_y$ and $B_z$ the first two non zero buckets, with $y < z$, the bucket $B_y$ is deleted from the summary after adding its count to $B_z$. The Algorithm \ref{ddsalg} provides the pseudo-code for inserting an item $x$ into the summary $\mathcal{S}$. 

\begin{algorithm}
  \caption{DDSketch-Insert($x, \mathcal{S}$)}
  \label{ddsalg}
  \begin{algorithmic}
  \Require {$x \in \mathbb R_{> 0}$}
  \State $i \leftarrow \ceil{\log_\gamma{x}}$
  \If{$B_i \in \mathcal{S}$}
  	\State $B_i \leftarrow B_i + 1$
  \Else
  	\State $B_i \leftarrow 1$
  	\State $\mathcal{S} \leftarrow \mathcal{S} \cup B_i$
  \EndIf
  \If{$\abs{\mathcal{S}} > m$}
  	\State let $B_y$ and $B_z$ be the first two buckets
    \State $B_z \leftarrow B_y + B_z$
    \State $\mathcal{S} \leftarrow \mathcal{S} \smallsetminus B_y$
  \EndIf
    \end{algorithmic}
\end{algorithm}
    
Due to its bucket collapsing strategy, \textsc{DDSketch} only offers a $\alpha$-accurate $(q_0, q_1)$-sketch for $q_0 > 0$ and $q_1 = 1$. The actual value of $q_0$ varies wildly, since many items may fall into the collapsed bucket greatly reducing the overall accuracy obtained \cite{uddsketch}. Specifically, the following proposition proved in \cite{Masson} specifies the necessary requirement for a quantile to be $\alpha$-accurate.

\begin{proposition}
\textsc{DDSKETCH} can $\alpha$-accurately answer a given $q$-quantile query if $x_1 \leq x_q \gamma^{m-1}$.
\end{proposition}

\subsection{Our \textsc{UDDSketch} algorithm}

Our deterministic algorithm \textsc{UDDSketch} \cite{uddsketch} is based on \textsc{DDSketch}. Therefore, the algorithm also allows removing items (i.e. it allows items with negative weight), thus its underlying model is the turnstile one.

\textsc{UDDSketch} improves upon its predecessor's accuracy by employing a novel and more precise collapse procedure. In particular, \textsc{UDDSketch} employs a uniform collapsing approach that provides far greater accuracy than \textsc{DDSketch}. Specifically, our algorithm provides a $\alpha$-accurate $(q_0, q_1)$-sketch for $q_0 = 0$ and $q_1 = 1$. As a consequence, the \textsc{UDDSketch} summary can answer any quantile query with $\alpha$ accuracy. The uniform collapsing is based on collapsing each bucket pair by pair. Given the indices $(i,  i+1)$ in which $i$ is an odd index and $B_i \neq 0$ or $B_{i+1} \neq 0$, we insert into the summary a new bucket whose index is $j = \lceil \frac{i}{2} \rceil$, with a counter value equal to the sum of $B_i$ and $B_{i+1}$. The new bucket $B_j$ replaces the collapsed buckets $B_i$ and $B_{i+1}$. The pseudocode for the \textsc{UniformCollapse} procedure is shown as Algorithm \ref{uddsketch-uniform-collapse}.

\begin{algorithm}
\caption{UniformCollapse($\mathcal{S}$)}
	\label{uddsketch-uniform-collapse}
 \begin{algorithmic}
	\Require {sketch $\mathcal{S} = \lbrace B_i \rbrace_i$}
	\ForAll{ $\lbrace i: B_i > 0 \rbrace$ }
		 \State $j \leftarrow  \lceil \frac{i}{2} \rceil$
		 \State $B'_{j} \leftarrow B'_{j} + B_{i}$
	\EndFor
	\State \Return $\mathcal{S} \leftarrow \lbrace B'_i \rbrace_i$
\end{algorithmic}
\end{algorithm}

The following Lemma, which we proved in  \cite{uddsketch}, is related to the result of applying the collapsing procedure to an $\alpha$-accurate $(0,1)$-quantile sketch.

\begin{lemma}
	The collapsing procedure applied to an $\alpha$-accurate $(0,1)$-quantile sketch produces an $\alpha^\prime$-accurate $(0,1)$-quantile sketch on the same input data with ${\alpha^\prime = \frac{2\alpha}{1+\alpha^2}}$. Moreover, an item $x$ falling in the bucket with index $i$ of a collapsing sketch, will fall in the bucket with index $\lceil i/2\rceil$ of the collapsed sketch.
	\label{lemma:3.2}  
\end{lemma}

After execution of the \textsc{UniformCollapse} procedure, the theoretical error bound related to the sketch increases from $\alpha$ to $\alpha^\prime$, which represents the new theoretical error bound. This means that we lose accuracy each time we perform a collapse. Anyway, it's easy to understand that, in practice, the collapsing procedure is not executed repeatedly so many times as to negatively impact the accuracy of the data structure. Indeed, the probability of collapsing reduces each time the \textsc{UniformCollapse} procedure is invoked, owing to the fact that collapsing the buckets also implies increasing the input interval which is actually covered by the $m$ available buckets; as a consequence, a few collapsing are enough to process input data streams with very large range of values. The reason for this is that, after a collapse, $\gamma$ increases to $\gamma\prime = \gamma^2$ and the interval covered by the $i$-th bucket $B_i$ becomes $(\gamma\prime^{i-1}, \gamma\prime^i]$. Correspondingly, $\alpha$ increases to $\alpha^\prime = \frac{\gamma^{\prime}-1}{\gamma^{\prime}+1}=\frac{\gamma^2-1}{\gamma^2+1}=\frac{2 \alpha}{1+\alpha^2}$.

In addition to this, we offer a theoretical bound on the level of accuracy that can be attained by \textsc{UDDSketch}. The next theorem, which we proved in \cite{uddsketch}, provides the details \footnote{In Theorem 3 of \cite{uddsketch} the value of $\tilde\gamma$ is incorrectly stated as $\sqrt[m]{\frac{x_{max}}{x_{min}}}$. \label{typo}}.

\begin{theorem}
\label{errorbound}
Assuming without loss of generality that the input items are drawn from a domain represented by the interval $[x_{min}, x_{max}] \in \mathbb R_{>0}$ and a space bound of $m$ buckets, the approximation error committed by \textsc{UDDSketch} when using the uniform collapse procedure is bounded by $\hat\alpha = \frac{\tilde\gamma^2-1}{\tilde\gamma^2+1}$, where $\tilde\gamma = \sqrt[m-1]{\frac{x_{max}}{x_{min}}}$. 
\end{theorem}

\section{Distributed setting}
\label{distributed}

A distributed message-passing system consists of a collection of processes, which by assumption collaborate toward a shared objective. In order to accomplish this objective, they may communicate only by exchanging \textit{messages}. As computing occurs separately on each node, adding additional nodes and functionality is often simple and affordable. Consequently, the system is efficient and generally unaffected by the failure of a single machine. 

Typically, the set of processes is expected to be static (however, the set of processes may be dynamic as well under different circumstances). Let $V = \{p_1,p_2,\dots,p_p\}$ be a set of $p$ different processes. Each process $p_i$ is assumed to be sequential. Communication among the processors happens by transmitting and receiving messages via a collection of \textit{channels} $E$. Each channel $e = (p_i, p_j)$ is reliable (it cannot produce, change, lose, or duplicate messages), bidirectional, and has an infinite capacity (so that it can deliver an arbitrary number of messages, each of any size). 

A distributed system is modeled by a connected undirected graph ${G = (V, E)}$. Given a process $p_i$, let $neighbours_i$ be the set of its neighbours: $neighbours_i = \{p_j: (p_i, p_j) \in E \}$. A \textit{distributed algorithm} is a collection of $n$ automata, one for each process. Formally, an automaton is a description of a finite state machine and its transitions, which correspond to the sequence of steps carried out by the  process modeled by the automaton. In addition, an automaton may transmit and receive messages on any channel.

A distributed \textit{synchronous} algorithm is one designed for execution on a synchronous distributed system. Such a system is controlled by an external global clock, and the processes, which are equipped with local clocks, execute a sequence of \textit{rounds}. The duration of a round corresponds to a tick of the external global clock. In synchronous distributed systems the relative speed of process execution, message transmission delays, and the drift rates of local clocks relative to the global clock have, by assumption, a known upper bound. Throughout a round, a process may send to each of its neighbours no more than one message. In synchronous systems, a message delivered by a process during a round $r$ is received during the same round $r$ by its target process. Therefore, by the time a process reaches round $r + 1$, it has received and already processed all of the  message sent to it during the round $r$.

The gossip-based protocol, described in \cite{Demers:1987}, is an example of a synchronous distributed algorithm that is made up of periodic rounds. A process, which in this context is referred to as a \textit{peer} or \textit{agent}, will choose one or more of its neighbours at random in each of the rounds, then it will exchange its local state with those neighbours, and will eventually update its local state. The information is spread across the network through the use of one of the following potential communication methods: (i) \textit{push}, (ii) \textit{pull} or (iii) \textit{push--pull}. The primary distinction between push and pull is that in push communication a peer will select peers at random to whom it will send its local state, whereas in pull, a peer will select peers at random from whom it will receive the local state. Last but not least, in the hybrid push-pull communication method a peer chooses at random both the peers to transmit the local state to and the peers from whom to receive the state. In this synchronous distributed model, it is assumed that updating the local state of a peer can be done in constant time, that is, with a worst-case time complexity of $O(1)$; furthermore, the duration of a round is set up in such a way that each peer can finish a push-pull communication within the round's allotted time frame. The effectiveness and performance of a protocol that is based on gossip can be evaluated by counting the number of rounds necessary to reach convergence.

Our distributed algorithm shall be based on the gossip-based protocol known as \textit{distributed averaging}, which can be thought of as a \textit{consensus} protocol. For the purpose of our theoretical analysis, we will make the assumption that neither peers nor communication links will ever fail, and that neither new peers nor existing peers will ever be able to join or leave the network (this is referred to as \textit{churning}). As a result, the graph denoted by $G = (V, E)$, in which $V$ represents the group of peers and $E$ models the available channels (thus completely characterizing the underlying network topology), does not change over time. However, it is important to point out that our distributed algorithm also works in time-varying graphs, in which the network can change as a result of failures or churning, and we shall show (through specific experiments carried out) evidence of that in the Section \ref{results}, in which we discuss the experimental results obtained, including those related to the effect of churning.

In order to simplify the notation, we shall refer to the process $p_i$ and its corresponding peer by using the index $i$. Therefore, given two peers $i$ and $j$ the notation $(i,j)$ stands for the channel connecting them. Regarding communication, in the so-called \textit{uniform gossiping} protocol a peer $i$ is allowed to select, randomly, another peer $j$ and to exchange information sending to it a message. Instead, communication among peers is restricted to neighbour peers in our scenario, i.e., the interaction between two arbitrary peers $i$ and $j$ is allowed only if the corresponding channel $(i,j)$ is available: it must hold that $(i,j) \in E$. In the sequel, we shall assume that if $(i,j) \in E$ then it also holds that $(j,i) \in E$, i.e. channels are bidirectional. At the beginning of the distributed averaging protocol, each peer $i$ is assigned or computes a real number $v_i$ and the aim is to let each peer compute in a distributed fashion the average $v_{{\rm avg}} = \frac{1}{p}\sum_{i=1}^p v_i$. However, in our stricter model the information held by a peer can be exchanged solely with its neighbours.  

Denoting by $v_i(r)$ the estimate of $v_{{\rm avg}}$ at round $r$ held by peer $i$, two peers $i$ and $j$ engaging in a gossip interaction exchange and update their estimates, therefore $v_{i}(r+1)=v_j(r+1) = \frac{1}{2}(v_i(r)+v_j(r))$ at round $r+1$. If a peer $i$ is not involved in any gossip interaction, then $v_i(r+1) = v_i(r)$ at round $r+1$.  It has been proved that distributed averaging reaches the desired value $v_{{\rm avg}}$ exponentially quickly. In the worst case, a peer is often only permitted to gossip with one other peer at a time. Our approach gives each peer the option to gossip with a user's defined number of neighbours. The number of neighbours a peer $i$ communicates with in a round is what we refer to as its  \textit{fan-out} ; as a result, ${1 \leq \text{\textit{fan-out}} \leq \left\vert \{j: (i, j) \in E\} \right\vert}$. Consequently, we permit two or more pairs of peers to gossip simultaneously, with the restriction that the pairs do not share a peer. In the definition that follows, we formalize this idea.

\begin{definition}
	Two gossip pairs of peers $(i,j)$ and $(x,y)$
	are \textit{noninteracting} if neither $i$ nor $j$ equals either $x$ or $y$.
\end{definition}

Our approach allows the simultaneous occurrence of many noninteracting pairs of permissible gossips. In the literature, non-interactivity is also known as the \textit{atomic}  push-pull communication style; it is necessary to maintain and guarantee the correctness of the results. Consider two peers $i$ and $j$ at round $r$, the atomic push-pull requirement implies that a push message sent from $i$ to $j$ must be followed by the corresponding pull message form $j$ to $i$: in practice, $i$ is not allowed to receive in round $r$ a push message by another peer $k$ in the interval between the initial push message sent to $j$ and the receipt of the corresponding pull message from $j$. We also note here that 
%our protocol does not explicitly require identifying the peers, and we do so to make the analysis simpler. Nevertheless, 
we do assume that each peer can tell its neighbours apart.

\subsection{Jelasity's averaging algorithm}
\label{jelasity}

The proposed averaging protocol is understood as a distributed variance reduction algorithm in the work of Jelasity et al. \cite{Jelasity2005}. This protocol is equivalent to a centralized algorithm computing the average of a set of values. Imagine a network with $p$ nodes and a variance measure denoted by $\sigma_r^2$, which is defined as follows:

\begin{equation}
\label{eq_sigma2}
\sigma_r^2 = \frac{1}{p-1} \sum_{l=1}^{p}\left(w_{r,l} - \bar{w}\right)^2.
\end{equation}

In the previous equation $w_{r,l}$ represents the estimate held by a peer $l$ after the execution of the gossip protocol for $r$ rounds, whilst $\bar{w} = \frac{1}{p}\sum_{l=1}^p w_{0,l}$ is the true value of the average. According to \cite{Jelasity2005}, the following theorem is true if each pair of nodes is uncorrelated and $\psi_l$ is a random variable representing the frequency with which a node $l$ is selected as a member of the pair of nodes exchanging their states during a round of the protocol.

\begin{theorem}{}
	\label{jelasity-thm}
	 If:
	\begin{enumerate}
		\item the random variables $\psi_1, \ldots , \psi_p$ are identically distributed (let $\psi$ denote a random variable with this common distribution);
		\item after a pair of nodes $(i, j)$ is selected, the number of times the nodes $i$ and $j$ shall be selected again have identical distributions;
	\end{enumerate}
	then:
	\begin{equation}
	\label{eq_conv_factor}
	\mathbb{E}[\sigma_{r+1}^2] \approx \mathbb{E}[2^{-\psi}] \mathbb{E}[\sigma_{r}^2].
	\end{equation}
\end{theorem}
\vspace{2mm}

In the previous Theorem, $\psi$ is a  random variable which strictly depends on the selection of the pair of nodes.  Equation \eqref{eq_conv_factor} allows defining the convergence factor as follows:

\begin{equation}
\frac{\mathbb{E}[\sigma_{r+1}^2]}{ \mathbb{E}[\sigma_{r}^2]} = \mathbb{E}[2^{-\psi}];
\end{equation}

Owing to this, the convergence factor is dependent on $\psi$ and, as a direct result of this, it is dependent on the pair selection process. Jelasity et al. have computed the convergence factor for several techniques of pair selection. Nevertheless, we are only interested in the method that can simulate the protocol for distributed gossip-based averaging. Drawing a random permutation of the nodes is the first step in this method. Next, for each node in the permutation, a random node is selected so that a pair can be formed. The convergence factor for this method of selection is $\gamma = \mathbb{E}[2^{-\psi}] = 1/(2\sqrt{e})$.

In a previous work \cite{CAFARO20191}, starting from Theorem~\ref{jelasity-thm} we proved the following proposition, whose usefulness lies in the provision of the error bound incurred by the distributed gossiping protocol.

\begin{proposition}
	\label{prop1}
	Let $\delta$ be a user-defined probability, $w_{r,l}$ the value held by peer $l$ after $r$ rounds of the averaging protocol, $p$ the number of peers participating in the protocol, $C = \mathbb{E}[2^{-\psi}] = 1/(2\sqrt{e})$ the convergence factor and $\bar{w}$ the mean of the initial vector of values $\boldsymbol{w}_0$, i.e. $\bar{w} = 1/p \sum_{l=1}^{p} w_{0,l}$. Then, with probability $1 - \delta$ it holds that, for any peer $l$:
	\begin{equation}
	\label{eq_error_due_to_gossip}
	\left|w_{r,l} - \bar{w}\right| < \sqrt{(p-1) \sigma_0^2} \sqrt{\frac{C^r }{\delta}}
	\end{equation}
\end{proposition}

\section{Distributed \textsc{UDDSketch}}
\label{duddsketch}

In this Section we present our Distributed \textsc{UDDSketch} protocol for tracking quantiles. Each peer has a local \textsc{UDDSketch} summary to process its underlying local dataset mapping the items to a set of buckets $\{B_i\}_i$ with a predefined user accuracy $\alpha$ and sketch size $m$ using the sequential \textsc{UDDSketch} algorithm. Then, the distributed set of \textsc{UDDSketch} summaries is aggregated using our gossip-based protocol. The peers exchange, through gossip interactions with their neighbours, their local summaries; moreover, they also exchange the number of items summarized in the local sketch and the estimated number of peers in the P2P overlay network.

The initialization is done as shown in pseudocode by Algorithm \ref{ch5.algo.dudd.init}. Each peer $l$, $l=1,\ldots,p$ must process its own local dataset $\mathcal{D}_l$, which can be interpreted as being either an online stream or a partition of a global dataset $D=\bigcup_{l=1}^{p} \mathcal{D}_l$, such that the size of $\mathcal{D}_l$ is $N_l$ and the total dataset size is $N = |D| = \sum_{l=1}^{p} N_l$. As an example, each peer may be assigned either $\lceil N/p \rceil$ or $\lfloor N/p \rfloor$ elements. 

Of course, in the case of an online stream the value of $N_l$ is initially zero and is incremented by one each time a new item arrives from the input stream. Finally, the remaining parameters are those required by \textsc{UDDSketch}, i.e. the desired accuracy $\alpha$ and the bound on the maximum number of buckets to be used, $m$.

\begin{algorithm}
    \caption{Distributed \textsc{UDDSketch}: Initialization procedure} \label{ch5.algo.dudd.init}
   \begin{algorithmic}[1]
      \Procedure{Initialization}{$l$, $\mathcal{D}_l$, $\alpha$, $m$}
    
        \Require{$\alpha$, the sketch's accuracy, $m$ the maximum number of bins in the sketch, $\mathcal{D}_l$ the dataset to process}  
       
       \State $r \gets 0$  \Comment Initialization of peer $l$
        \If{$l == 1$}
          \State $\tilde{q}_{r,l} \gets 1;$
        
           \Else
              \State $\tilde{q}_{r,l} \gets 0;$
        \EndIf

       \State  $(\mathcal{S}_{r,l}, \tilde{N}_{r,l}) \gets \textsc{UDDSketch}(\mathcal{D}_l,\alpha,m)$
        \State $state_{r,l} \gets (S_{r,l}, \tilde{N}_{r,l}, \tilde{q}_{r,l})$
        
        \EndProcedure

      \Procedure {UDDSketch}{$\mathcal{D}$,$\alpha$,$m$}
            \State \textsc{Init}($S$);    
             \State  $N \gets 0;$ \Comment{$N$ counts the total number of processed items}
            \ForAll{$x_i \in \mathcal{D}$}
                 \State \textsc{UDDSketch-Insert}($x_i, \mathcal{S}$);
                 \State $N \gets N+1;$
            \EndFor
           \State \textbf {return} $(\mathcal{S}, N);$
        \EndProcedure
 \end{algorithmic}
\end{algorithm}

To correctly initialize each peer, the round $r$ is set to zero. The protocol does not require explicitly $p$, the number of peers. Instead, this number can be estimated during the execution. For this purpose, each peer sets $\tilde{q}_{r,l}$, which represents its estimate of the network size: in particular, this value shall converge to $1/p$, from which the value of $p$ can be easily obtained. Estimating $p$ requires that one peer, peer $l=1$ in our case, sets $\tilde{q}_{r,l} = 1$. All of the remaining peers, instead, set $\tilde{q}_{r,l} = 0$. We remark here that $l$, the unique identifier of a peer ($1 \leq l \leq p$), is one of the parameters of the initialization procedure, so that the statements shown in lines 3 - 6 can be performed by each peer without the need for a separate leader election protocol to determine which peer starts with a value equal to 1 in order to estimate $p$.

Next, each peer invokes the sequential \textsc{UDDSketch} algorithm to process its local dataset $\mathcal{D}_l$ of size $N_l$. The state $state_{r,l}$ of a peer at round $r$ includes the peer summary $\mathcal{S}_l$, the estimate of the average number of items $\tilde{N}_{r,l}$ (which estimates the value $\overline{N} = \frac{1}{p}\sum_{l=1}^{p} N_l$, and is set initially to $N_l$) and the estimate $\tilde{q}_{r,l}$. When the protocol terminates, with high probability it holds that  $\tilde{q}_{r,l} \approx 1/p,\ \forall l$. Gossiping is shown as pseudo-code in Algorithm \ref{ch5.algo.dudd.gossip}.

\begin{algorithm}
    \caption{Distributed \textsc{UDDSketch}: Gossip procedure} \label{ch5.algo.dudd.gossip}
  \begin{algorithmic}[1]
    \Procedure{Gossip}{}
        \For{$r$ = 0 to $R$}
            \State $neighbours \gets$ select $fan-out$ random neighbours;
		\ForAll{$i \in neighbours$}
                      \State \textsc{send}($push,i,state_{r,l}$);
	            \EndFor
	\EndFor
     \EndProcedure   

    \Procedure{OnReceive}{$msg$}
        \State $type \gets msg.type$;
        \State $j \gets msg.sender$;
        \State $state \gets msg.state$;
        
        \If{type == push}
            \State $state_{r+1,l} \gets \textsc{update}(state,state_{r,l})$;
            \State \textsc{send}($pull,j,state_{r+1,l}$);
        \EndIf
        
        \If{type == pull}
            \State $state_{r+1,l} \gets state$;
        \EndIf

    \EndProcedure

        \Procedure{Update}{$state_j,state_l$}

            \State $(\mathcal{S}_l, \tilde{N}_l, \tilde{q_l}) \gets state_l$;
            \State $(\mathcal{S}_j, \tilde{N}_j, \tilde{q_j}) \gets state_j$;
            
            \State $\mathcal{S} \gets \textsc{Merge}(\mathcal{S}_l,\mathcal{S}_j)$;

            \State $\tilde{N} \gets \frac{\tilde{N}_l+\tilde{N}_j}{2}$;
            \State $\tilde{q} \gets \frac{\tilde{q}_l+\tilde{q}_j}{2}$;
            \State $state \gets (\mathcal{S}, \tilde{N}, \tilde{q})$;
            \State \textbf{return} $state$;

 \EndProcedure
\end{algorithmic}
\end{algorithm}

The peer $l$ exchanges its current local state with its neighbors via a \emph{push} message. When a message is received, the peer $l$ extracts the sender, message type, and remote state information. If the message is a \emph{pull} communication, then $l$ updates its local state by setting it to the remote state received; on the contrary, if the message is a \emph{push} communication, $l$ updates the local state using the remote state (via an averaging procedure to be discussed later), and then sends back a \emph{pull} message with the updated state. Averaging, shown in pseudo-code as the \textsc{Update} procedure, combines the two summaries $S_{r,l}$ and $S_{r,j}$ by merging them into a new summary $S$.

The \textsc{Merge} procedure, shown in pseudo-code as Algorithm \ref{ch5.algo.merge}, simply scans the two input summaries computing the average values of corresponding buckets. When the scan terminates, we may be left we more than $m$ buckets in the merged summary. In this case, we collapse the merged summary $S$ by invoking the \textsc{UniformCollapse} procedure. As stated in Section \ref{distributed}, in the synchronous distributed model it is assumed that the worst-case time complexity of updating the local state of a peer is $O(1)$, so that the only relevant performance metric is the number of rounds required to converge. Since the number of buckets $m$ is a constant, the worst-case complexity of the \textsc{Merge} procedure is $O(1)$ and the assumption is respected anyway - even though this is not strictly required by the distributed model.

It is worth recalling here that merging two summaries require their $\alpha$ values to be equal. If this condition is not verified, the summary whose $\alpha$ value is the smallest is repeatedly collapsed until the condition is met. Moreover, if collapsing is required to maintain the space bound, the $\alpha$ value of the merged summary $S$ may be greater than the $\alpha$ values of the input summaries.

\begin{algorithm}
    \caption{Distributed \textsc{UDDSketch}: \textsc{Merge}($\mathcal{S}_1, \mathcal{S}_2$)}
	\label{ch5.algo.merge}
\begin{algorithmic}[1]
    \Require{$\mathcal{S}_1 = \lbrace B^1_i \rbrace_i, \mathcal{S}_2 = \lbrace B^2_j \rbrace_j$: sketches to be merged}; the $\alpha$ values of $\mathcal{S}_1$ and $\mathcal{S}_2$ must be equal
    \Ensure{$\mathcal{S}_M \leftarrow \lbrace B^M_k \rbrace_k$: merged sketch}
    \State \textsc{Init}($\mathcal{S}_M$);
     \For{\textbf{each}$\lbrace i: B^1_i > 0 \lor B^2_i > 0 \rbrace$ }
        \State $B^M_i \gets \frac{B^1_i + B^2_i}{2} $;
    \EndFor
    \While{$|\mathcal{S}_M| > m$}
        \State \textsc{$UniformCollapse(\mathcal{S}_M)$};
    \EndWhile

    \State{return} {$\mathcal{S}_M$};
\end{algorithmic}
\end{algorithm}

When the gossip-based protocol ends, it is possible to issue any quantile query to an arbitrary peer, since they all converged to the same summary. Thus, the distributed algorithm allows computing an estimated $\tilde{x}_q$ over the whole dataset $D$ distributed among the peers. Algorithm \ref{duddsketch-query} reports the query procedure.

%TODO insert query pseudocode and comment.
\begin{algorithm}
	\caption{Distributed \textsc{UDDSketch}: \textsc{Query}($q$)}
		\label{duddsketch-query}
	 \begin{algorithmic}
		\Require{$state_l$: status of queried peer $l$}
		\Ensure{$\tilde{x}_q$: estimated quantile value}
		\State $(\lbrace \tilde{B}_i \rbrace_i, \tilde{N}_l, \tilde{q}_l) \gets state_l$;
		\State $\tilde{p} \leftarrow \lceil 1/\tilde{q}_l \rceil$
		\State $\tilde{N} \leftarrow \lceil \tilde{p}  \tilde{N}_l \rceil$
		\State $i_0 \leftarrow \min{(\{j: \tilde{B}_j > 0\})}$
		\State $count \leftarrow \lceil \tilde{B}_{i_0} * \tilde{p} \rceil$
		\While{ $count \leq \lfloor 1 + q(\tilde{N}-1) \rfloor$} 
			 \State $i \leftarrow  \min{(\{j: \tilde{B}_j > 0 \land j > i\})}$
			 \State $count \leftarrow count + \lceil \tilde{B}_{i} * \tilde{p} \rceil$
		\EndWhile
		\State  $\tilde{x}_q \leftarrow 2 \gamma^i/(\gamma + 1)$
		\State \Return $\tilde{x}_q$
	\end{algorithmic}
\end{algorithm}

\section{Correctness}
\label{correctness}
In this Section, we shall show that each peer's loacal sketch converges to the sketch we would obtain processing the union of the peer's local stream by a sequential UDDSketch with same initial parameters $\alpha$ and $m$. In the following, we will refer to this sketch as the global sketch.

From Lemma 1 in \cite{CAFARO2023155}, we already know that UDDSketch is permutation invariant with regard to insertion-only streams, i.e., it produces the same sketch regardless of the order in which the input items are inserted or the order in which the collapses are executed. Therefore, without loss of generality, we can assume that all of the collapsing operations are postponed until the end of the gossip phase
%until we obtain the final sketch 
and simply show that when each peer processes its local stream with a common initial value of $\alpha$ and a sketch whose size can grow unbounded, during the gossip phase the peers' local sketches will converge to the sketch we would obtain processing the union of the peer's local streams by a sequential UDDSketch with the same initial $\alpha$ and sketch size unbounded. 
%After convergence the two sketches will remain the same also after the collapsing operations needed to bound their size. 

In this setting, the merge operation reduces to the sum of counters with same index and the sequential global sketch is equal to the sum of the  local sketches at round zero, before engaging in the gossip phase. 

We will represent a sketch with minimum bucket index $i_m$ and maximum bucket index $i_M$, as a set of $i_M-i_m+1$ counters with contiguous indices from $i_m$ to $i_M$, leaving equal to zero the counters of empty buckets.

Let $\mathcal{S} = \{B_{i}\}^{i_M}_{i_m}$ be the sketch produced by a sequential UDDSketch on the union of the local streams and denote by $\tilde{\mathcal{S}}_{l,r} = \{\tilde{B}_{l,i,r}\}^{i_M}_{i_m}$ the sketch held by peer $l$ at round $r$. We have that 
$$B_{i} = \sum_{l=1}^{p} \tilde{B}_{l,i,0} \text{ for } i = i_m, \ldots, i_M.$$

 Recalling that the averaging protocol is applied to all of the counters in the sketches exchanged by two peers, from eq. \eqref{eq_error_due_to_gossip} we have that at each round $r$ and for each peer $l$ the following relation holds:

\begin{equation}
\label{counter-bound}
\left|\tilde{B}_{l,i,r} - \bar{S_i}\right| < \sqrt{(p-1) \sigma_0^2} \sqrt{\frac{C^r }{\delta}}\end{equation}

\noindent where $\bar{S_i}= \frac{1}{p} \sum_{l=1}^p \tilde{B}_{l,i,0}$, for $i = i_m, \ldots, i_M$ is the average of the buckets with index $i$ in the sketch at round zero, i.e. after each peer processes its local stream through \textsc{UDDSketch}.

From eq. \eqref{eq_error_due_to_gossip} we can also derive similar relations for the value $\tilde{q}_{l,r}$ which approximates $1/p$, and the value $\tilde{N}_{l,r}$ which estimates $\bar{N} = \sum_{l=1}^p \tilde{N}_{l,0}$, that is the average length of the local input streams.

Therefore, from Theorem \ref{jelasity-thm} and Proposition \ref{prop1}, noticing that the term $C^r$ goes to zero when $r$ grows,  we have that for each peer $l$ and $r \rightarrow \infty$, it holds that $\tilde{B}_{l,i,r} \rightarrow \bar{S_i}$, $\tilde{q}_{l,r} \rightarrow 1/p$ and $\tilde{N}_{l,r} \rightarrow \bar{N}$. As a consequence, once the convergence is reached, each peer can successfully reconstruct the global sketch $\mathcal{S}$ and be able to answer a quantile query on the global dataset with the requested accuracy.

In the next section we shall show experimentally that in almost all of the cases considered, only a few rounds are required for our algorithm to converge.

\section{Experimental results}
\label{results}

We present and discuss the results of the experiments carried out for the distributed version of \textsc{UDDSketch}, showing that the local sketch of a peer engaged in the gossip-based distributed averaging protocol converges to the sketch that would be obtained by running \textsc{UDDSketch} on the union of the local datasets (or sub-streams) held by the peers.

We do not compare our algorithm versus others, since, to the best of our knowledge, no other published algorithm works exactly in the same setting (i.e. gossip based P2P quantile tracking protocol with relative value error).

The distributed protocol is based on a simulator which has been implemented in C++; the source code has been compiled using the GCC compiler v9.4 on Ubuntu Linux version 20.04.1 and is freely available\footnote{https://github.com/cafaro/DUDDSketch\label{fnsourcecode}}.  \\
The tests have been carried out on a machine equipped with two exa-core Intel Xeon-E5 2620 CPUs at 2.0 GHz and 64 GB of main memory. 

The experiments have been executed varying the number of peers, the number of rounds and three different churning models. Table \ref{params} reports the default settings for the parameters.

The tests consist in the comparison of the estimated quantiles computed by the distributed gossip-based protocol versus the corresponding estimated quantiles computed sequentially by \textsc{UDDSketch} processing the whole dataset $D = \cup_i D_i$, $\forall i=1,\ldots, p$; The set of quantiles in Table \ref{params} was evaluated and for each estimated quantile the Relative Error between the sequential algorithm estimation $\hat{x}_q$ and the value estimated by each peer through the distributed algorithm, $\tilde{x}_{q,i}$, has been computed.

The tests have been carried out over two different kinds of graphs, by using the Barab\'{a}si-Albert (denoted by $BA$) and the 
Erd\H{o}s-R\'{e}nyi (denoted by $ER$) random graph models (generated by  the iGraph Library version 0.7.1). In the Barab\'{a}si-Albert model undirected graphs with $p$ vertices are generated with both the power of preferential attachment and the constant attractiveness of vertices set to $1$, while the number of outgoing edges generated for each vertex is set to 5.
In the Erd\H{o}s-R\'{e}nyi model, undirected graphs with $p$ vertices are generated such that every possible edge is included in the graph with probability $\frac{10}{p}$.

The experiments showed no appreciable differences between the two random graph models with reference to the behaviour of the distributed \textsc{UDDSketch} algorithm, and in the following we report only the Barab\'{a}si-Albert random graph plots.

\subsection{Synthetic datasets}
\label{synthetic-datasets}

The tests have been performed on 4 synthetic datasets, whose properties are summarized in Table \ref{datasets}. The \textit{adversarial} dataset has been built as follows: data have been drawn uniformly at random in the range $(1, 10^2)$, peers have been partitioned in groups of at most one hundred peers, so that if two peers belong to different groups, they are assigned as input items falling into disjoint intervals, ensuring that they also fall into disjoint sets of sketch buckets. Therefore, at the end of the local computation the sketches of each group of peers will not have buckets in common, which is the worst case scenario for the distributed averaging consensus protocol (for additional details, please see the source code, footnote \ref{fnsourcecode} on page \pageref{fnsourcecode}). 
The \textit{uniform}, \textit{exponential} and \textit{normal} datasets were extracted from the respective distributions, the parameters of which were chosen independently and uniformly at random by each peer in the ranges specified for each parameter in Table \ref{datasets}.

\begin{table*}[h]
	\centering

	\begin{tabular}{@{}ll@{}}
		\textbf{Dataset} & \textbf{Distribution}\\
		\hline
		adversarial (see Section \ref{synthetic-datasets}) & $\textit{Uniform}(1, 10^2)$\\
		uniform &  $\textit{Uniform}([1, 10^5], [10^6, 10^7])$ \\
		exponential  & $\textit{Exp}([0.1, 3.5])$ \\
		normal & $\textit{N}([10^6, 10^7], [10^5, 10^6])$ \\
		\hline
	\end{tabular}
	\caption{Synthetic datasets} \label{datasets}
\end{table*}

\begin{table*}[h]
	\centering
	\begin{tabular}{@{}ll@{}}
		\textbf{Parameter} & \textbf{Values}\\
		\hline
		$\alpha$ &  $0.001$ \\
		quantiles &   $\{0.01,0.1, 0.2, 0.3, 0.4, 0.5, 0.6, 0.7, 0.8, 0.9, 0.99\}$ \\
		number of buckets  & $m=1024$ \\
		number of peers $P$ & $\{1000, 5000, 10000, 15000\}$\\
		number of rounds $R$ & $\{5, 10, 15, 20, 25\}$ \\
		fan-out & $1$ \\
		items/peer & $100000$ \\
		\hline
	\end{tabular}
	\caption{Default settings of the parameters} \label{params}
\end{table*}

The number of items that each peer processes in its local dataset is set to $100000$. 

\begin{figure*}[htb]
    \centering
    \begin{tabular}{cc}

		\subfloat[]{
		    \includegraphics[width=0.42\textwidth]{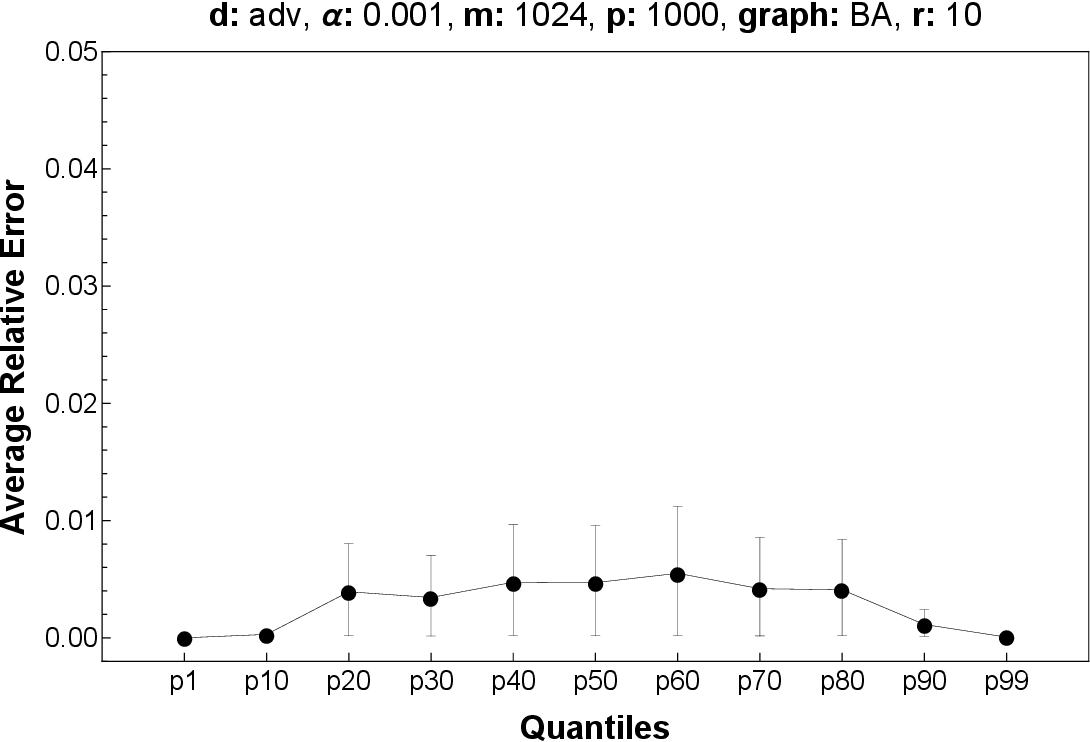}
		} &	
		
		\subfloat[]{
		    \includegraphics[width=0.42\textwidth]{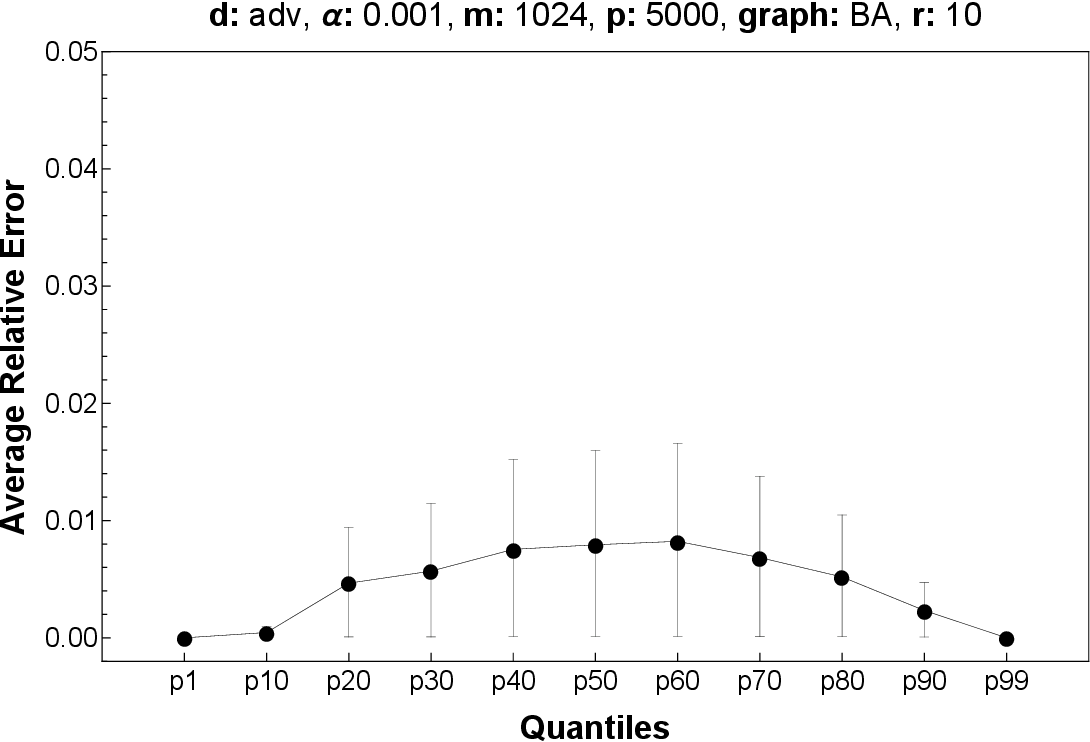}
		} \\	

        \subfloat[]{
		    \includegraphics[width=0.42\textwidth]{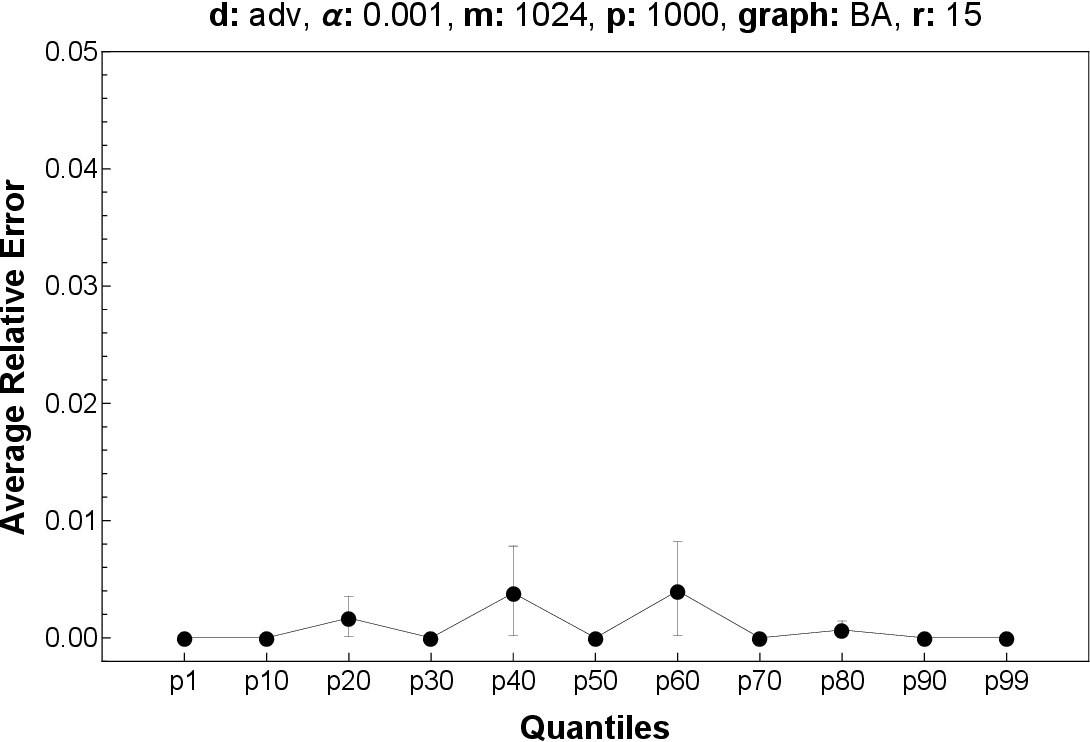}
		} &
		
		\subfloat[]{
		    \includegraphics[width=0.42\textwidth]{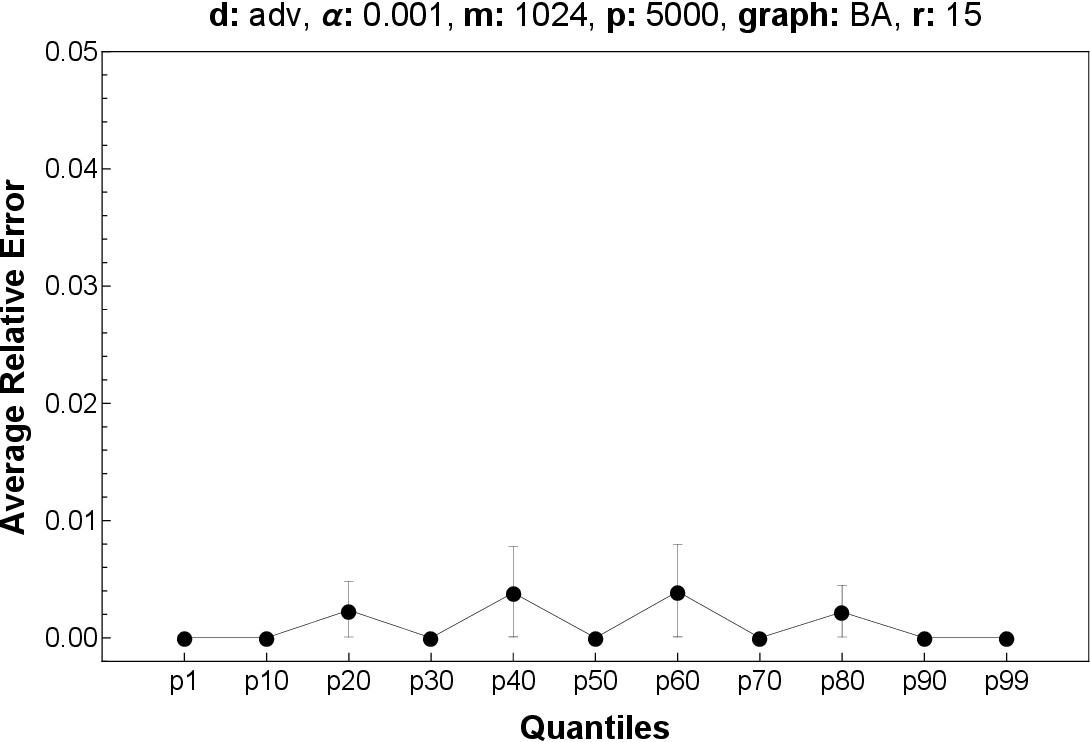}
		} \\

        \subfloat[]{
		    \includegraphics[width=0.42\textwidth]{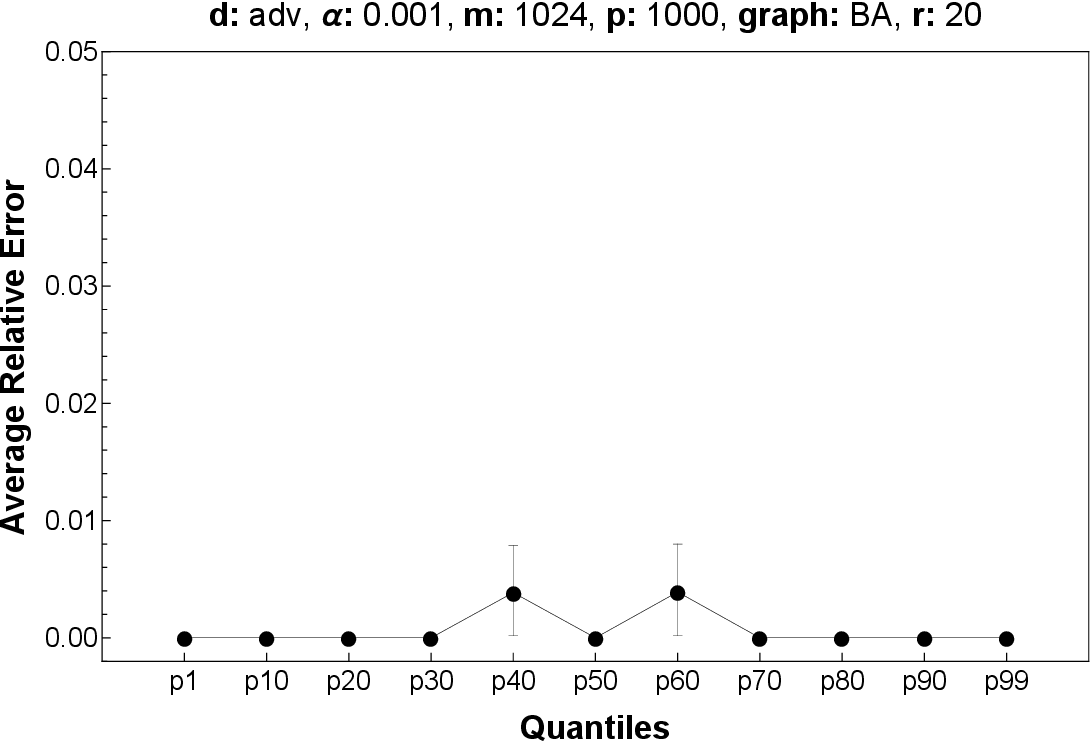}
		} &

		\subfloat[]{
		    \includegraphics[width=0.42\textwidth]{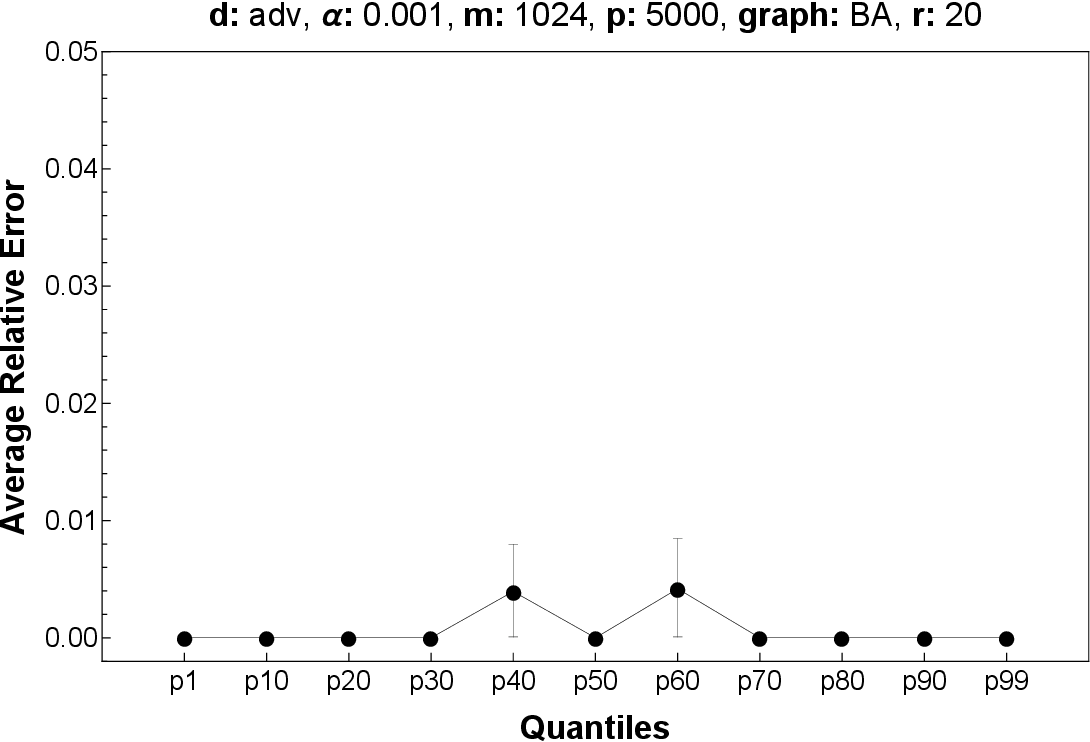}
		} \\
		
        \subfloat[]{
		    \includegraphics[width=0.42\textwidth]{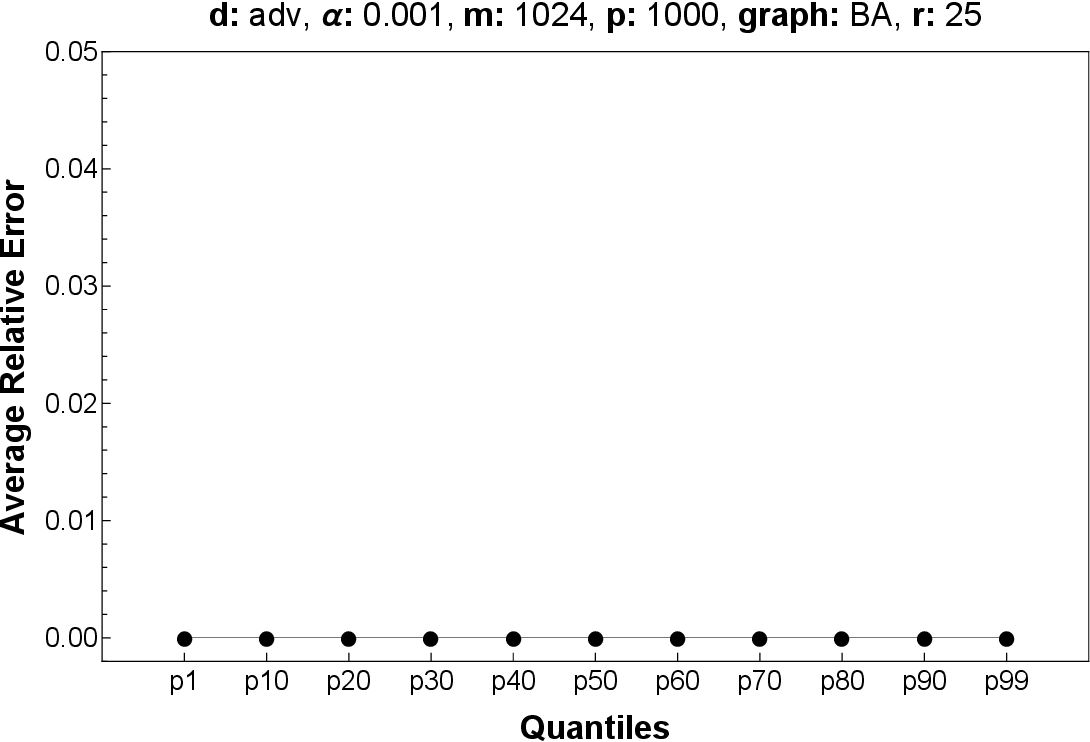}
		} &
		
        \subfloat[]{
		    \includegraphics[width=0.42\textwidth]{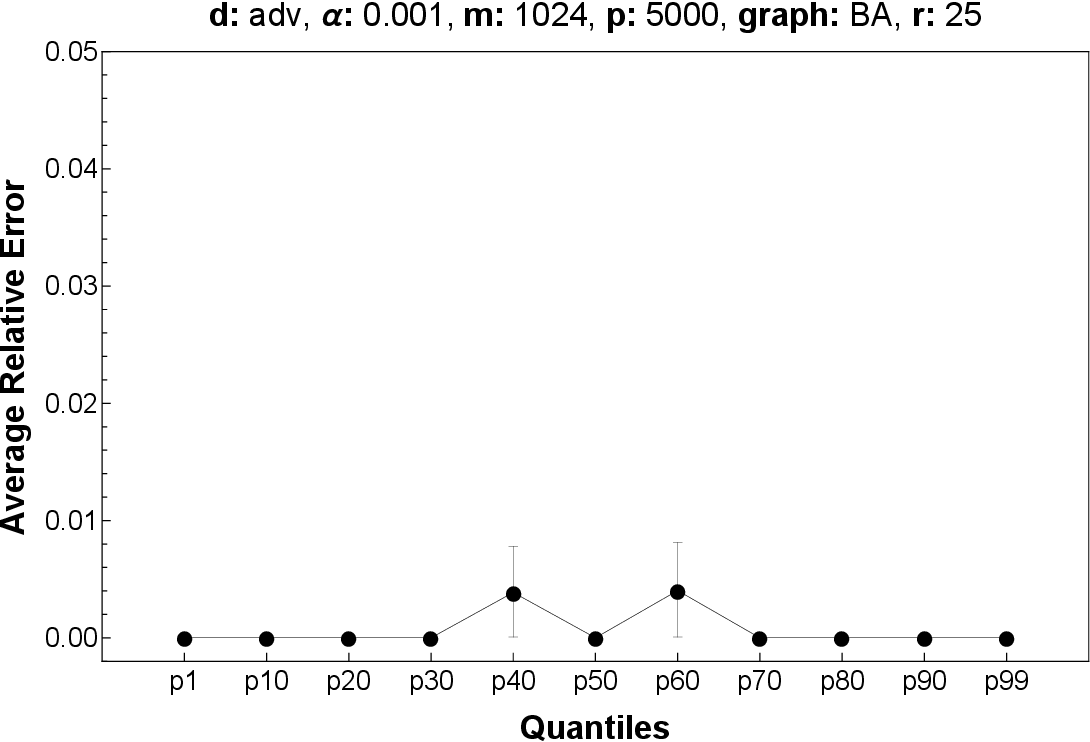}
		} \\

	\end{tabular}	
	\caption{Protocol convergence varying the number of rounds with 1000 peers (column on the left) and 5000 peers (column on the right) on a Bar\'abasi-Albert graph, with a sketch size of $1024$ buckets, initial $\alpha=0.001$ over the adversarial input.} 
	\label{fig.peers.ba.adv1}
\end{figure*}

\begin{figure*}[htb]
    \centering
    \begin{tabular}{cc}

		\subfloat[]{
		    \includegraphics[width=0.42\textwidth]{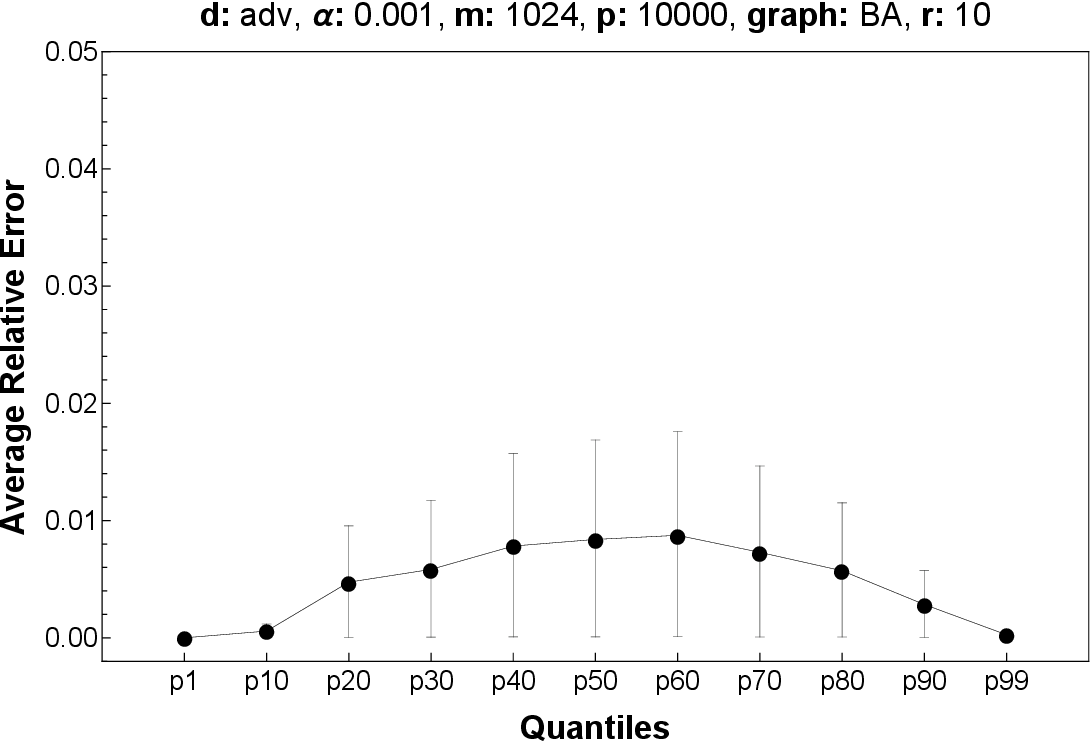}
		} &	
		
		\subfloat[]{
		    \includegraphics[width=0.42\textwidth]{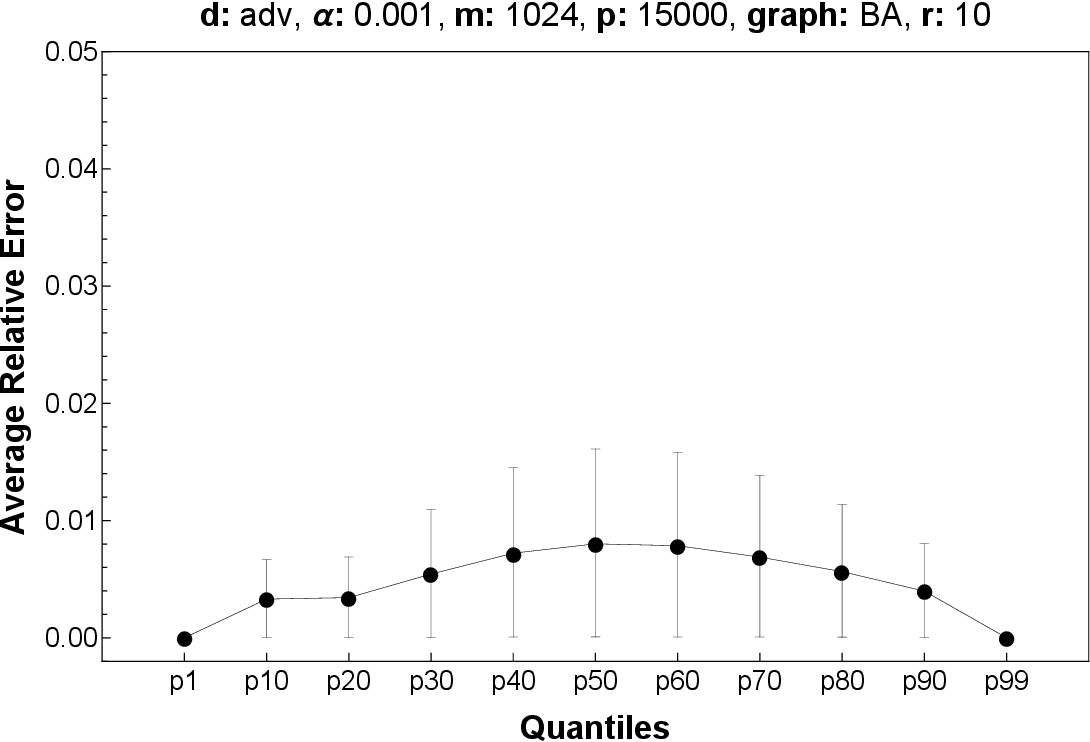}
		} \\	

        \subfloat[]{
		    \includegraphics[width=0.42\textwidth]{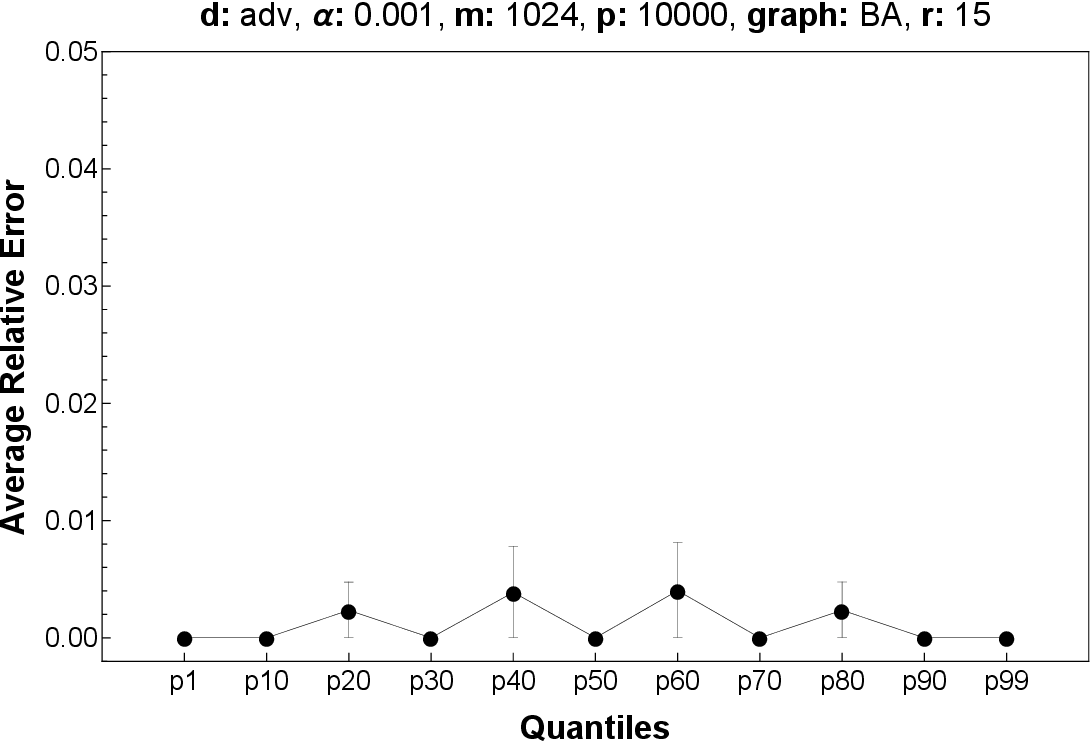}
		} &
		
		\subfloat[]{
		    \includegraphics[width=0.42\textwidth]{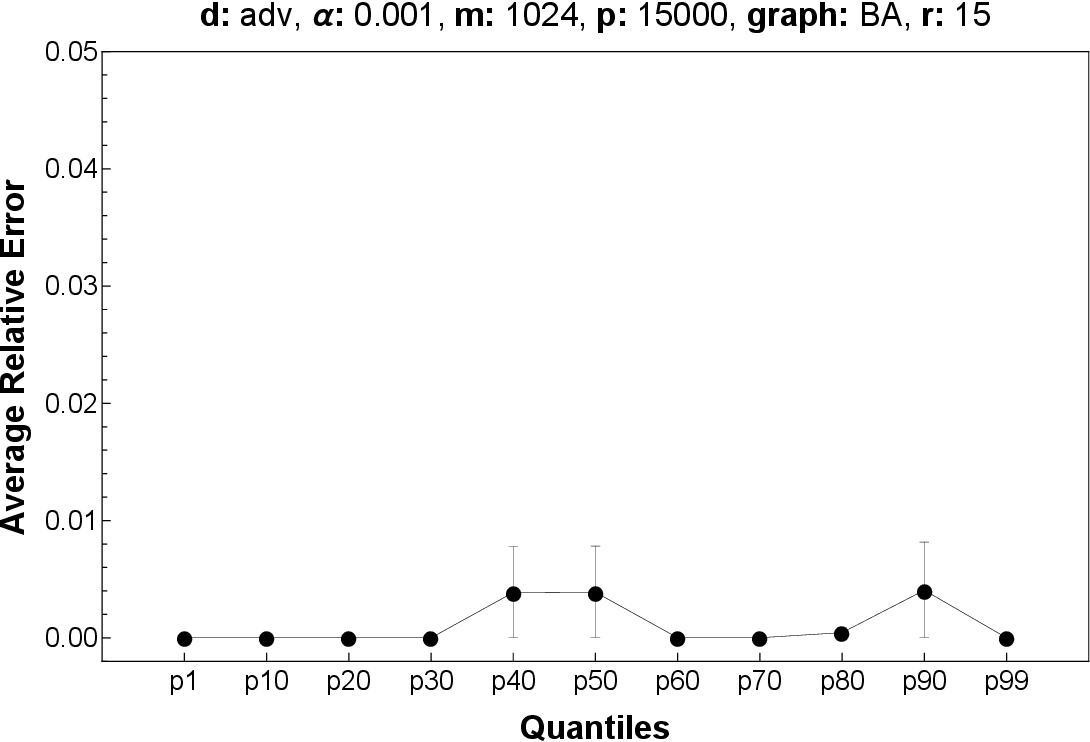}
		} \\

        \subfloat[]{
		    \includegraphics[width=0.42\textwidth]{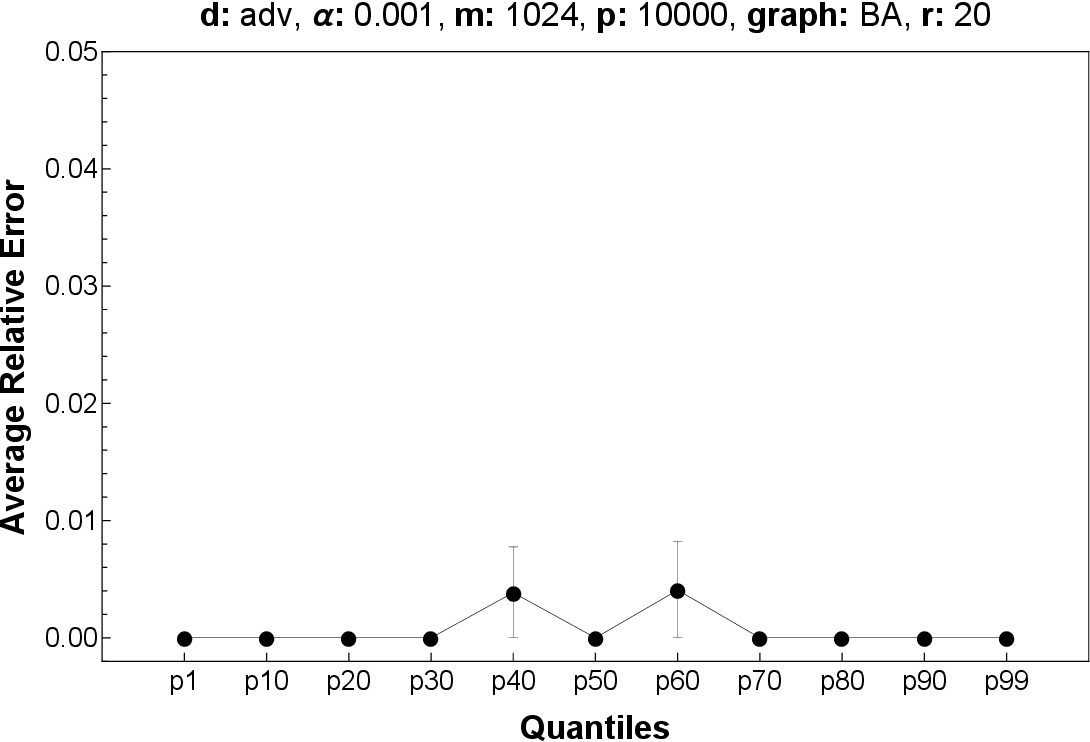}
		} &

		\subfloat[]{
		    \includegraphics[width=0.42\textwidth]{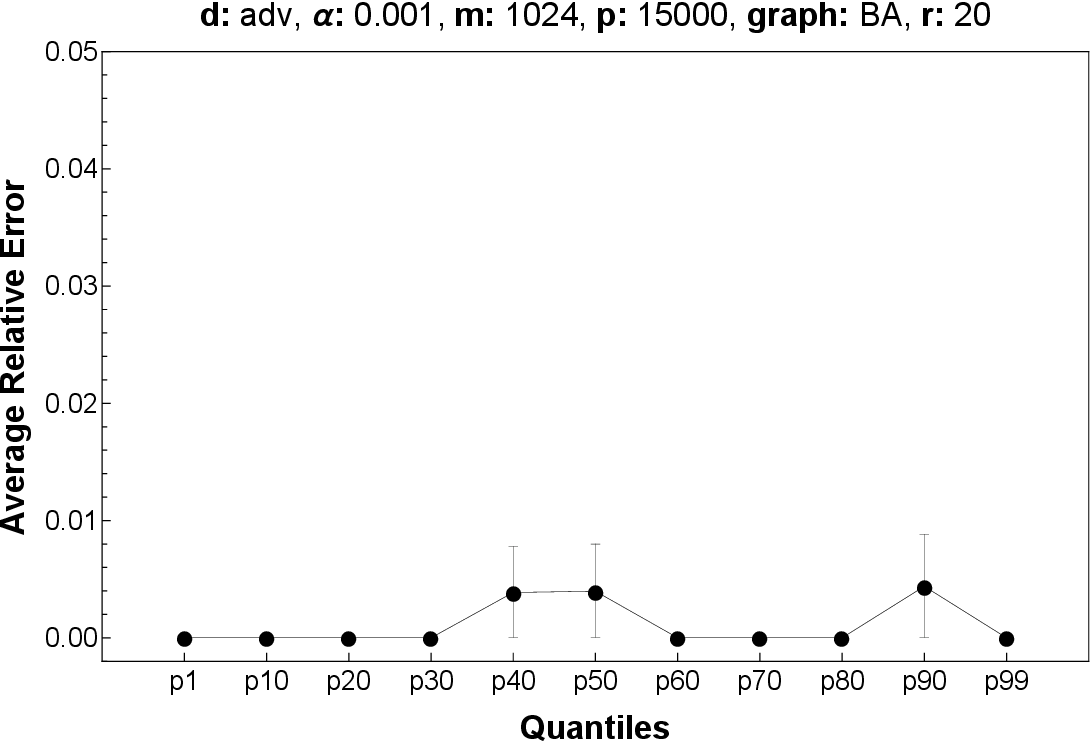}
		} \\
		
        \subfloat[]{
		    \includegraphics[width=0.42\textwidth]{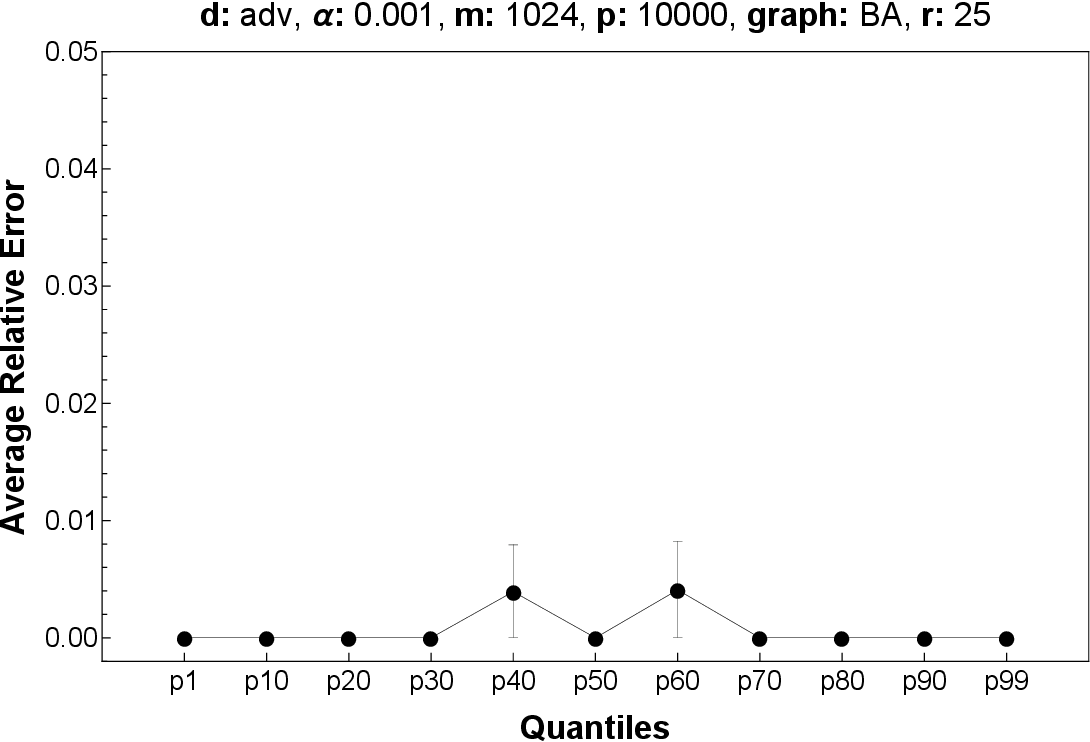}
		} &
		
        \subfloat[]{
		    \includegraphics[width=0.42\textwidth]{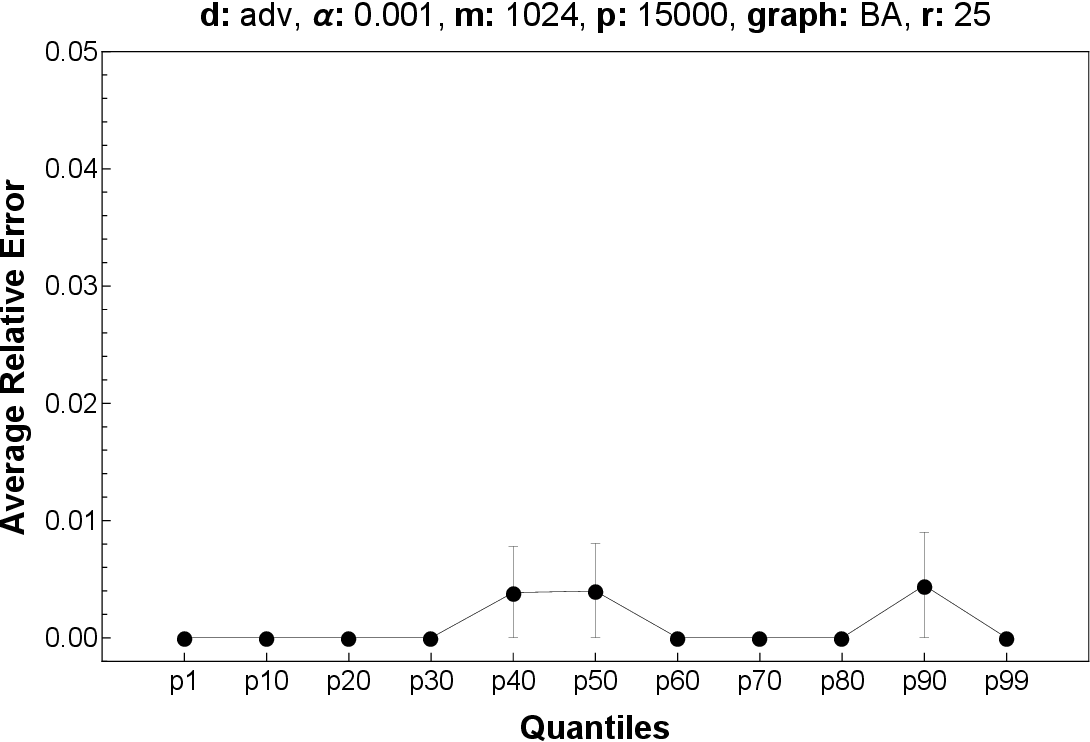}
		} \\

	\end{tabular}	
	\caption{Protocol convergence varying the number of rounds with 10000 peers (column on the left) and 15000 peers (column on the right) on a Bar\'abasi-Albert graph, with a sketch size of $1024$ buckets, initial $\alpha=0.001$ over the adversarial input.} 
	\label{fig.peers.ba.adv2}
\end{figure*}

Figures \ref{fig.peers.ba.adv1}-\ref{fig.peers.ba.adv2} highlight the convergence speed of the protocol when varying the number of peers in the network: tests have been carried out over networks of different sizes, running up to $25$ rounds. The sketches have been initialized with $\alpha=0.001$, the maximum number of buckets has been set to $m=1024$, the fan-out value used is $1$ and the network topology has been generated through a \textit{BA} model. \\
In particular, plots in each column of Figures \ref{fig.peers.ba.adv1}-\ref{fig.peers.ba.adv2} refer to different network sizes: $1000$ and $5000$ peers in Figure \ref{fig.peers.ba.adv1}, $10000$ and $15000$ peers in Figure \ref{fig.peers.ba.adv2}. In both figures, rows refer to the tests for $10$, $15$, $20$ and $25$ rounds respectively.

The box-and-whisker plots represent the distribution of relative errors committed by peers in the computation of quantile estimates compared to the estimates of the sequential algorithm. As shown, as soon as the number of rounds starts approaching those required for convergence, Relative Errors go to zero. For an \textit{adversarial} input, about $25$ rounds are enough to guarantee an Relative Error almost zero for all of the quantiles, independently of the number of peers.

\begin{figure*}[htb]
    \centering
    \begin{tabular}{cc}

		\subfloat[]{
		    \includegraphics[width=0.45\textwidth]{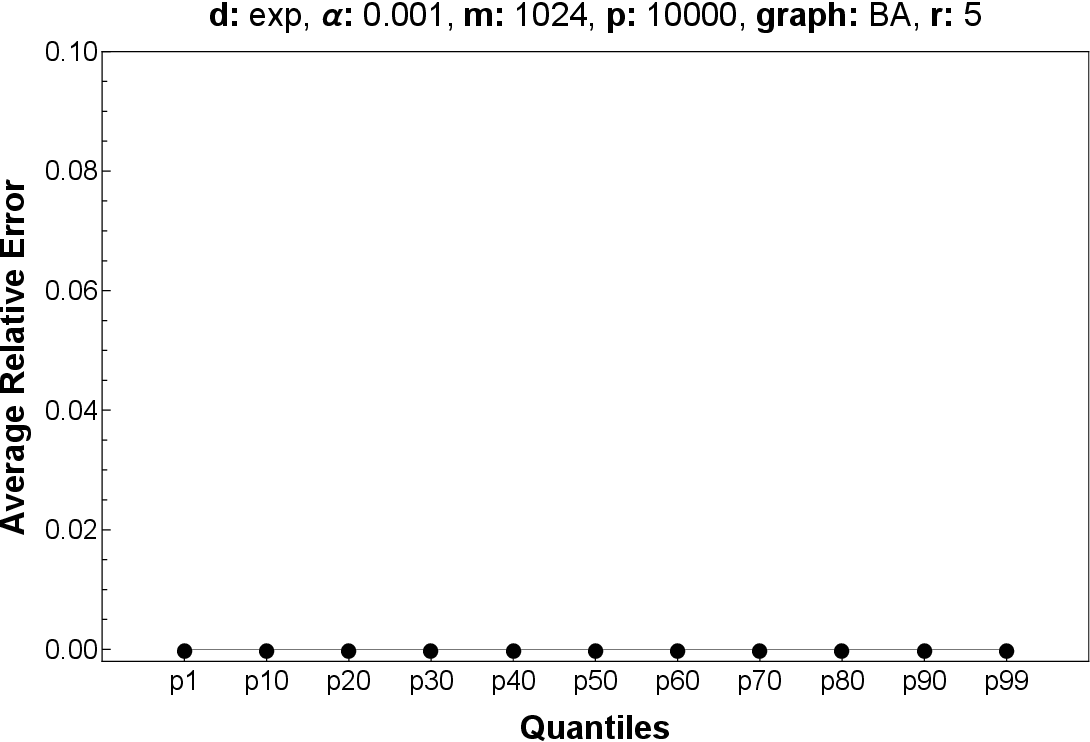}
		} &	
		
		\subfloat[]{
		    \includegraphics[width=0.45\textwidth]{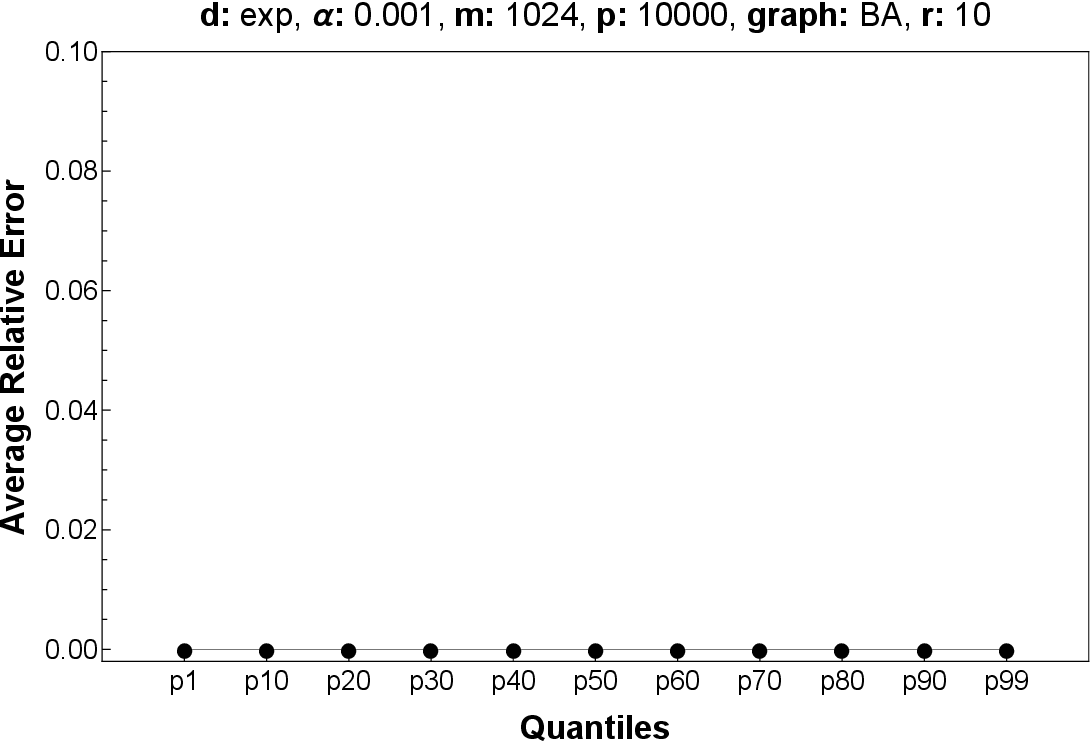}
		} \\	

        \subfloat[]{
		    \includegraphics[width=0.45\textwidth]{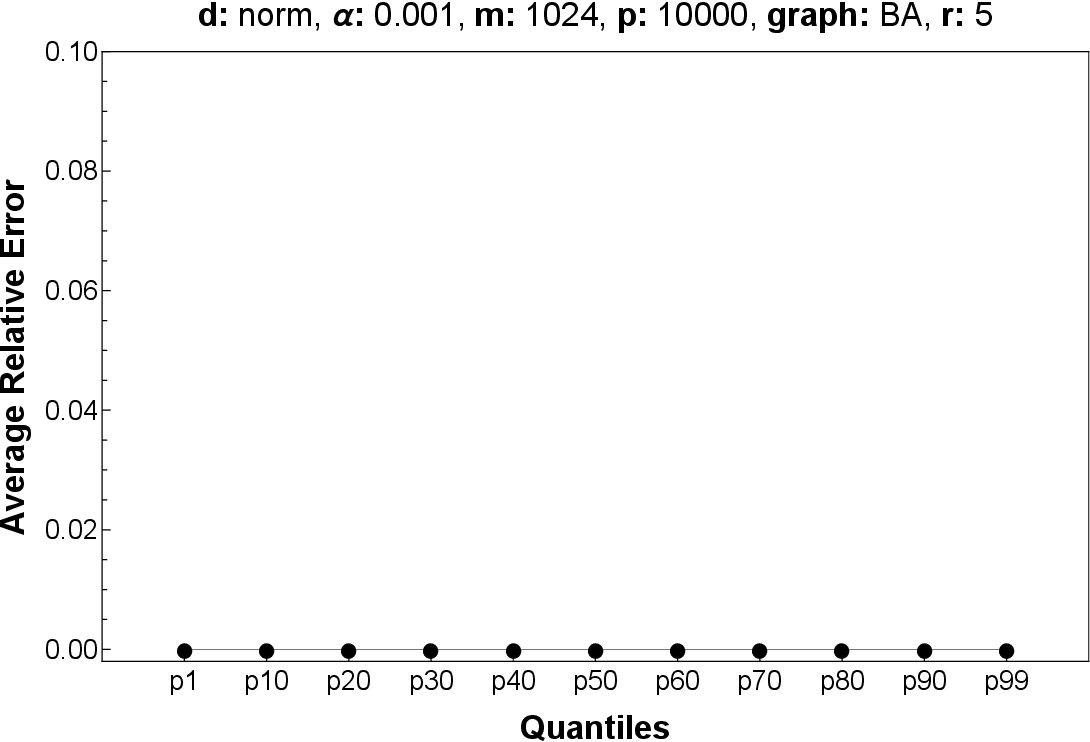}
		} &
		
		\subfloat[]{
		    \includegraphics[width=0.45\textwidth]{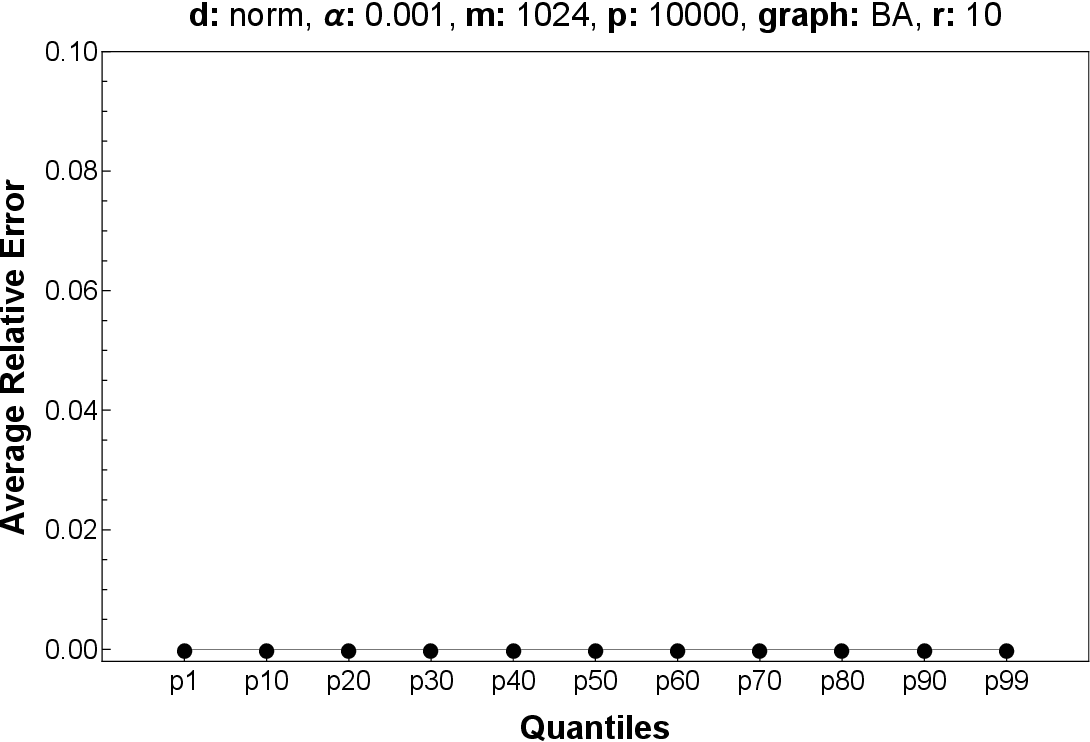}
		} \\

        \subfloat[]{
		    \includegraphics[width=0.45\textwidth]{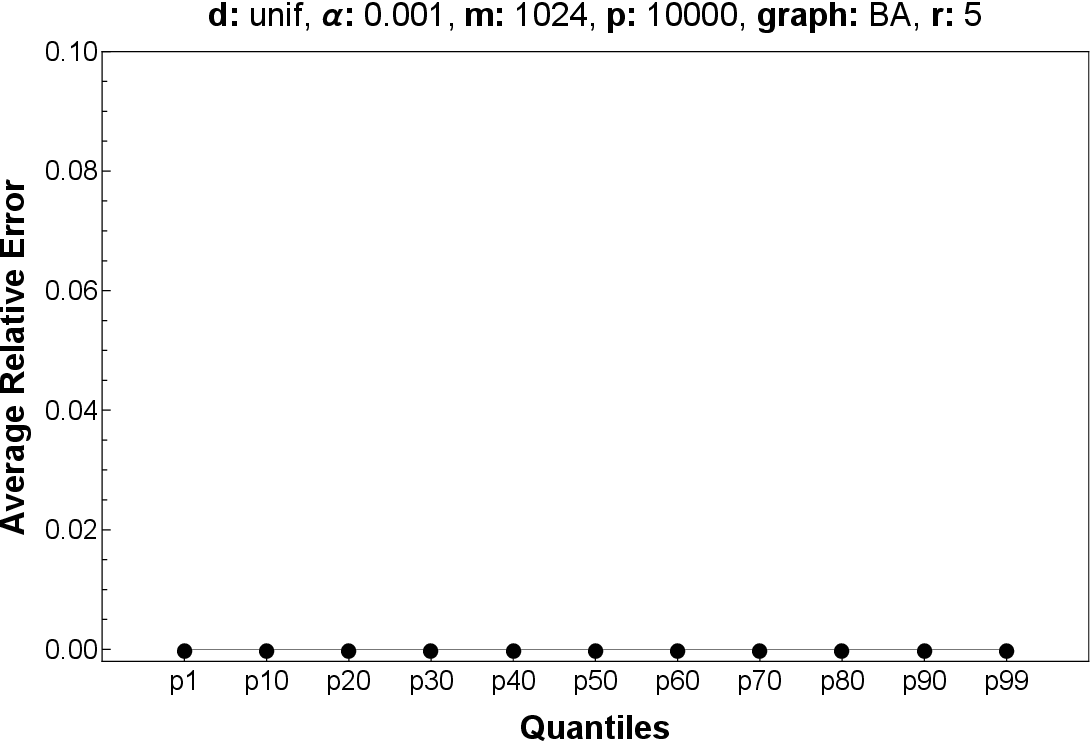}
		} &

		\subfloat[]{
		    \includegraphics[width=0.45\textwidth]{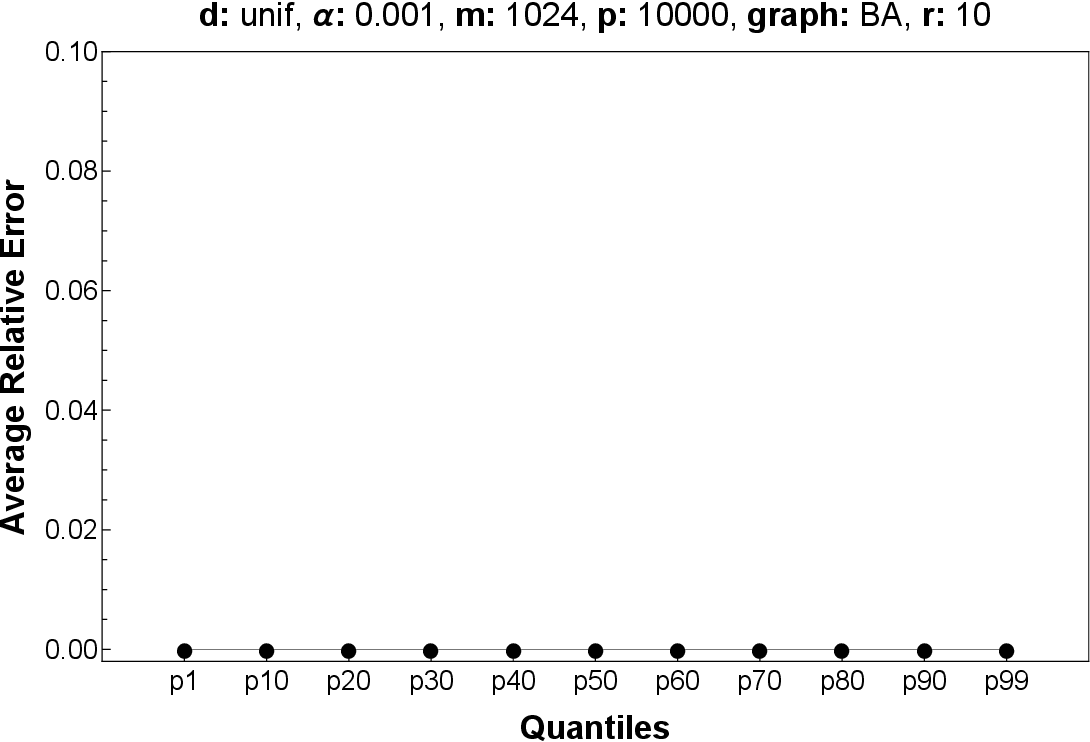}
		} \\

	\end{tabular}	
	\caption{Protocol convergence at 5 (left column) and 10 rounds (right column) for a network of 10000 peers on a Bar\'abasi-Albert graph, with a sketch size of $1024$ buckets, initial $\alpha=0.001$ over an exponential (a,b), normal (c,d) and uniform (e,f) inputs.} 
	\label{fig.peers.ba.oth1}
\end{figure*}

\begin{figure*}[htb]
    \centering
    \begin{tabular}{cc}

		\subfloat[]{
		    \includegraphics[width=0.45\textwidth]{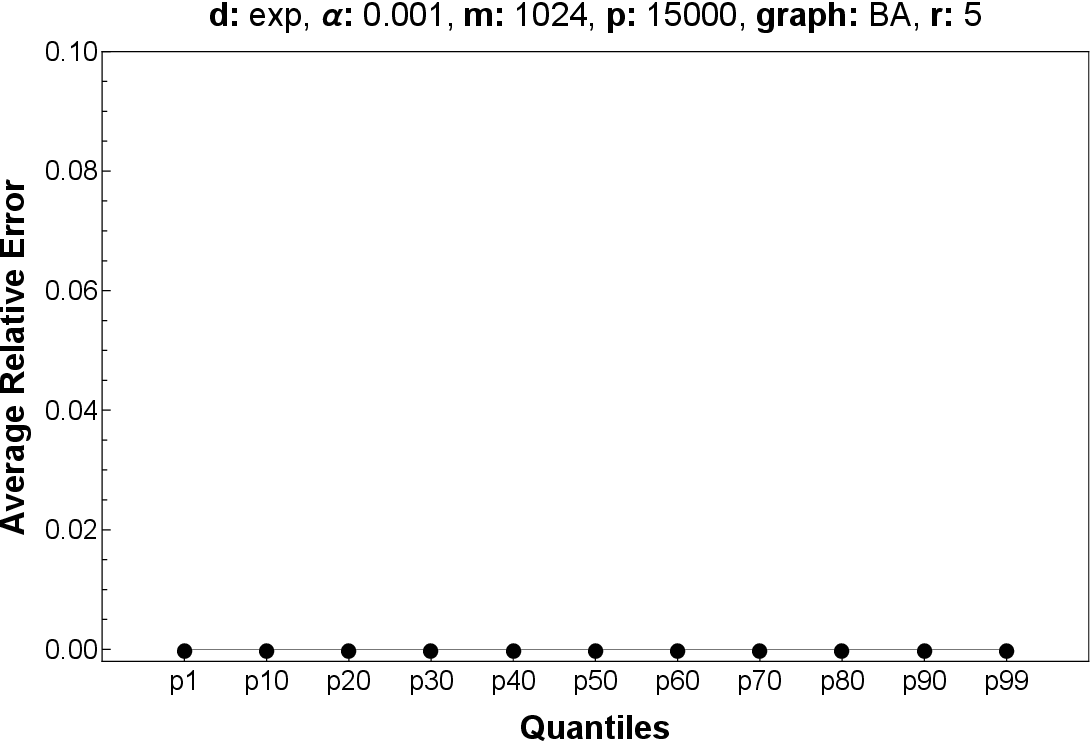}
		} &	
		
		\subfloat[]{
		    \includegraphics[width=0.45\textwidth]{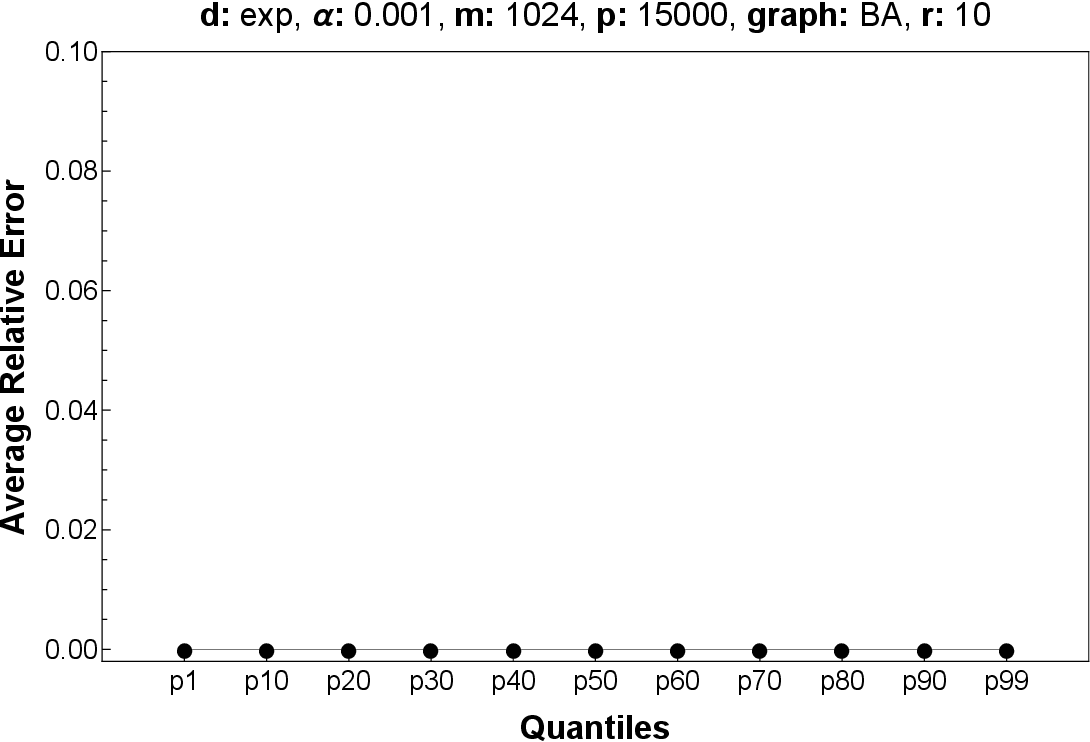}
		} \\	

        \subfloat[]{
		    \includegraphics[width=0.45\textwidth]{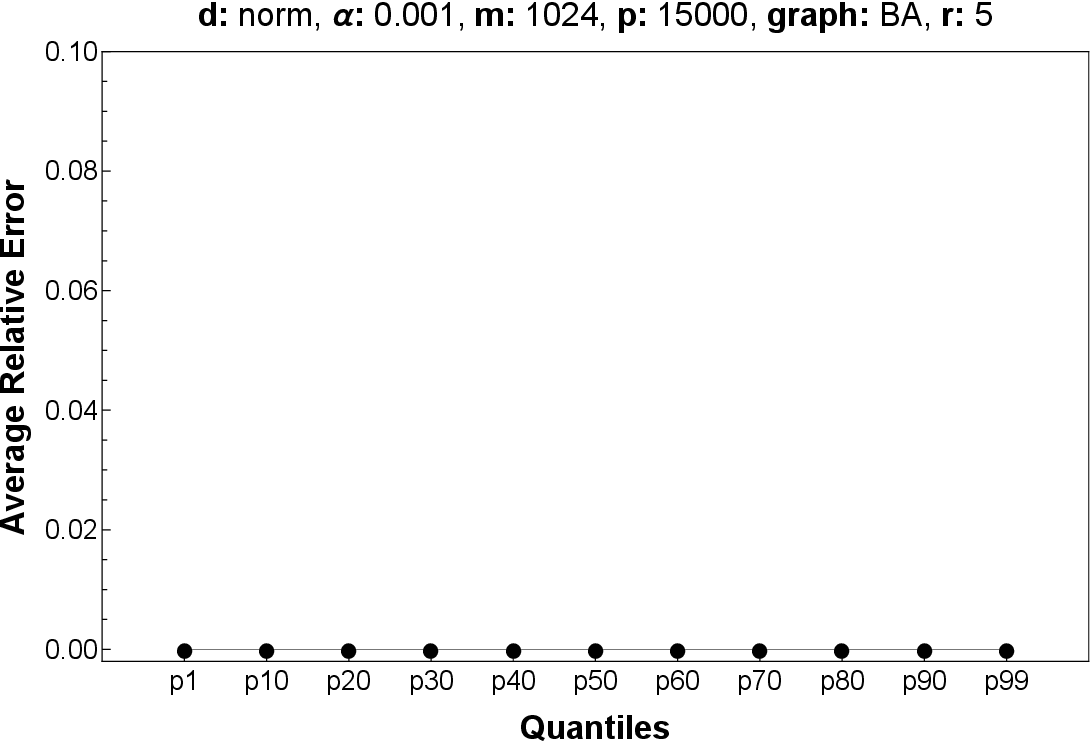}
		} &
		
		\subfloat[]{
		    \includegraphics[width=0.45\textwidth]{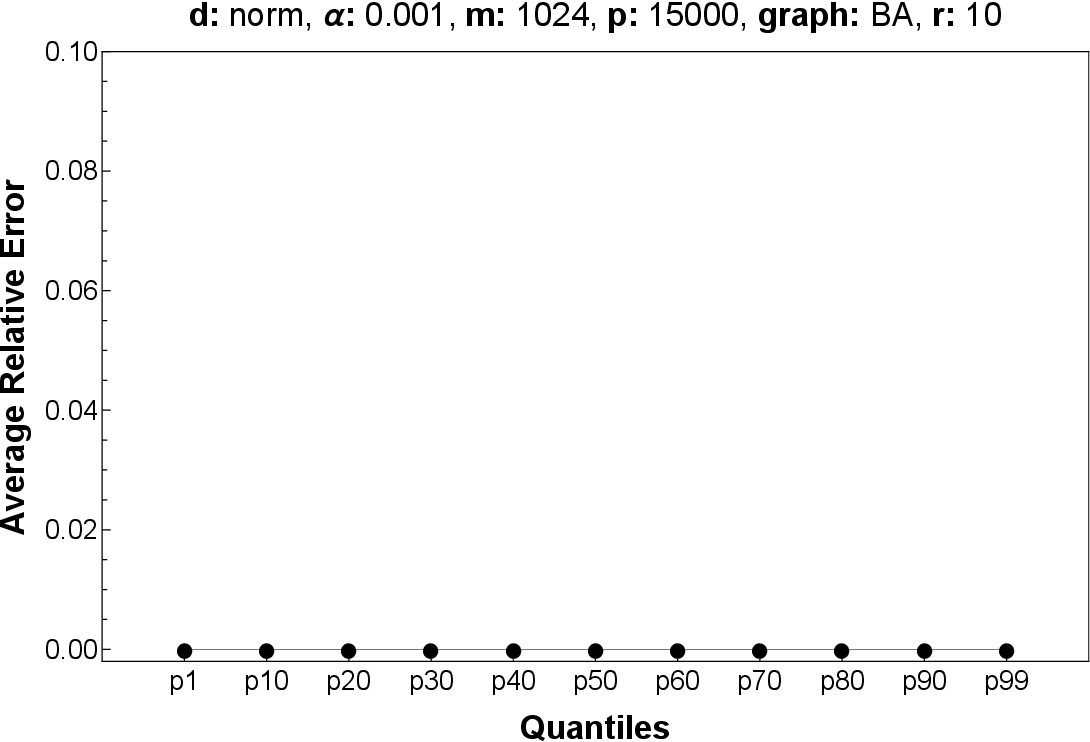}
		} \\

        \subfloat[]{
		    \includegraphics[width=0.45\textwidth]{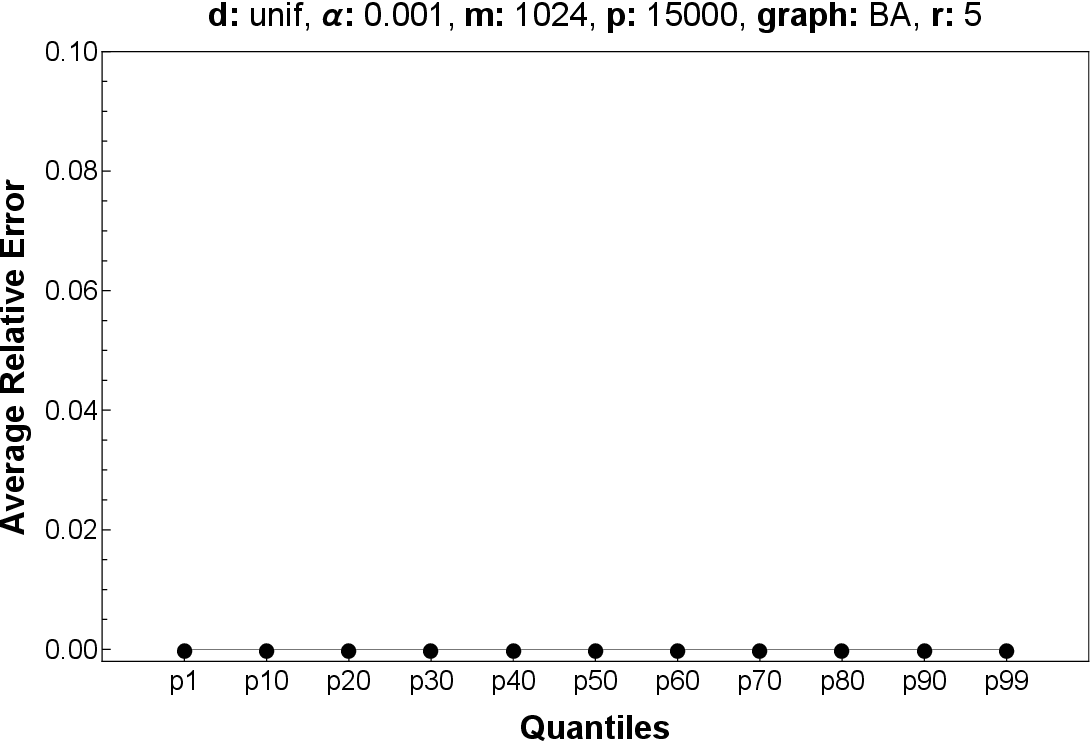}
		} &

		\subfloat[]{
		    \includegraphics[width=0.45\textwidth]{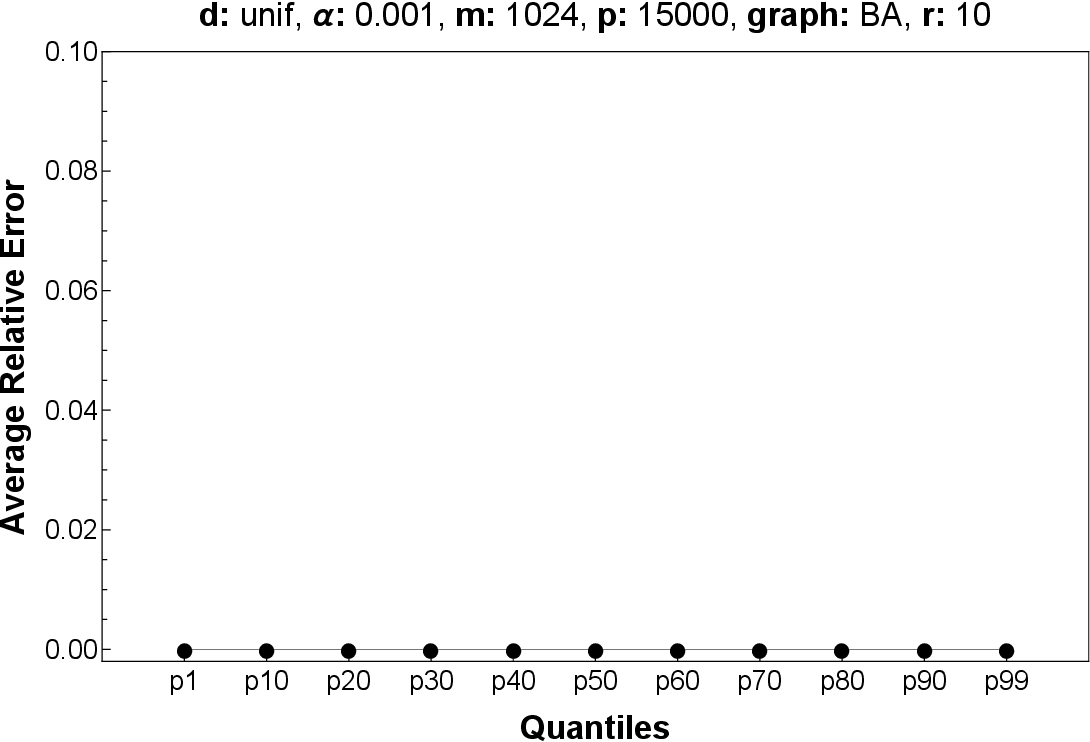}
		} \\

	\end{tabular}	
	\caption{Protocol convergence at 5 (left column) and 10 rounds (right column) for a network of 15000 peers on a Bar\'abasi-Albert graph, with a sketch size of $1024$ buckets, initial $\alpha=0.001$ over an exponential (a,b), normal (c,d) and uniform (e,f) inputs.} 
	\label{fig.peers.ba.oth2}
\end{figure*}

Results are even better when the input is not drawn from adversarial random distributions: Figures \ref{fig.peers.ba.oth1} and \ref{fig.peers.ba.oth2} refer to experiments carried out on inputs following \textit{uniform}, \textit{exponential} and \textit{normal} distributions, respectively for network sizes of $10000$ and $15000$ peers. As shown, $10$ rounds are enough to drive to almost zero the errors in comparison to the sequential \textsc{UDDSketch} algorithm. We do not report plots at $15$ rounds and above, because in all of the experiments relative errors go to zero at $15$ rounds.

%adv 1

%adv 2

%others1

%others2

\subsection{Convergence in presence of churning}

To verify the efficiency of the proposed Distributed \textsc{UDDSketch} protocol under more realistic operating conditions, tests that simulate churning events in the overlay network have been performed.
We implemented three different churning models: the \emph{Fail \& Stop} model and the \emph{Yao} model in its two variants based respectively on the shifted Pareto and the Exponential distribution \cite{yao2006modeling}.

In the Fail \& Stop model, a peer could leave the network with a given failure probability and the failed peers cannot join the network anymore. In the experiments we set the failure probability to $0.01$.%For the Fail \& Stop model, the set of failure probabilities used in the tests has been {0.01, 0.05, 0.1}.

\begin{figure*}[htb]
    \centering
    \begin{tabular}{cc}

		\subfloat[]{
		    \includegraphics[width=0.42\textwidth]{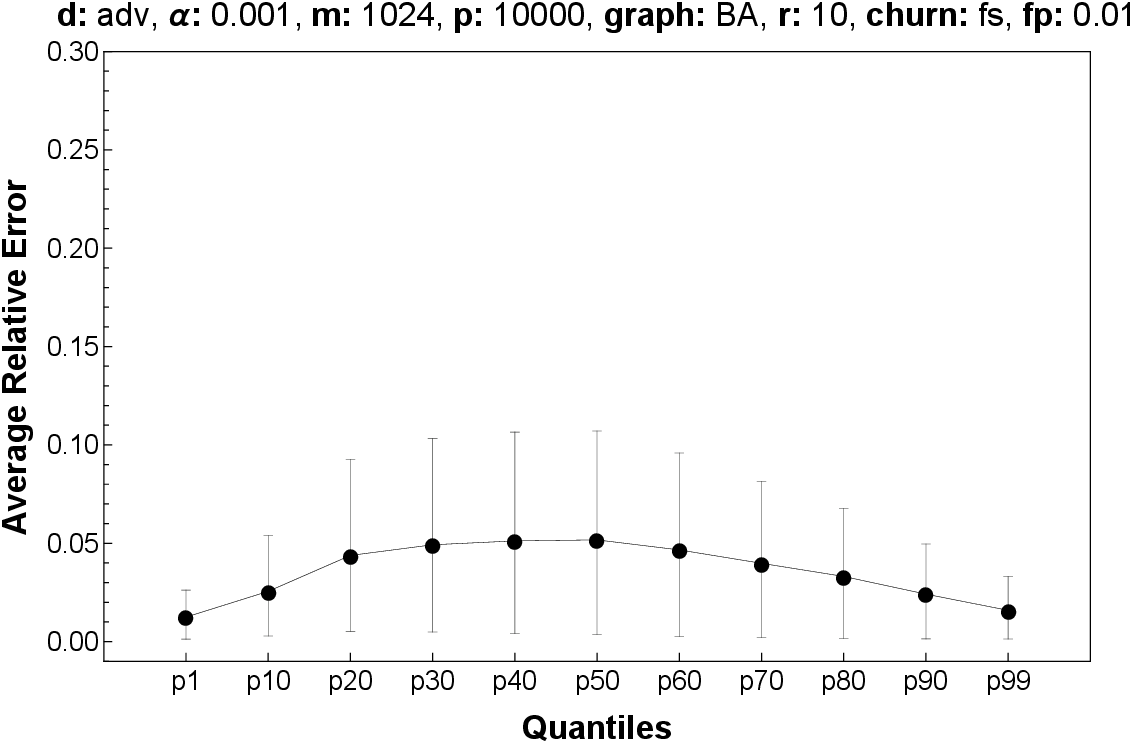}
		} &	
		
		\subfloat[]{
		    \includegraphics[width=0.42\textwidth]{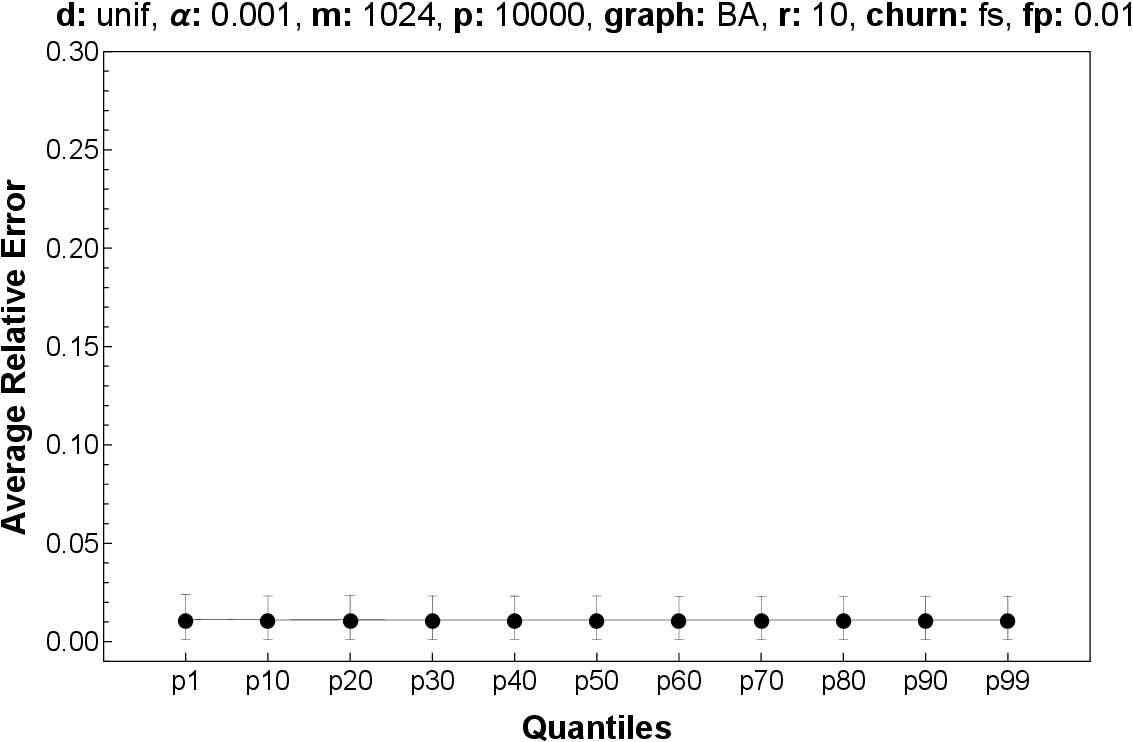}
		} \\	

        \subfloat[]{
		    \includegraphics[width=0.42\textwidth]{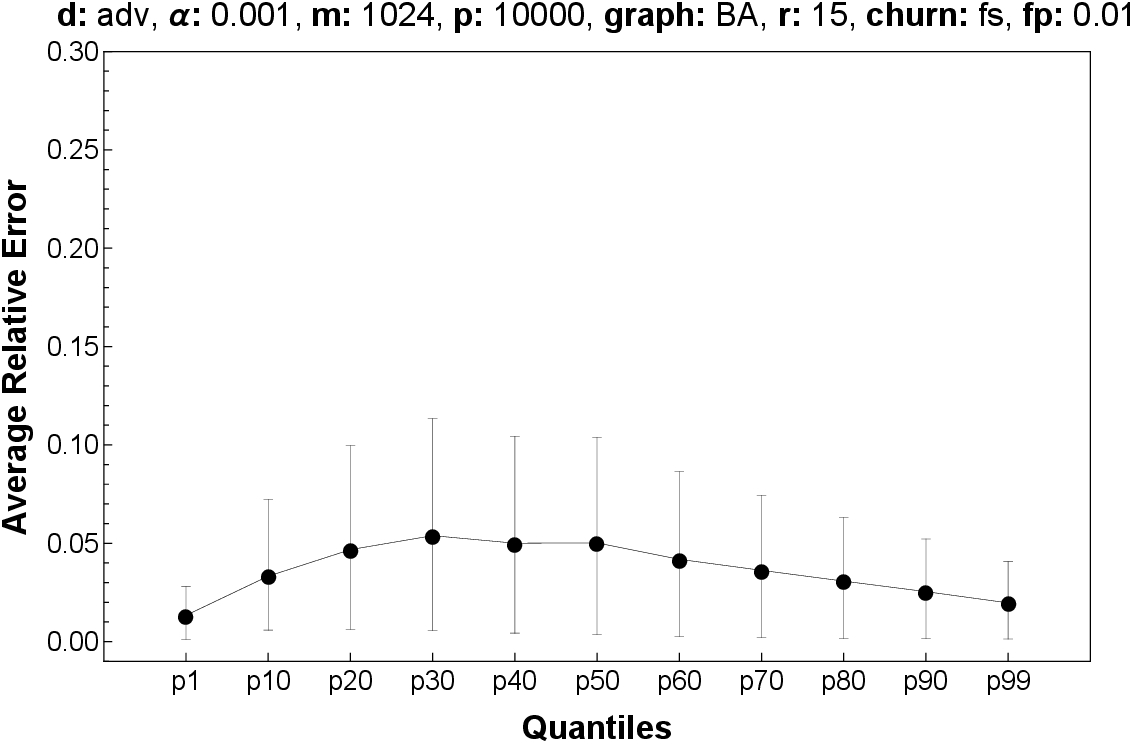}
		} &
		
		\subfloat[]{
		    \includegraphics[width=0.42\textwidth]{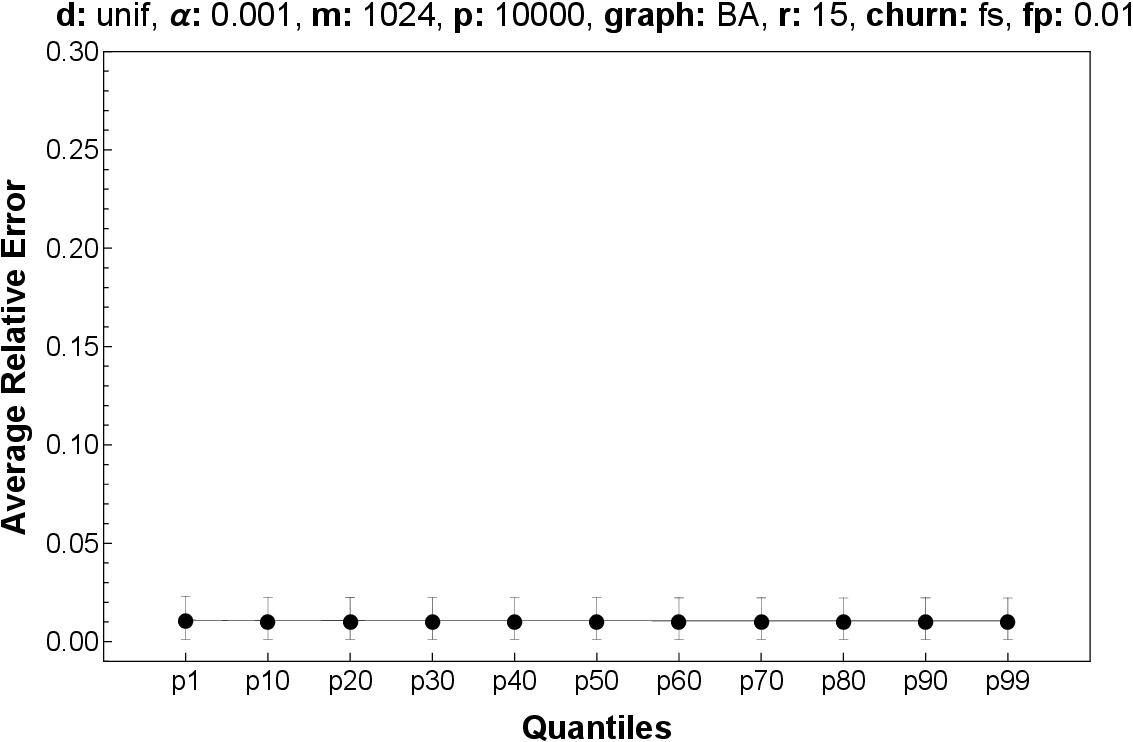}
		} \\

        \subfloat[]{
		    \includegraphics[width=0.42\textwidth]{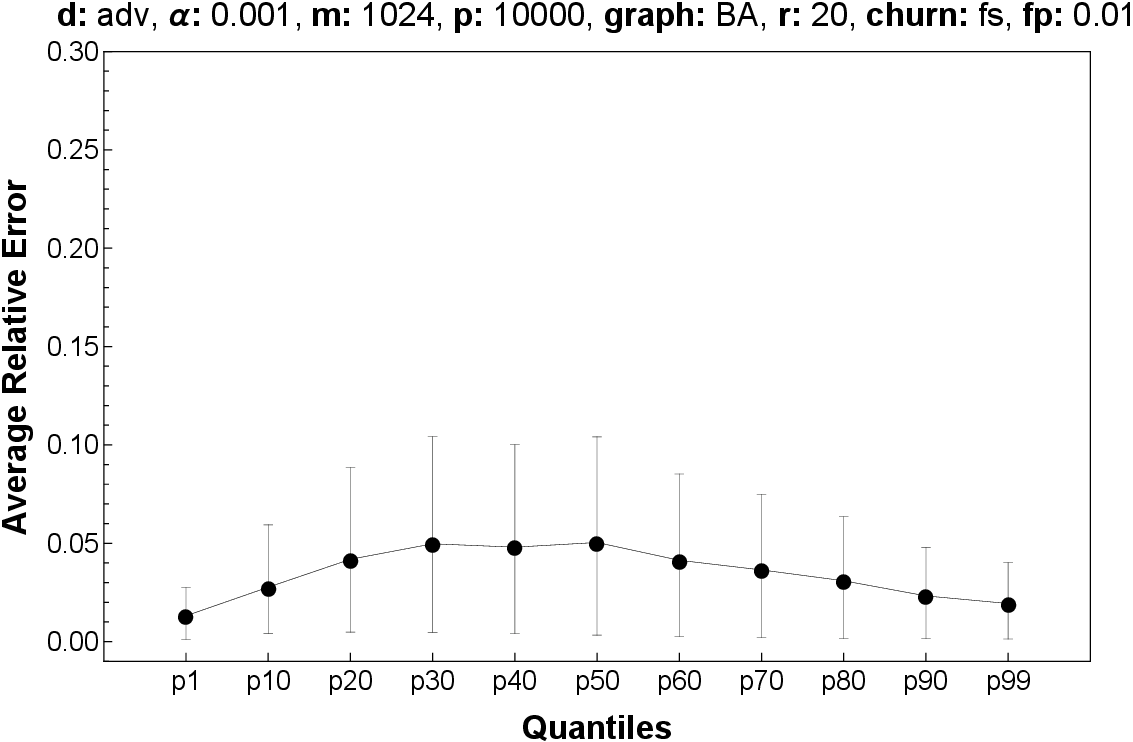}
		} &

		\subfloat[]{
		    \includegraphics[width=0.42\textwidth]{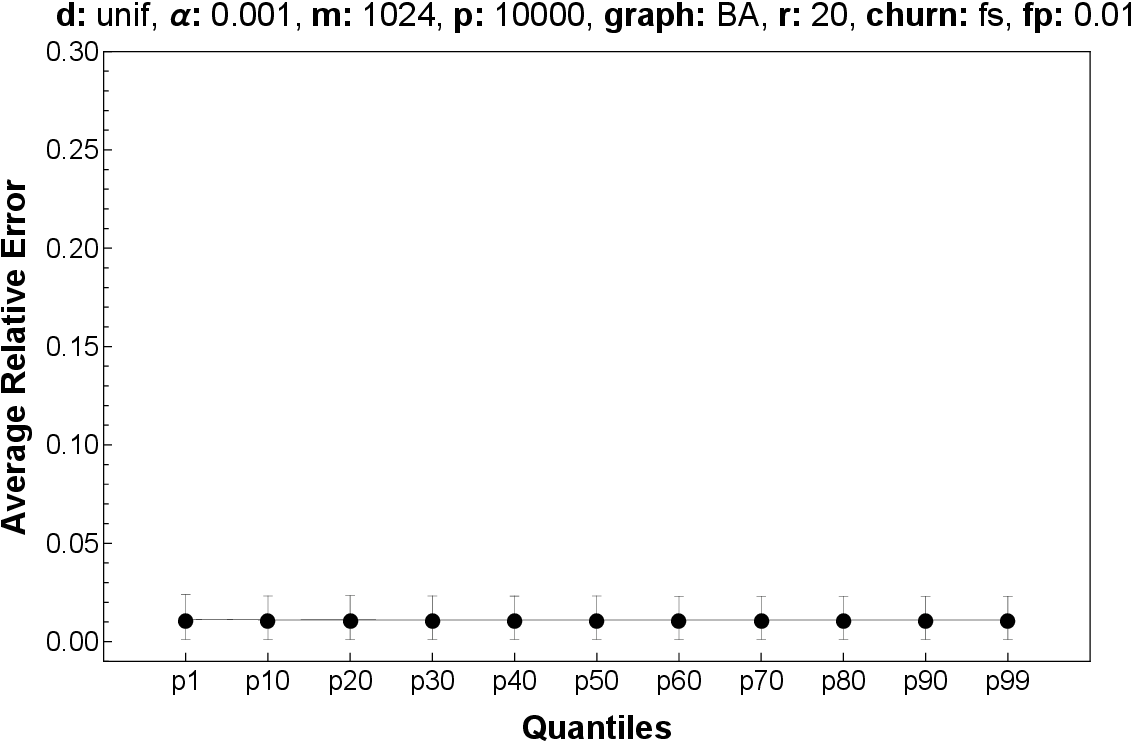}
		} \\
		
        \subfloat[]{
		    \includegraphics[width=0.42\textwidth]{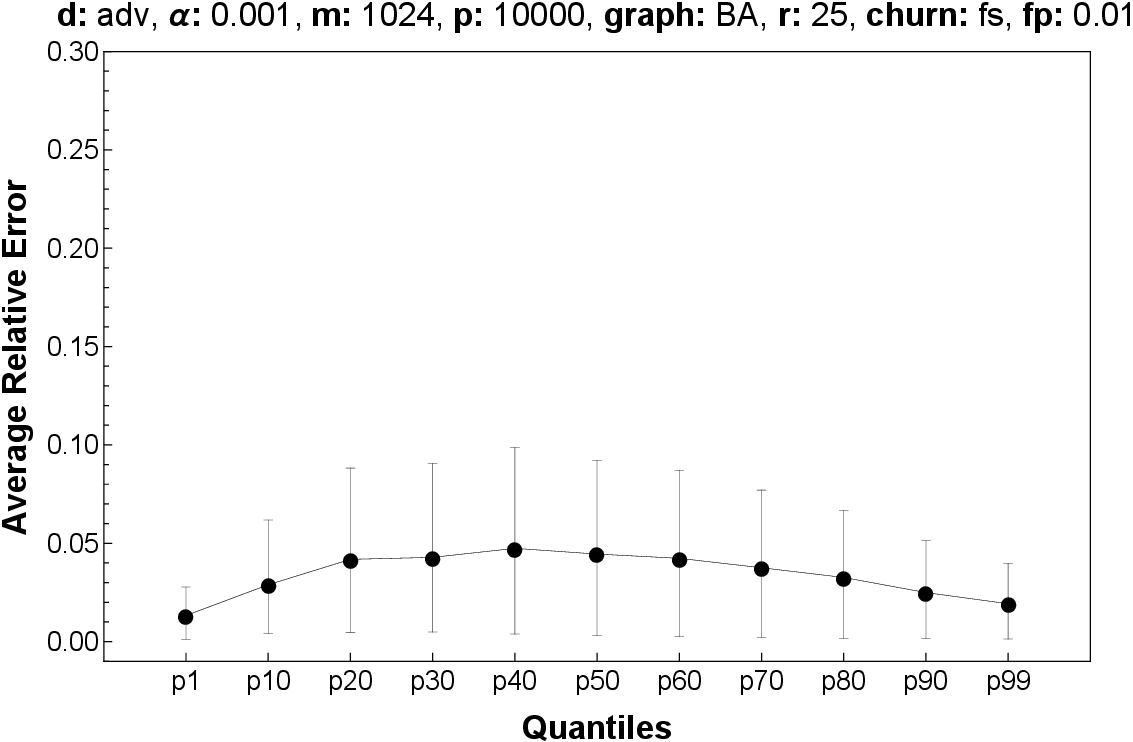}
		} &

		\subfloat[]{
		    \includegraphics[width=0.42\textwidth]{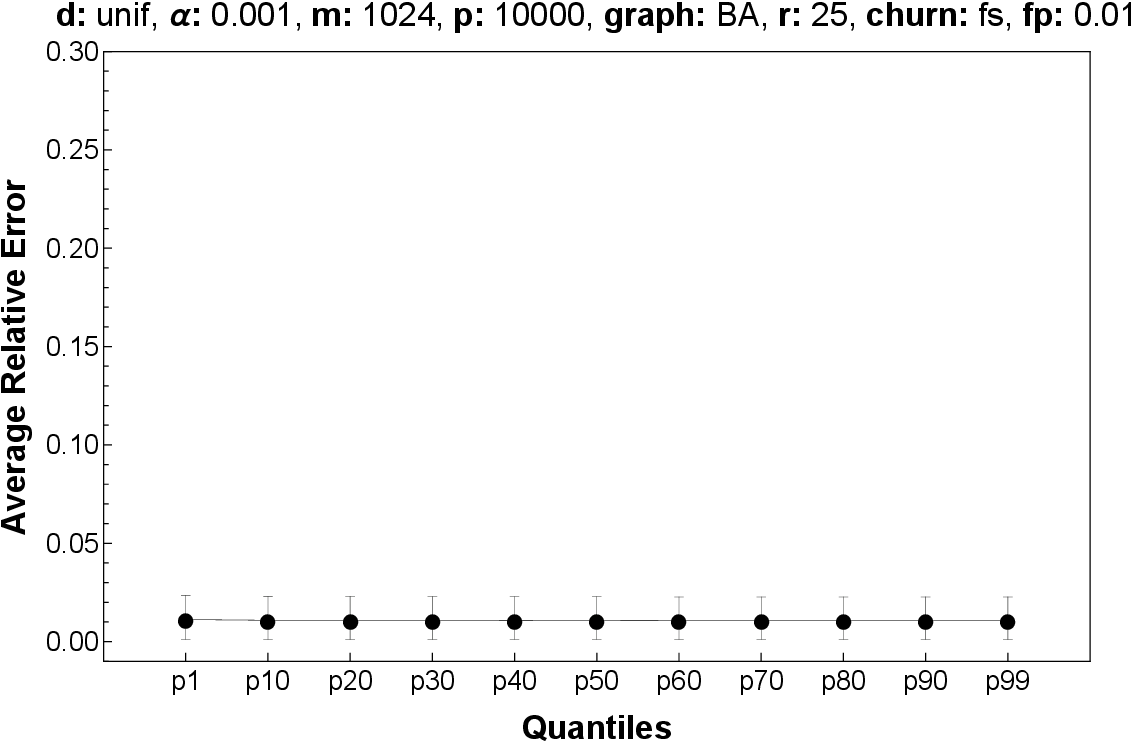}
		} \\
		
	\end{tabular}	
	\caption{Protocol convergence varying the number of rounds, in presence of \emph{Fail \& Stop} churning with probability of failure $0.01$, adversarial (left column) and uniform input (right column), $\alpha=0.001$, $m=1024$, over a network of $10000$ peers on a Bar\'abasi-Albert random graph.} 
	\label{fig.peers.ba.fs1}
\end{figure*}

\begin{figure*}[htb]
    \centering
    \begin{tabular}{cc}

		\subfloat[]{
		    \includegraphics[width=0.42\textwidth]{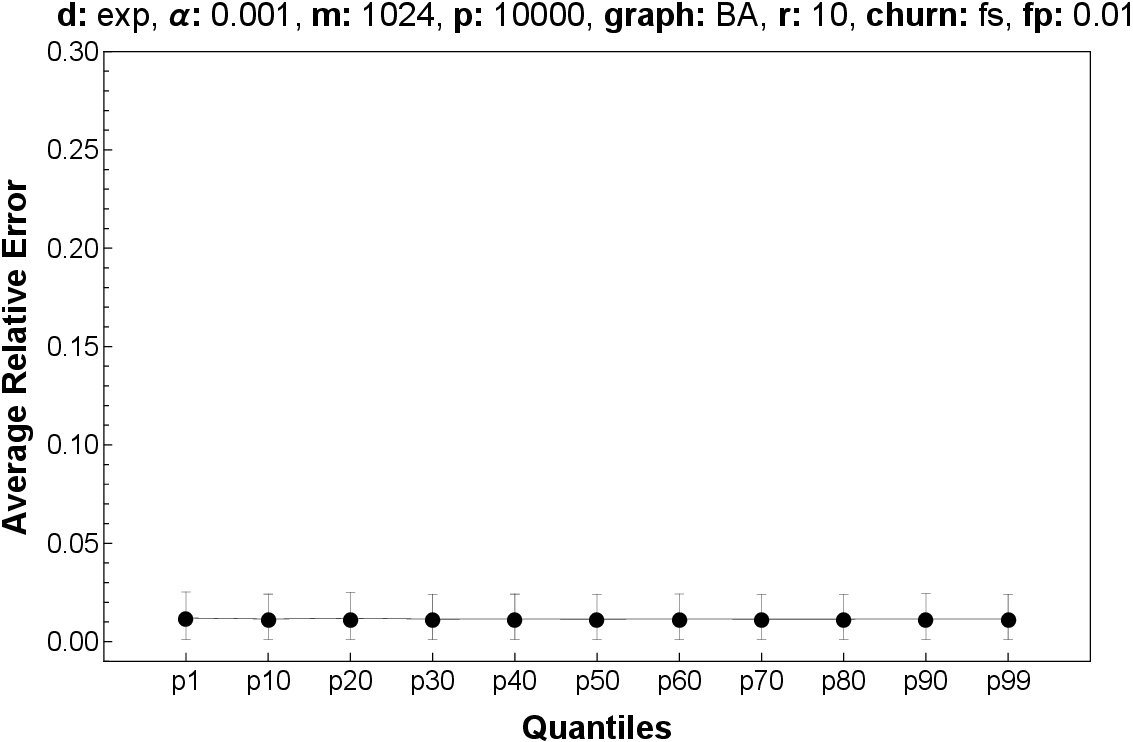}
		} &	
		
		\subfloat[]{
		    \includegraphics[width=0.42\textwidth]{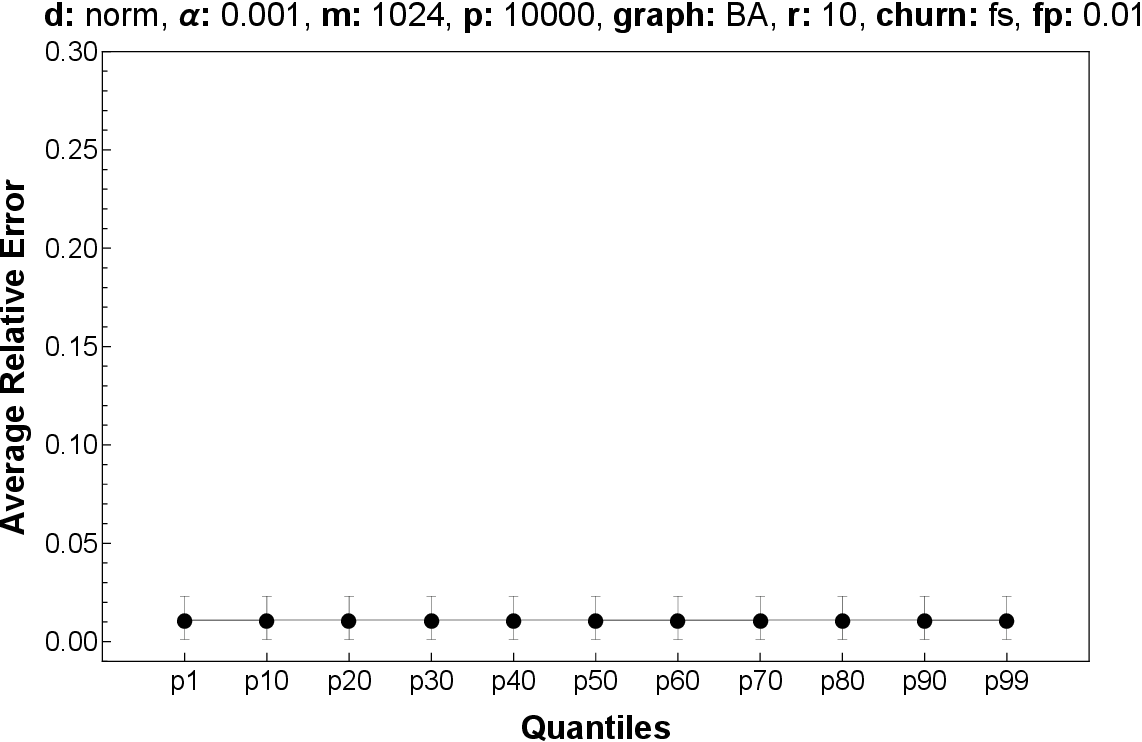}
		} \\	

        \subfloat[]{
		    \includegraphics[width=0.42\textwidth]{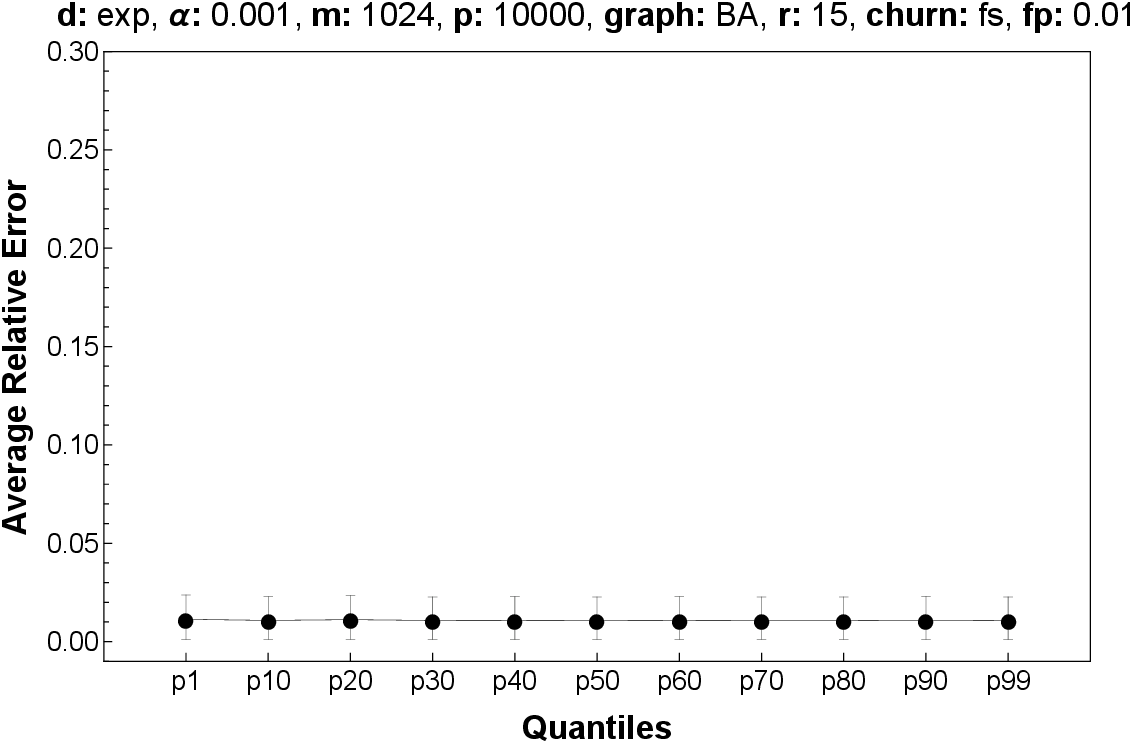}
		} &
		
		\subfloat[]{
		    \includegraphics[width=0.42\textwidth]{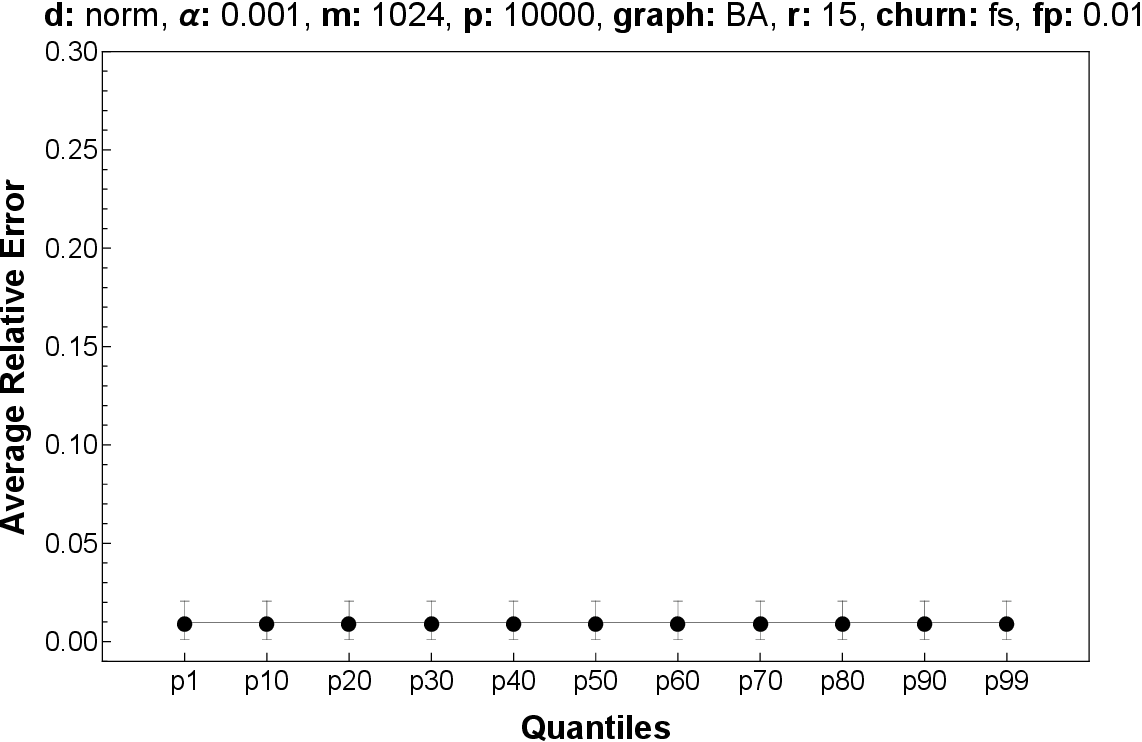}
		} \\

        \subfloat[]{
		    \includegraphics[width=0.42\textwidth]{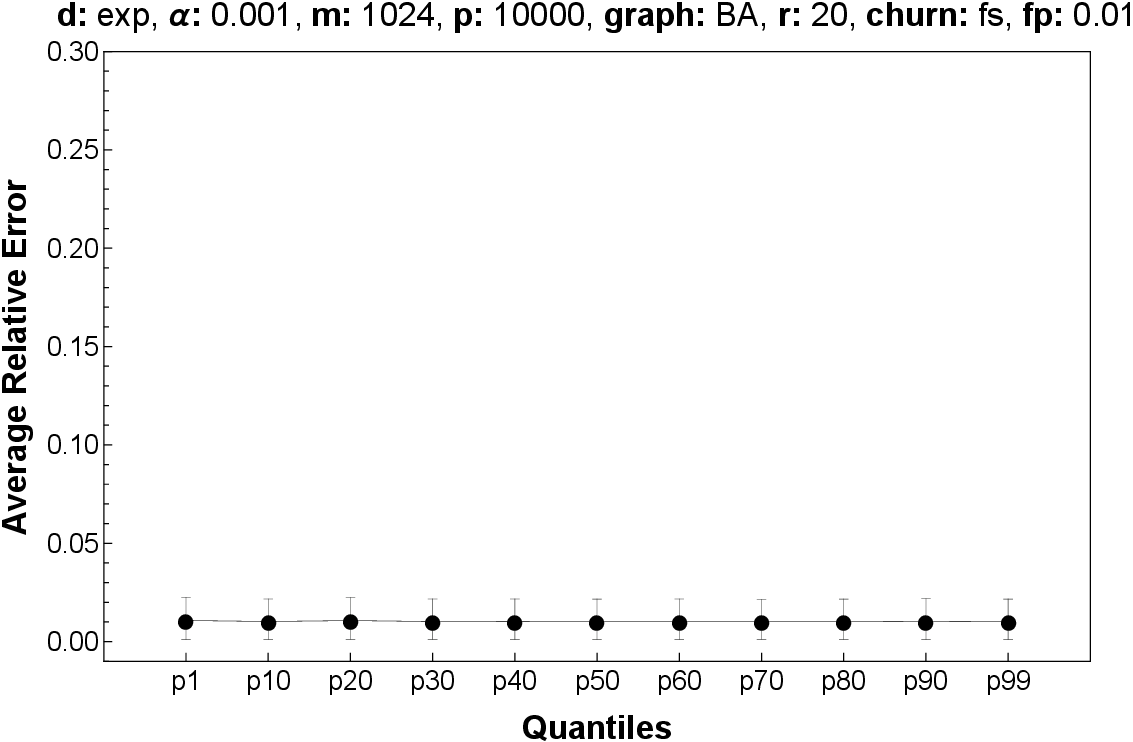}
		} &

		\subfloat[]{
		    \includegraphics[width=0.42\textwidth]{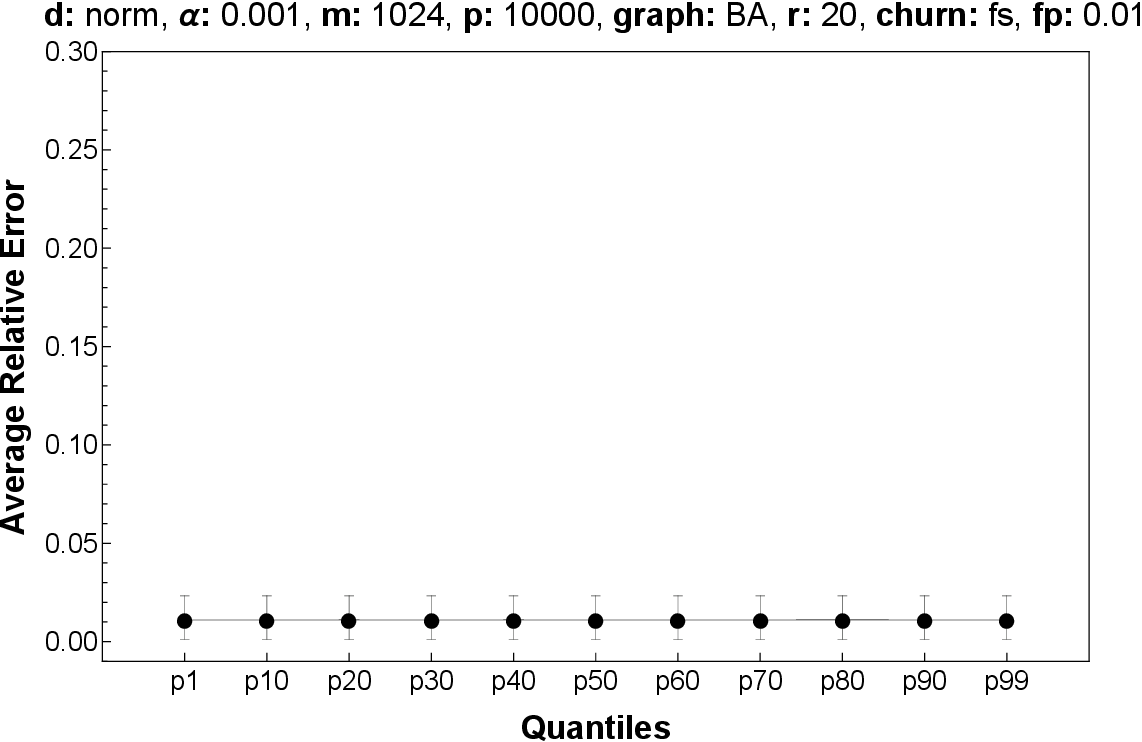}
		} \\
		
        \subfloat[]{
		    \includegraphics[width=0.42\textwidth]{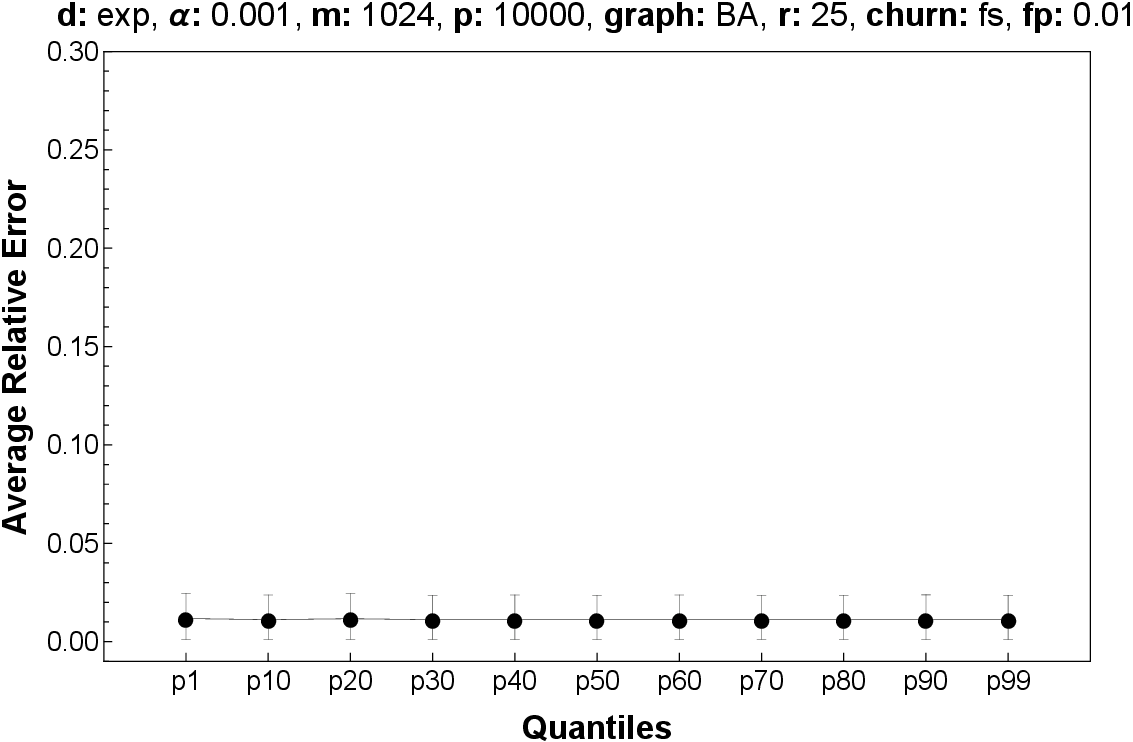}
		} &

		\subfloat[]{
		    \includegraphics[width=0.42\textwidth]{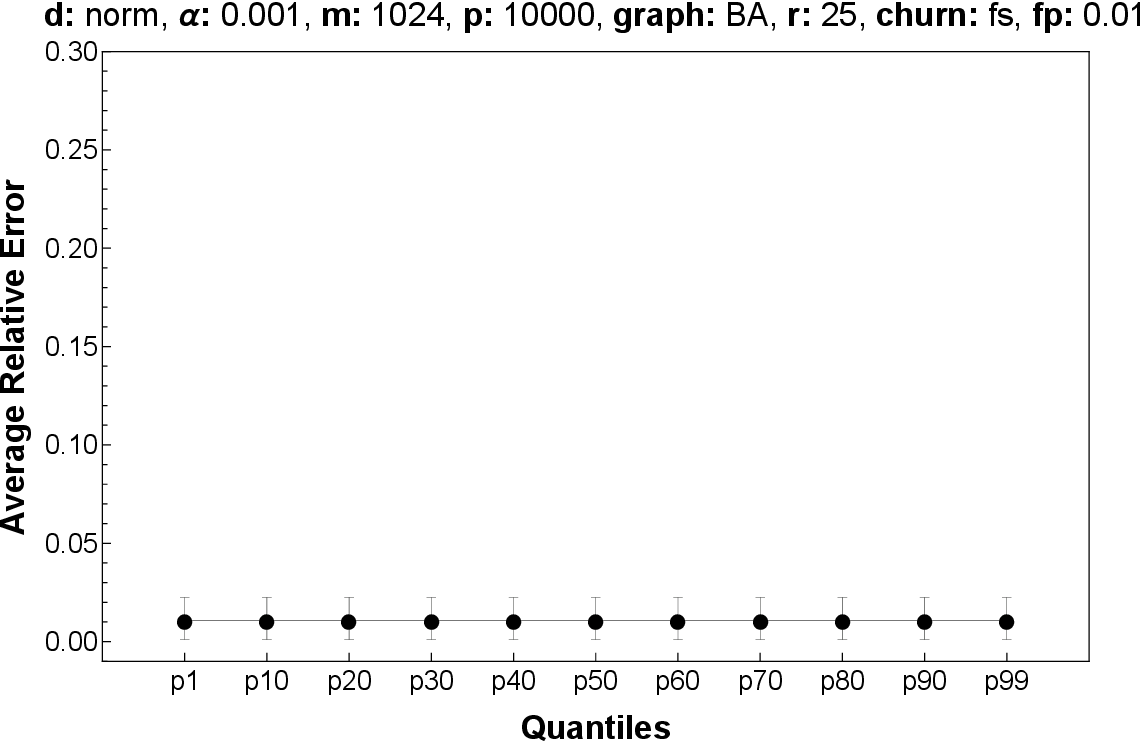}
		} \\
		
	\end{tabular}	
	\caption{Protocol convergence varying the number of rounds, in presence of \emph{Fail \& Stop} churning with probability of failure $0.01$, exponential (left column) and normal input (right column), $\alpha=0.001$, $m=1024$, over a network of $10000$ peers on a Bar\'abasi-Albert random graph.} 
	\label{fig.peers.ba.fs2}
\end{figure*}

In the Yao model, the peers randomly join and leave the network. 
For each peer $i$, a random average \emph{lifetime} duration $l_i$ has been  generated from a shifted Pareto distribution with parameters $\alpha = 3$, $\beta = 1$ and $\mu = 1.01$. Similarly, a random average \emph{offline} duration $d_i$ has been generated from a shifted Pareto distribution with the same $\alpha$ and $\mu$ values and with $\beta=2$.

The values $l_i$ and $d_i$ have been used to set for each peer $i$ two shifted Pareto distributions: whenever the state of a peer changed, the duration of its current status has been drawn from one these two distributions. If at a given round the peer's state was online, then the peer could fail and remain offline for a period of time drawn from the corresponding shifted Pareto distribution. Instead, if  the peer's state was offline, then it could join again the network with a probability drawn from a shifted Pareto or from an exponential distribution (with $\lambda = 1/l_i$). 

\begin{figure*}[htb]
    \centering
    \begin{tabular}{cc}

		\subfloat[]{
		    \includegraphics[width=0.42\textwidth]{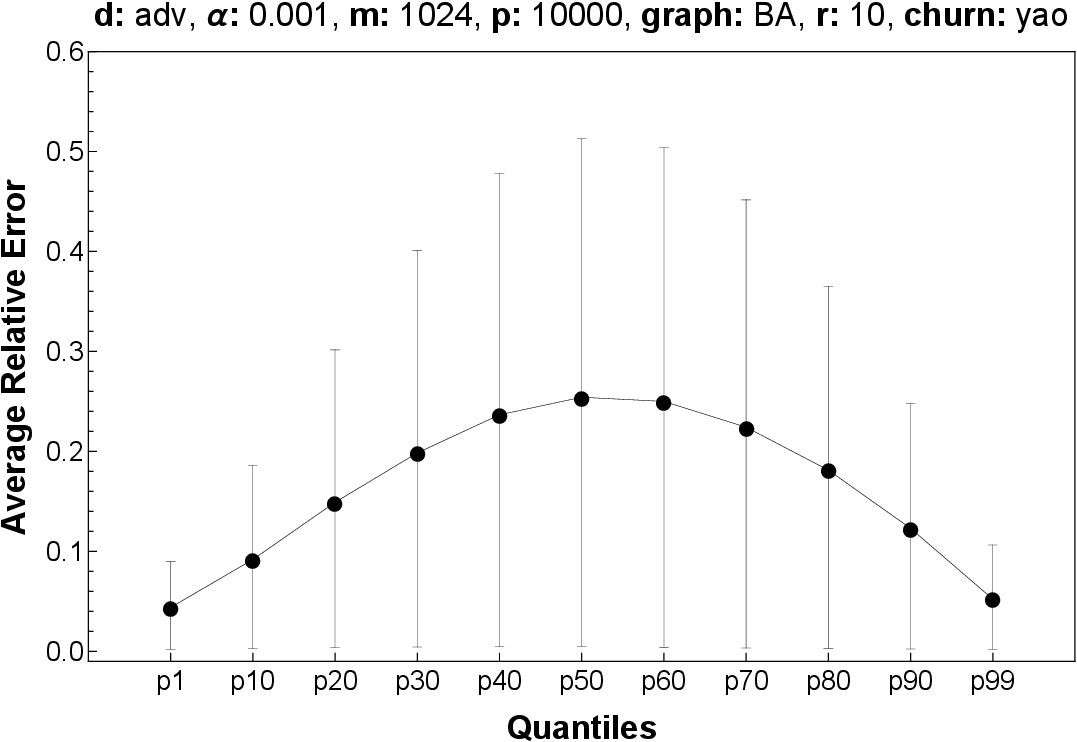}
		} &	
		
		\subfloat[]{
		    \includegraphics[width=0.42\textwidth]{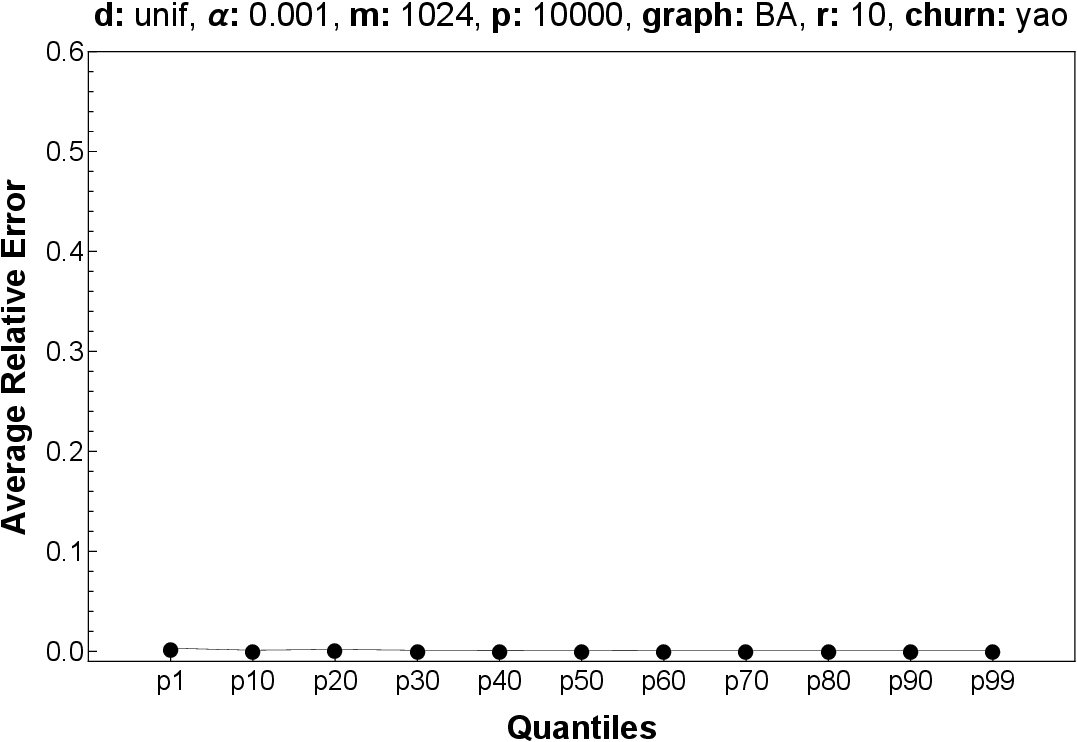}
		} \\	

        \subfloat[]{
		    \includegraphics[width=0.42\textwidth]{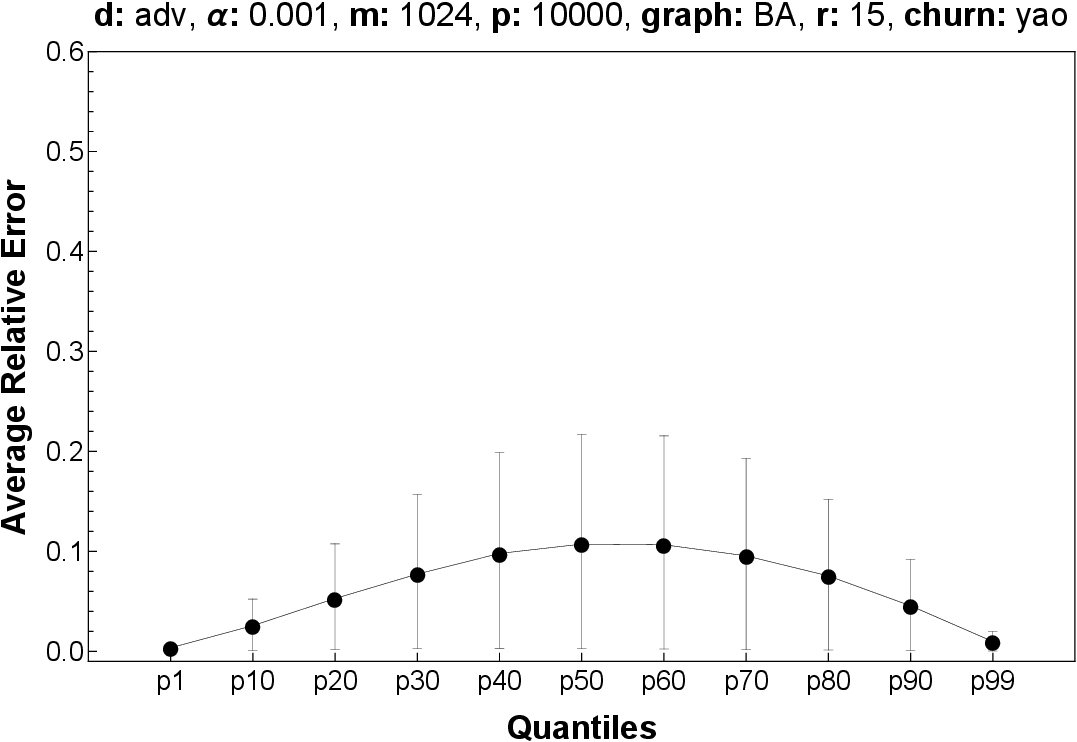}
		} &
		
		\subfloat[]{
		    \includegraphics[width=0.42\textwidth]{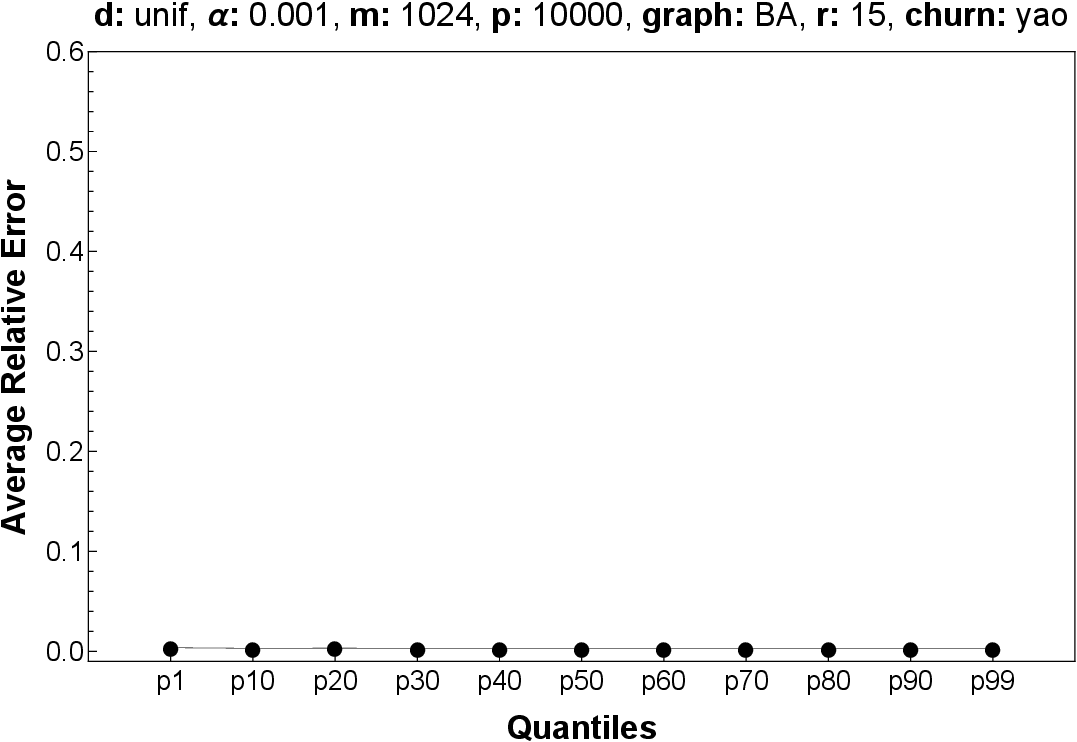}
		} \\

        \subfloat[]{
		    \includegraphics[width=0.42\textwidth]{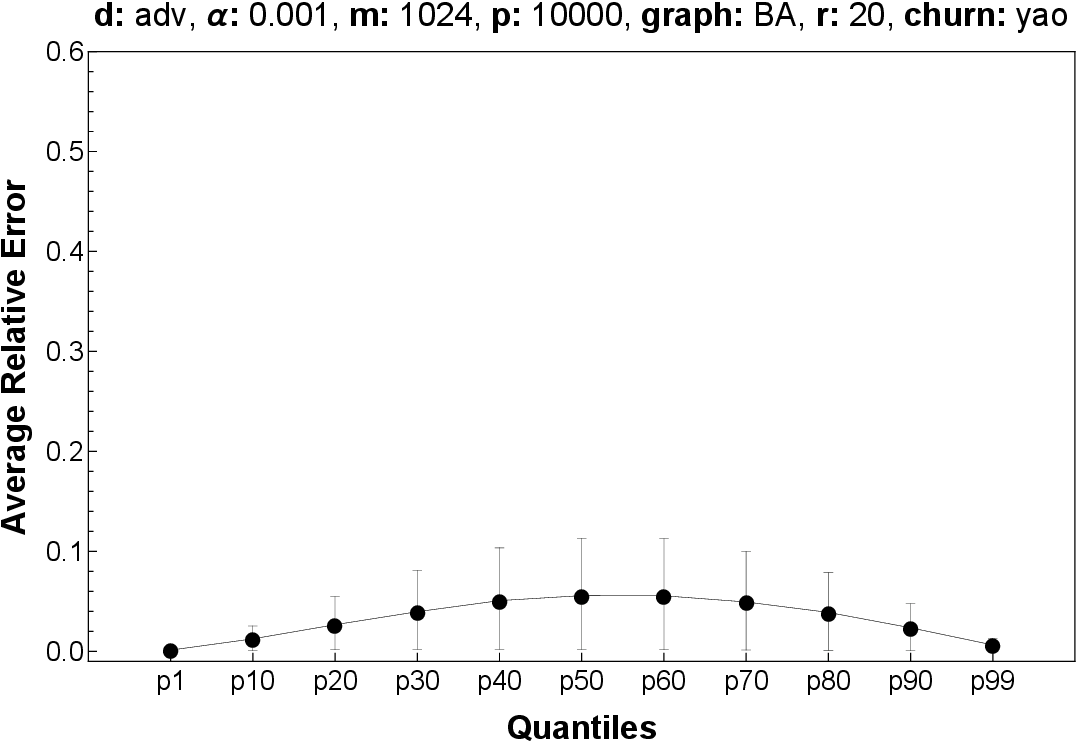}
		} &

		\subfloat[]{
		    \includegraphics[width=0.42\textwidth]{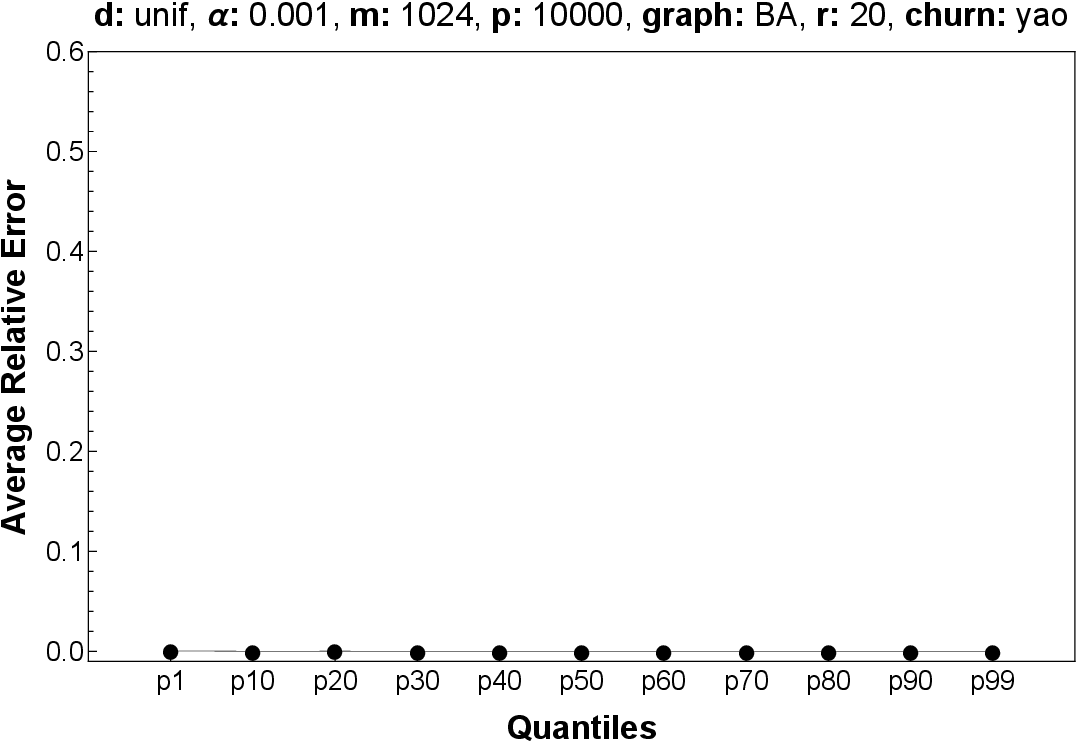}
		} \\
		
        \subfloat[]{
		    \includegraphics[width=0.42\textwidth]{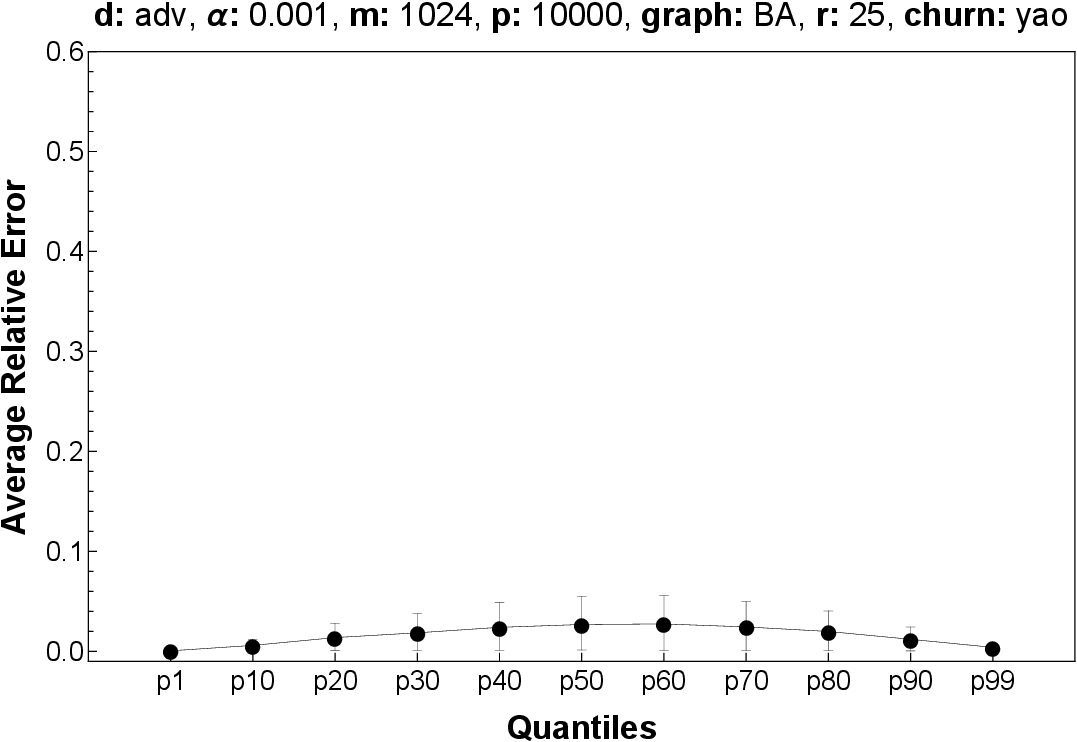}
		} &

		\subfloat[]{
		    \includegraphics[width=0.42\textwidth]{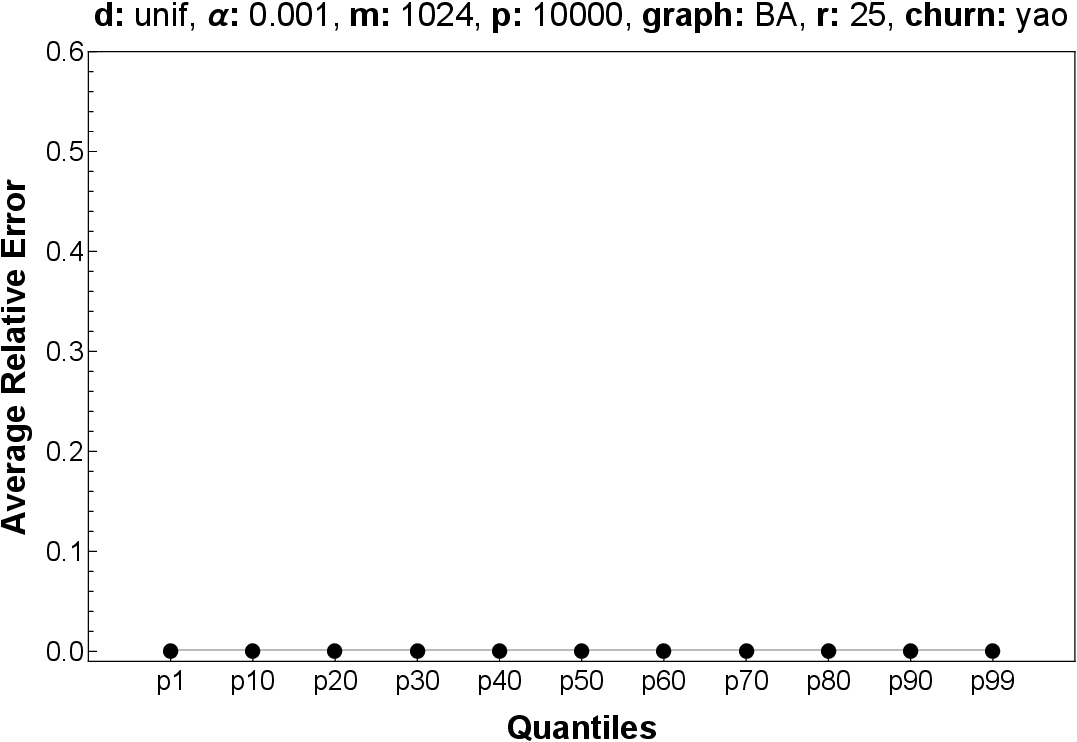}
		} \\
		
	\end{tabular}	
	\caption{Protocol convergence varying the number of rounds, in presence of \emph{Yao} churning, adversarial (left column) and uniform input (right column), $\alpha=0.001$, $m=1024$, over a network of $10000$ peers on a Bar\'abasi-Albert random graph.} 
	\label{fig.peers.ba.yao1}
\end{figure*}

\begin{figure*}[htb]
    \centering
    \begin{tabular}{cc}

		\subfloat[]{
		    \includegraphics[width=0.42\textwidth]{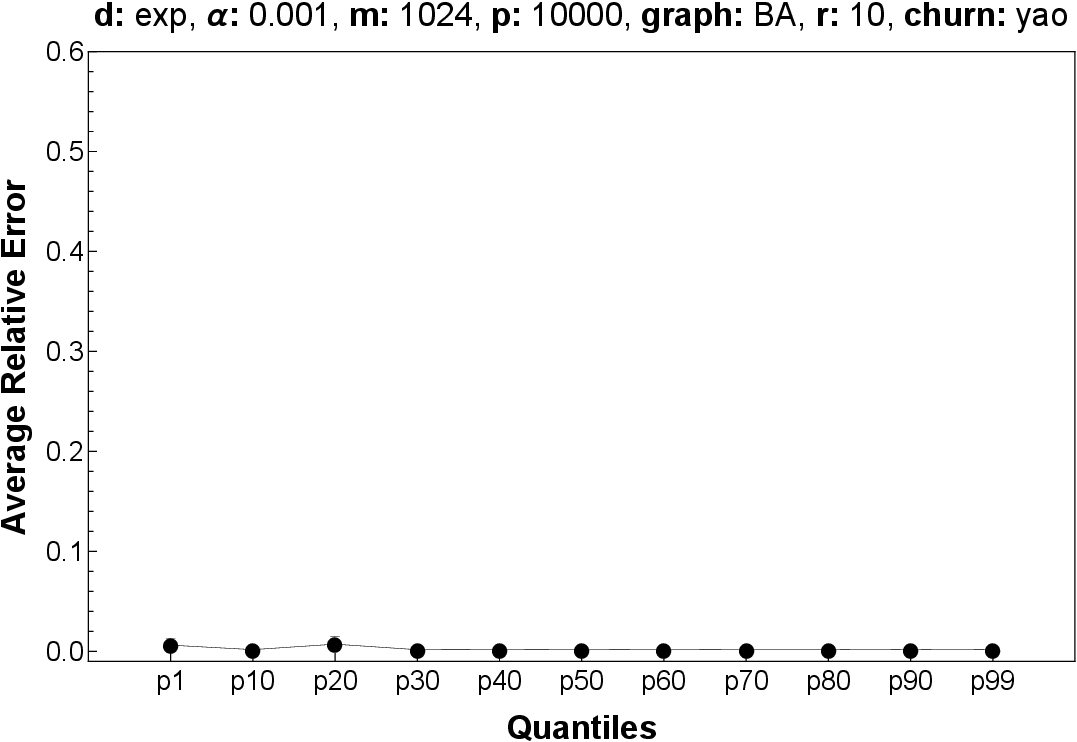}
		} &	
		
		\subfloat[]{
		    \includegraphics[width=0.42\textwidth]{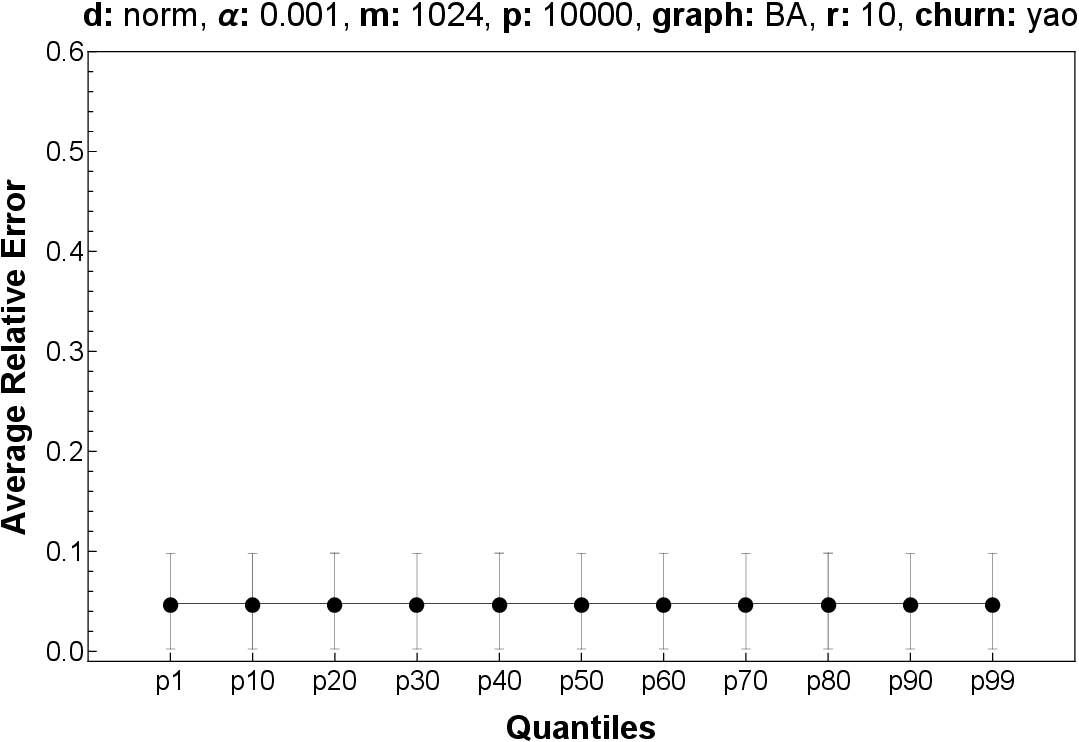}
		} \\	

        \subfloat[]{
		    \includegraphics[width=0.42\textwidth]{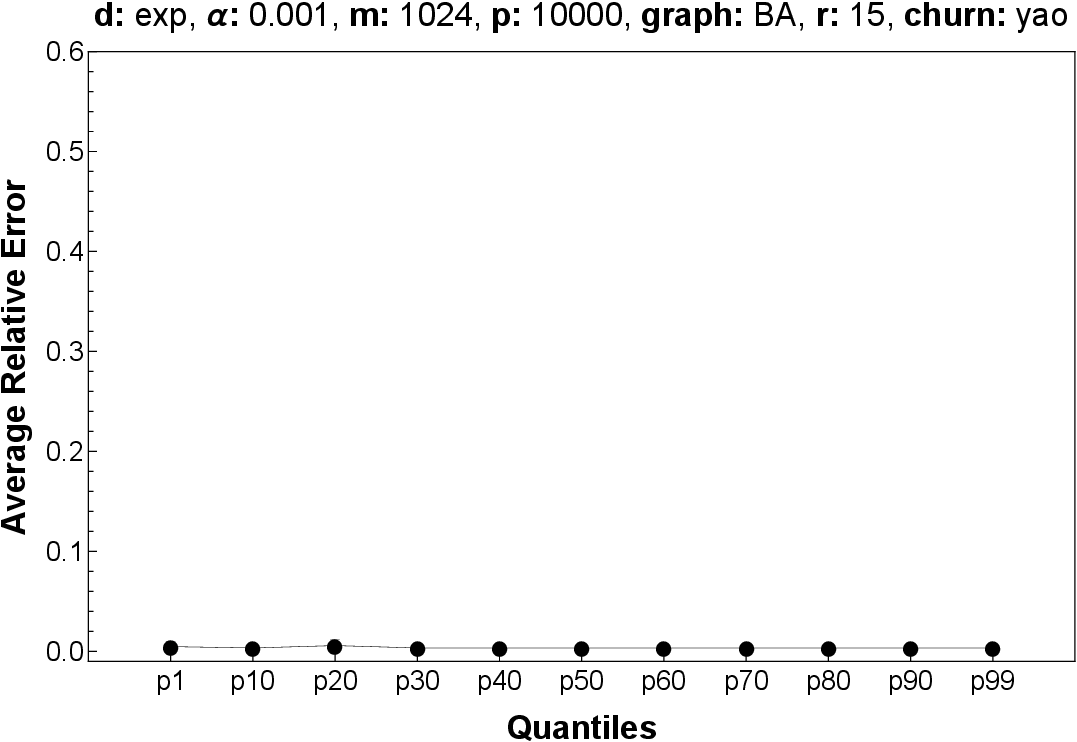}
		} &
		
		\subfloat[]{
		    \includegraphics[width=0.42\textwidth]{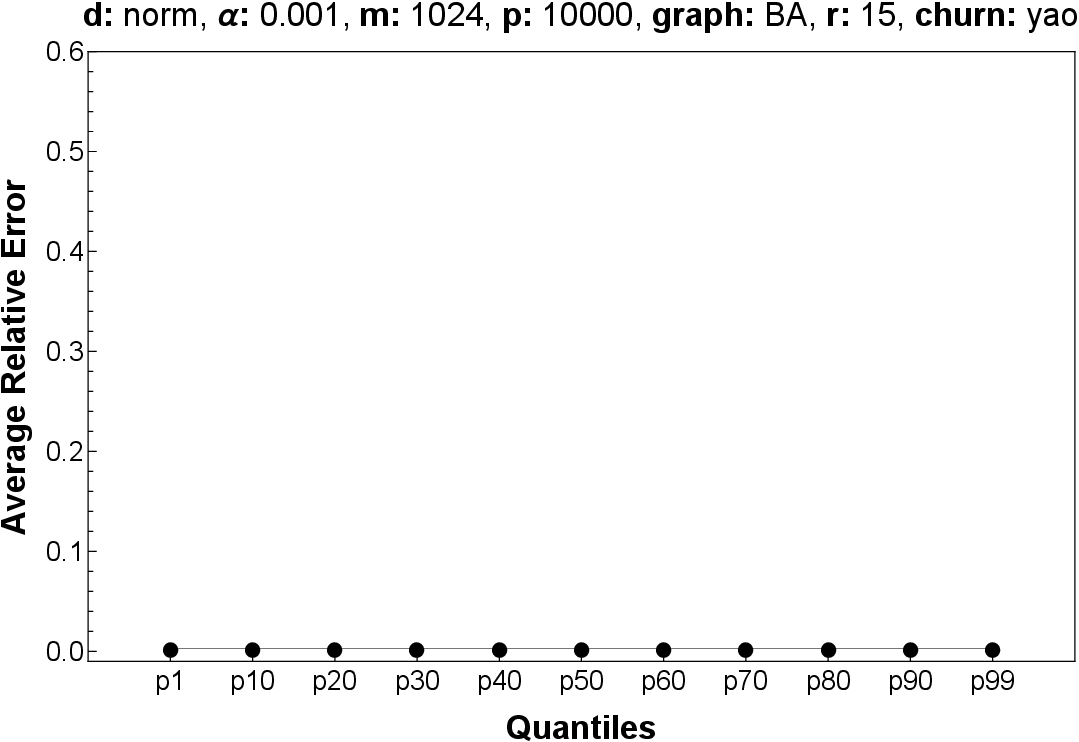}
		} \\

        \subfloat[]{
		    \includegraphics[width=0.42\textwidth]{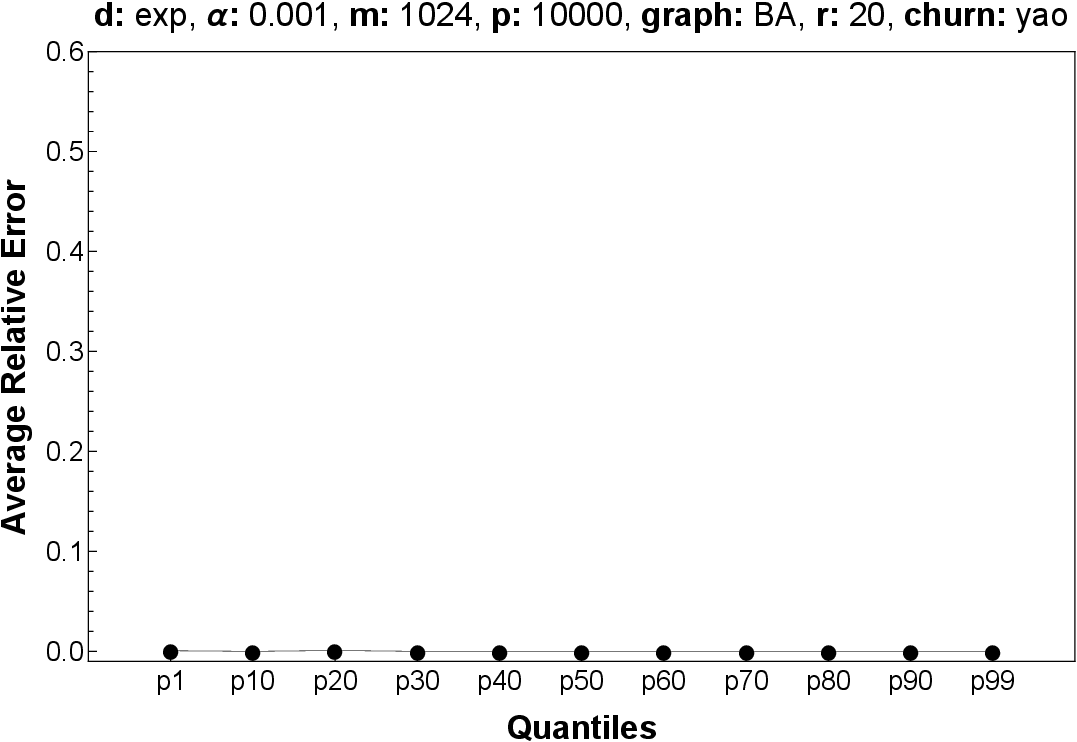}
		} &

		\subfloat[]{
		    \includegraphics[width=0.42\textwidth]{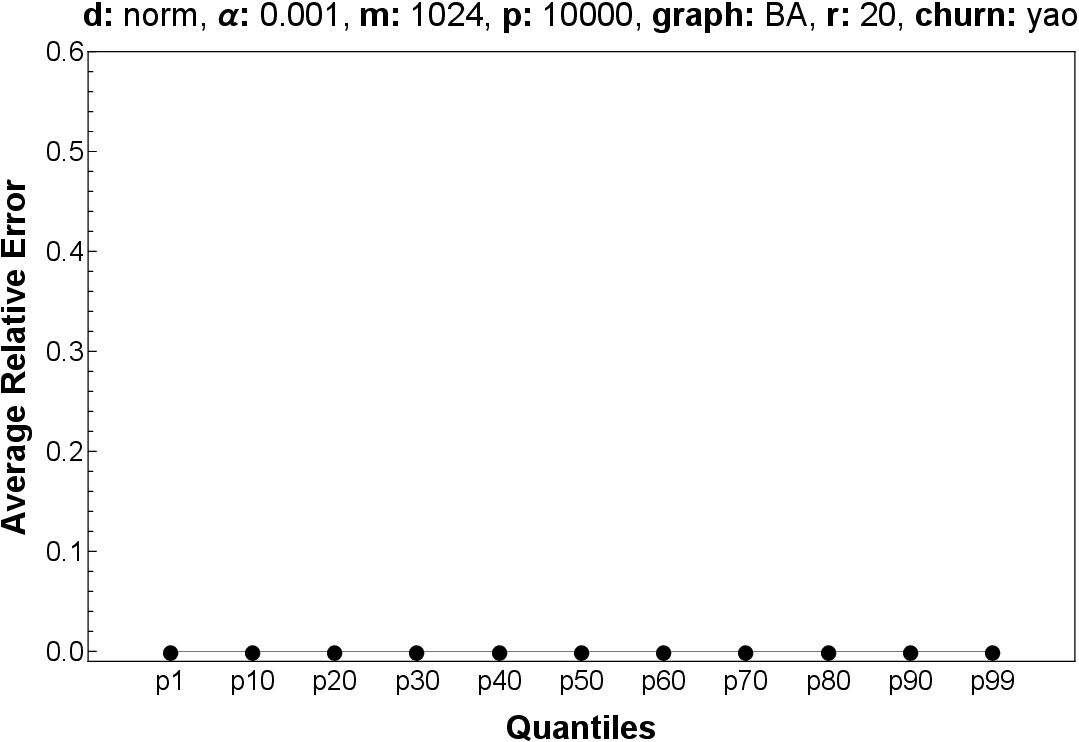}
		} \\
		
        \subfloat[]{
		    \includegraphics[width=0.42\textwidth]{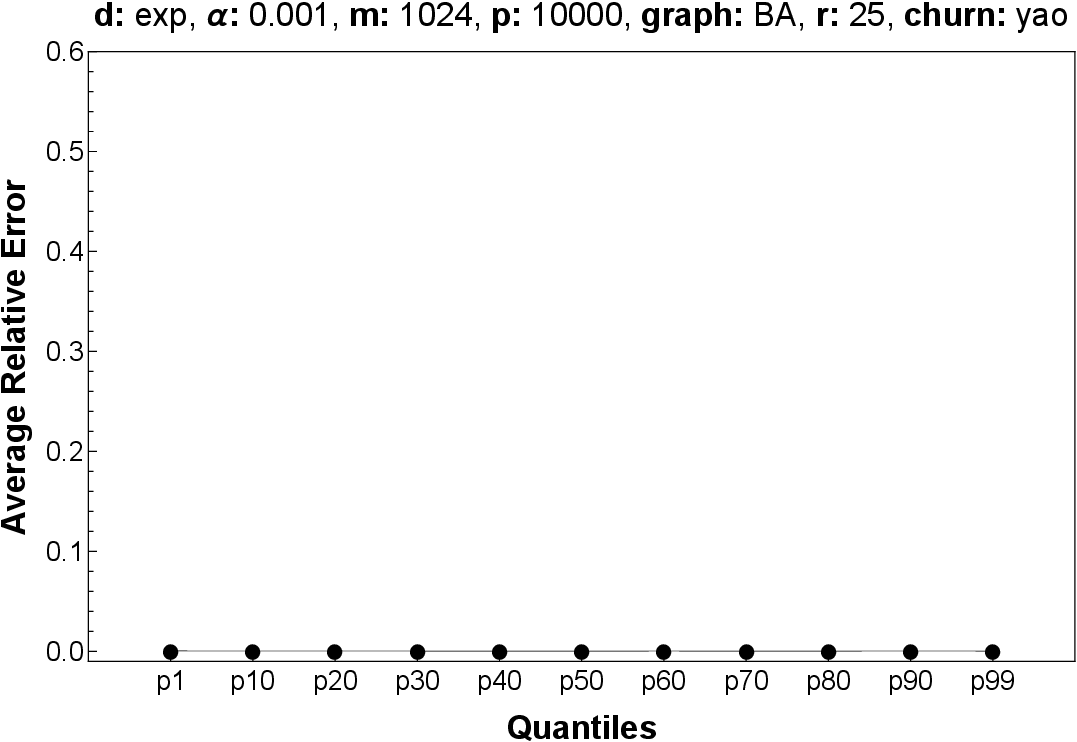}
		} &

		\subfloat[]{
		    \includegraphics[width=0.42\textwidth]{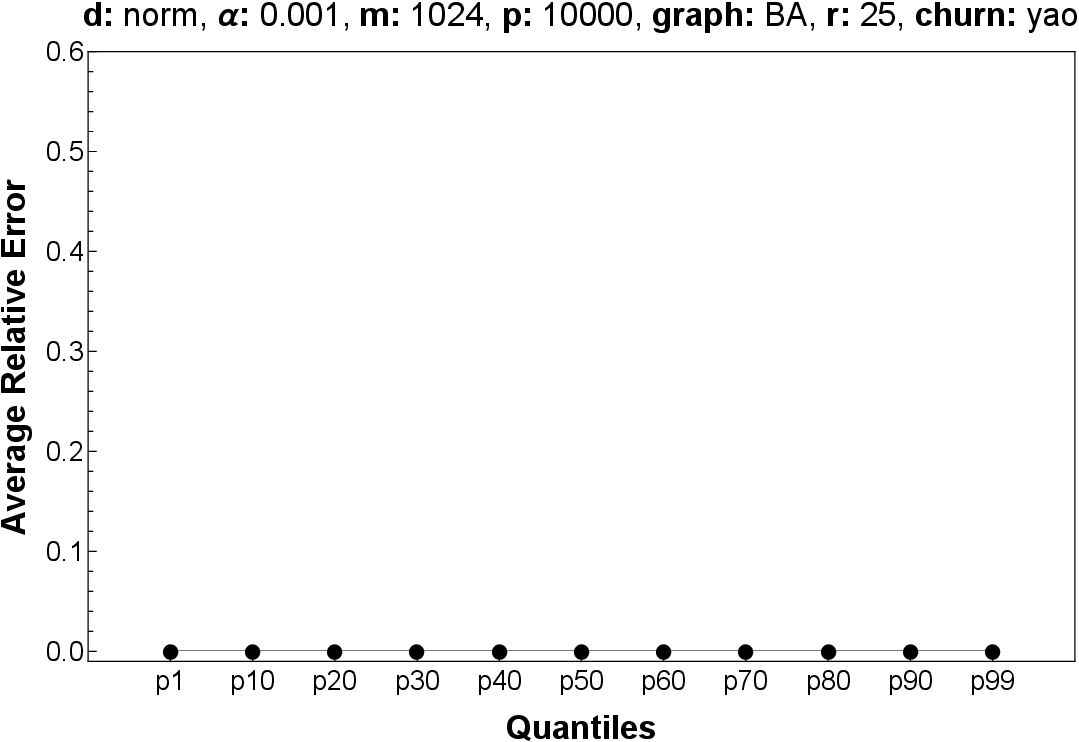}
		} \\
		
	\end{tabular}	
	\caption{Protocol convergence varying the number of rounds, in presence of \emph{Yao} churning, exponential (left column) and normal input (right column), $\alpha=0.001$, $m=1024$, over a network of $10000$ peers on a Bar\'abasi-Albert random graph.} 
	\label{fig.peers.ba.yao2}
\end{figure*}

\begin{figure*}[htb]
    \centering
    \begin{tabular}{cc}

		\subfloat[]{
		    \includegraphics[width=0.42\textwidth]{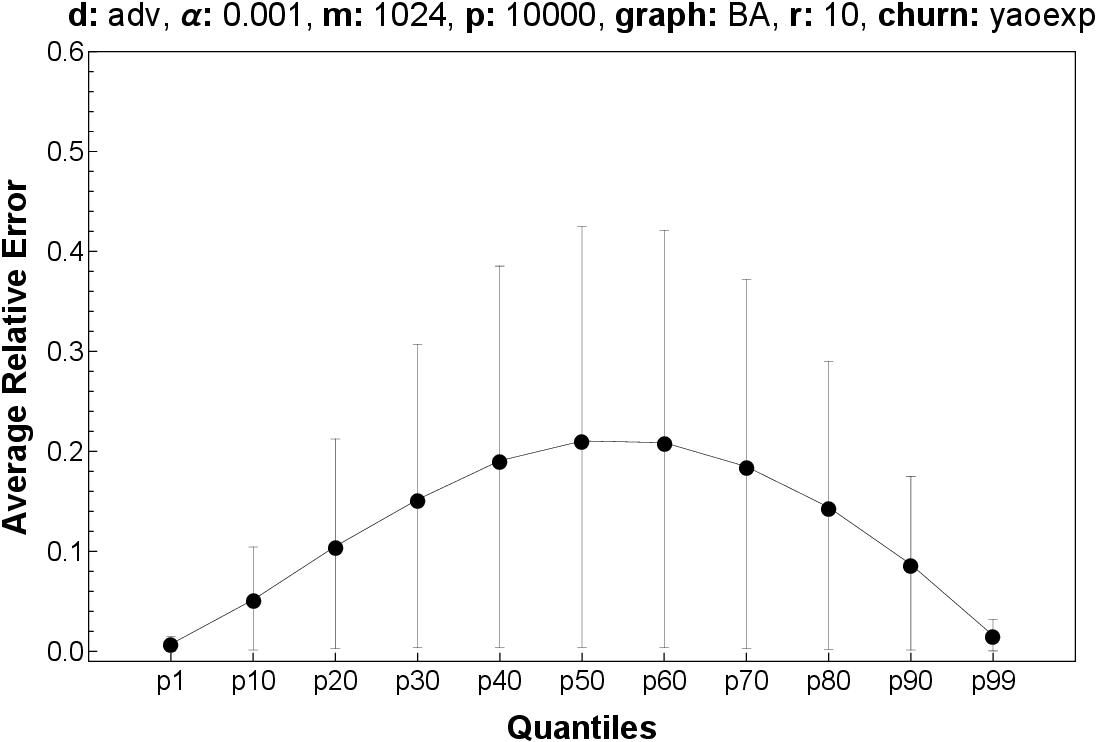}
		} &	
		
		\subfloat[]{
		    \includegraphics[width=0.42\textwidth]{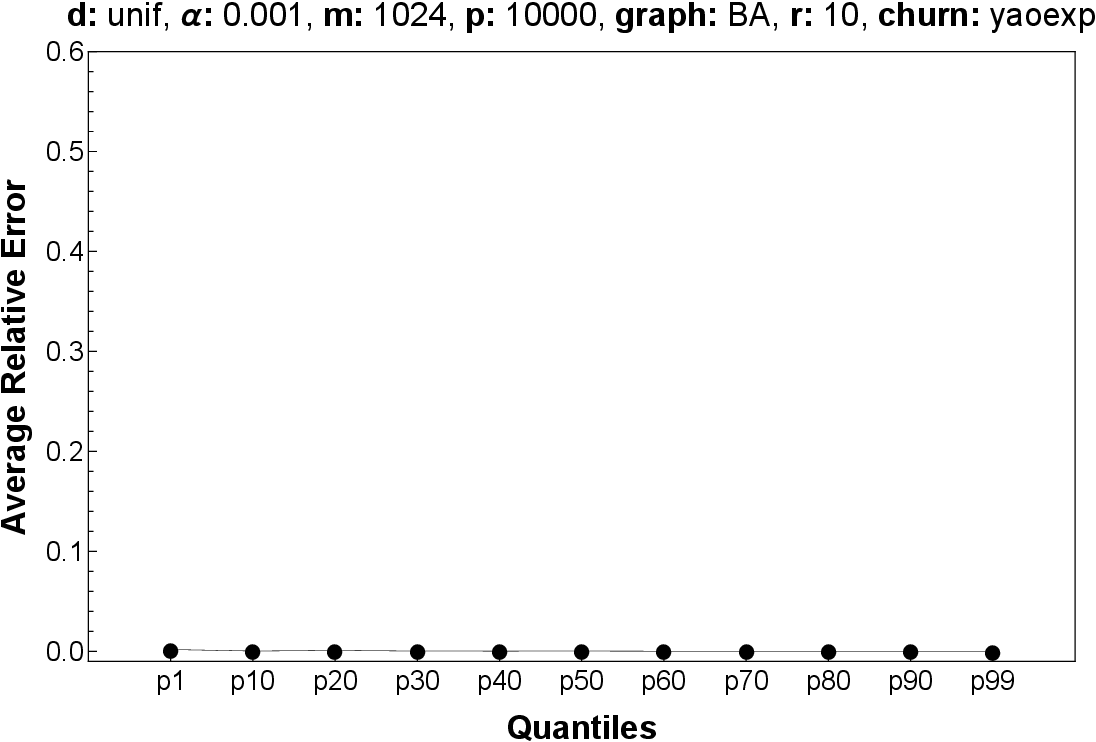}
		} \\	

        \subfloat[]{
		    \includegraphics[width=0.42\textwidth]{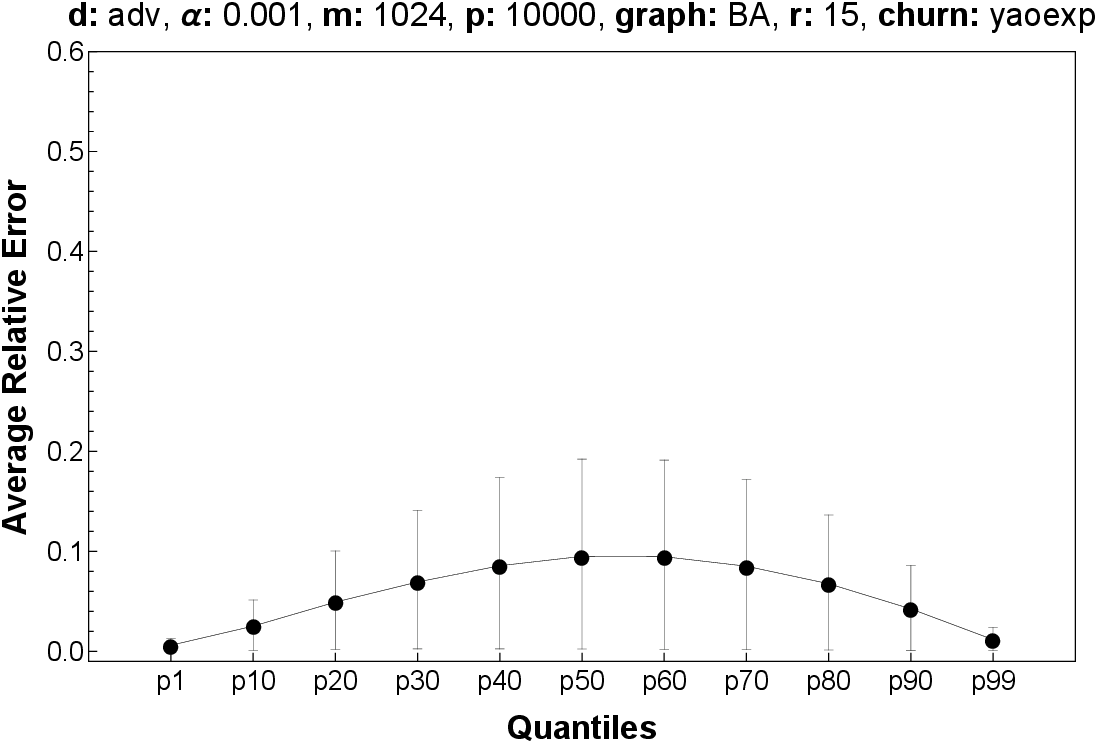}
		} &
		
		\subfloat[]{
		    \includegraphics[width=0.42\textwidth]{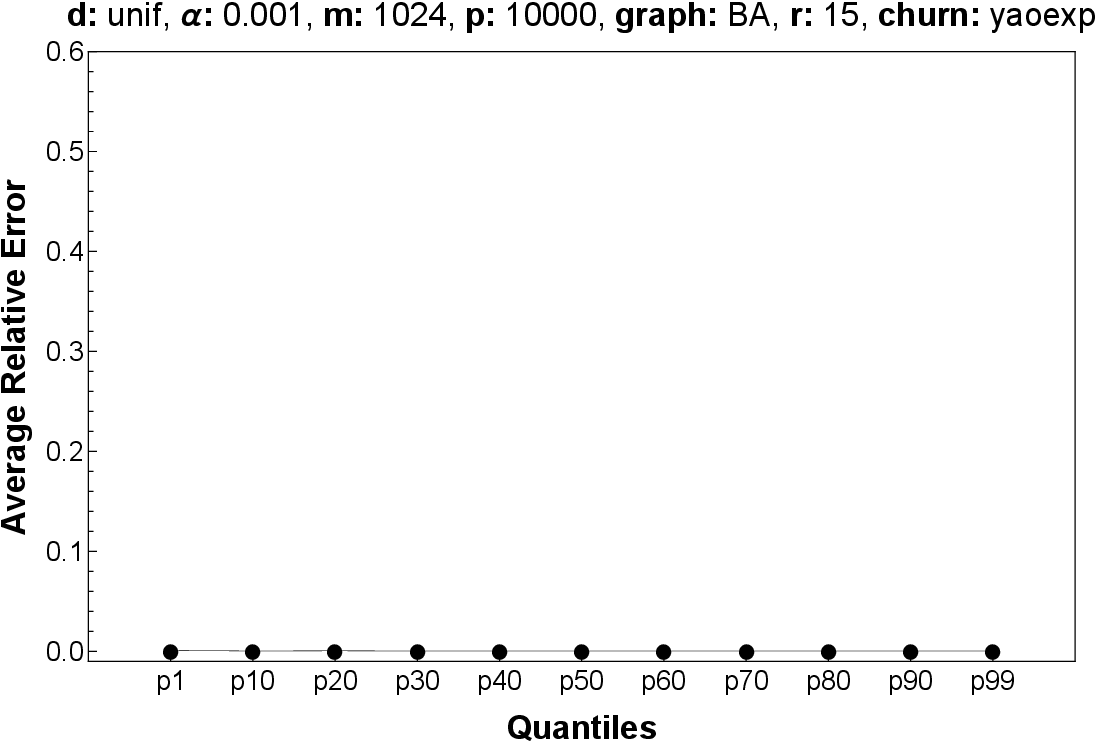}
		} \\

        \subfloat[]{
		    \includegraphics[width=0.42\textwidth]{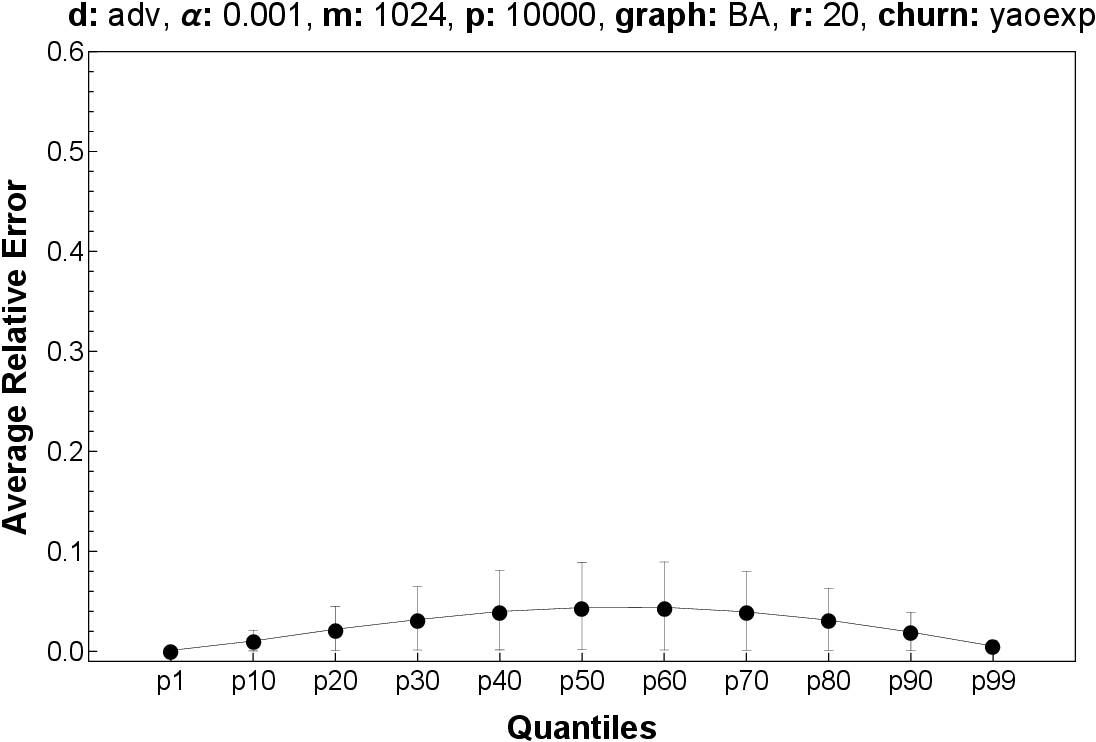}
		} &

		\subfloat[]{
		    \includegraphics[width=0.42\textwidth]{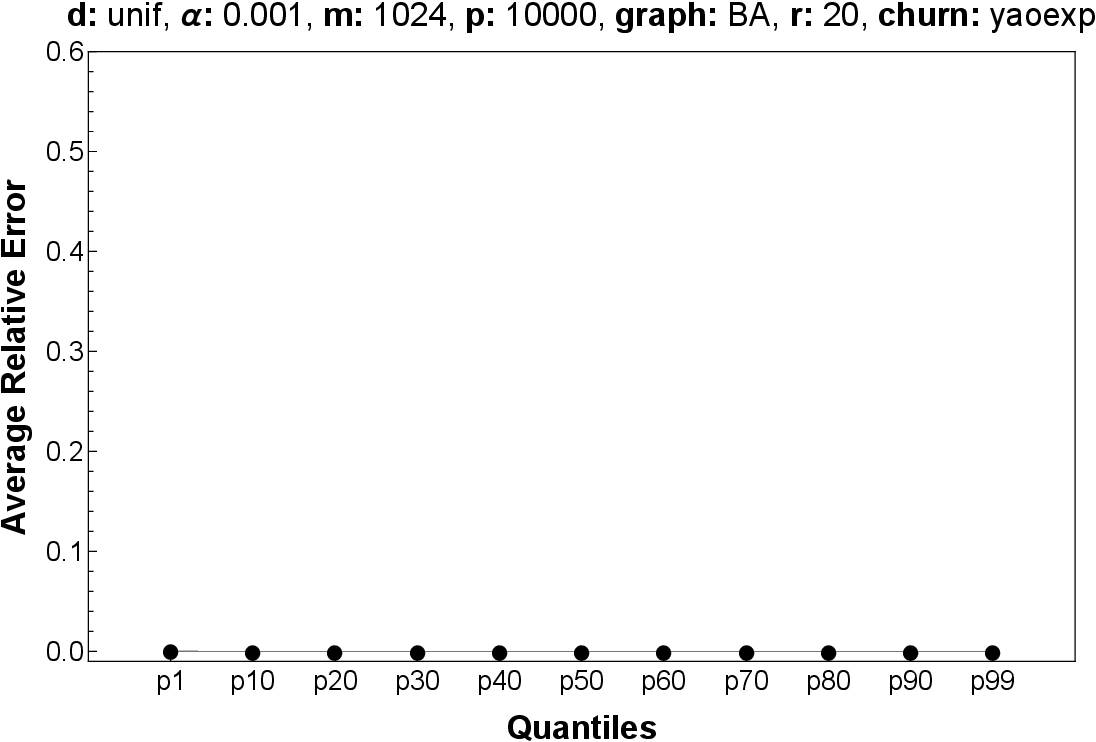}
		} \\
		
        \subfloat[]{
		    \includegraphics[width=0.42\textwidth]{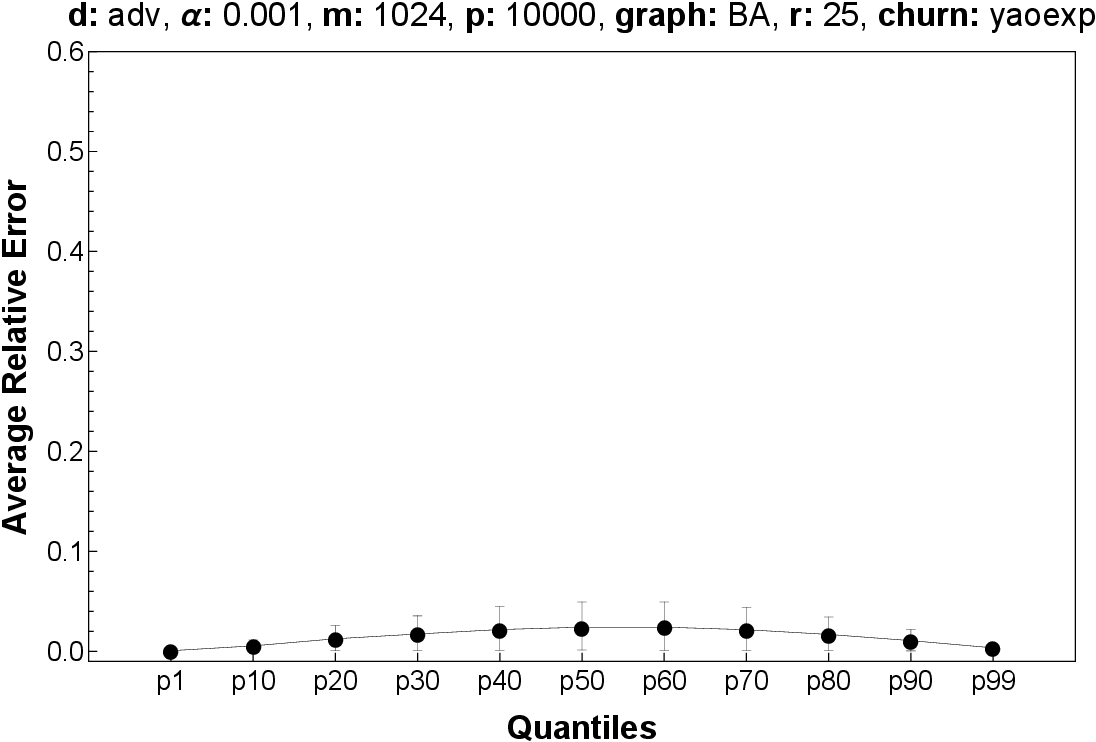}
		} &

		\subfloat[]{
		    \includegraphics[width=0.42\textwidth]{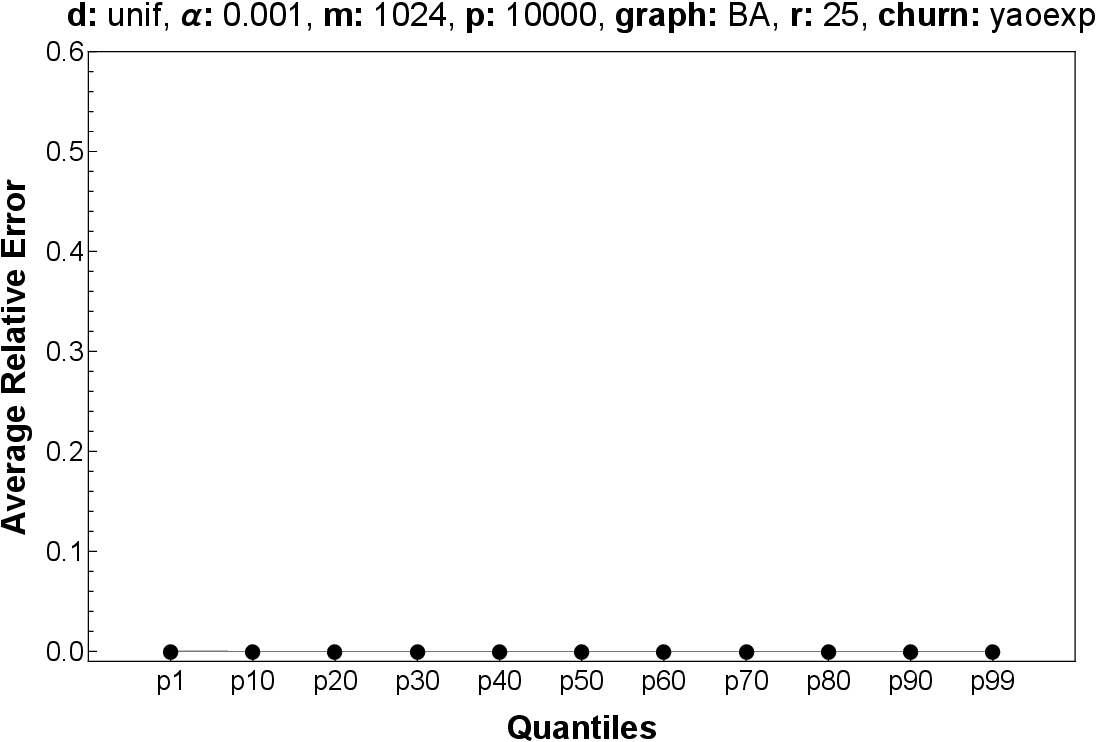}
		} \\
		
	\end{tabular}	
	\caption{Protocol convergence varying the number of rounds, in presence of \emph{Yao exponential} churning, adversarial (left column) and uniform input (right column), $\alpha=0.001$, $m=1024$, over a network of $10000$ peers on a Bar\'abasi-Albert random graph.} 
	\label{fig.peers.ba.yaoexp1}
\end{figure*}

\begin{figure*}[htb]
    \centering
    \begin{tabular}{cc}

		\subfloat[]{
		    \includegraphics[width=0.42\textwidth]{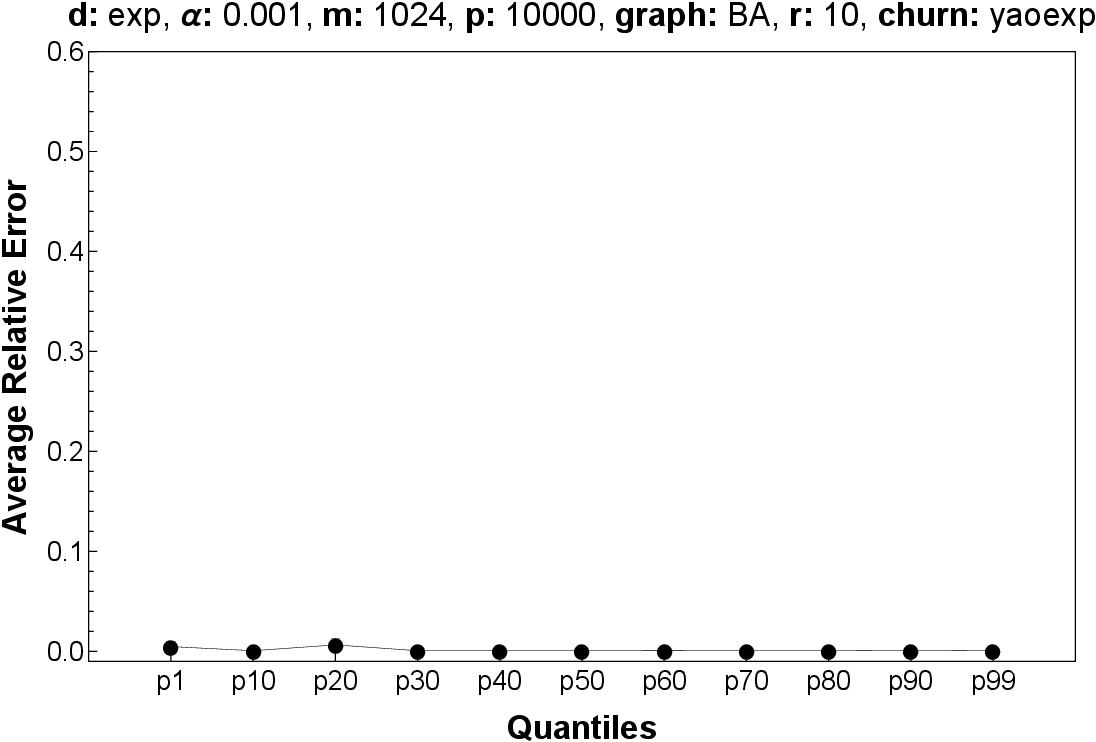}
		} &	
		
		\subfloat[]{
		    \includegraphics[width=0.42\textwidth]{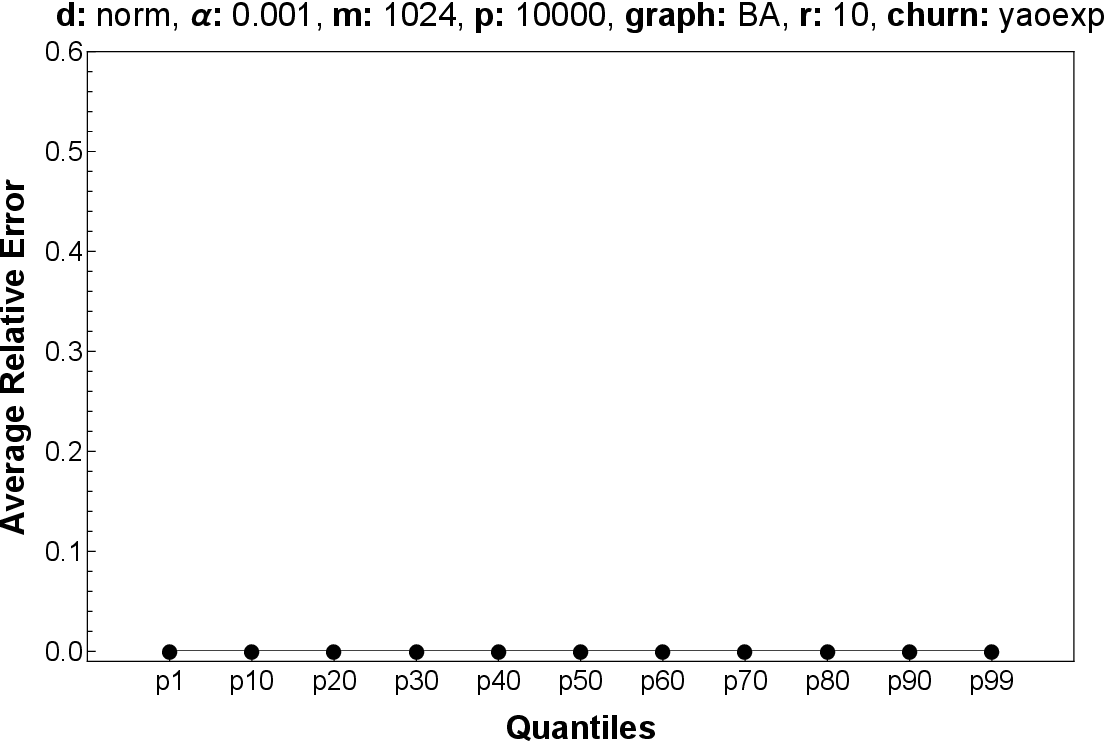}
		} \\	

        \subfloat[]{
		    \includegraphics[width=0.42\textwidth]{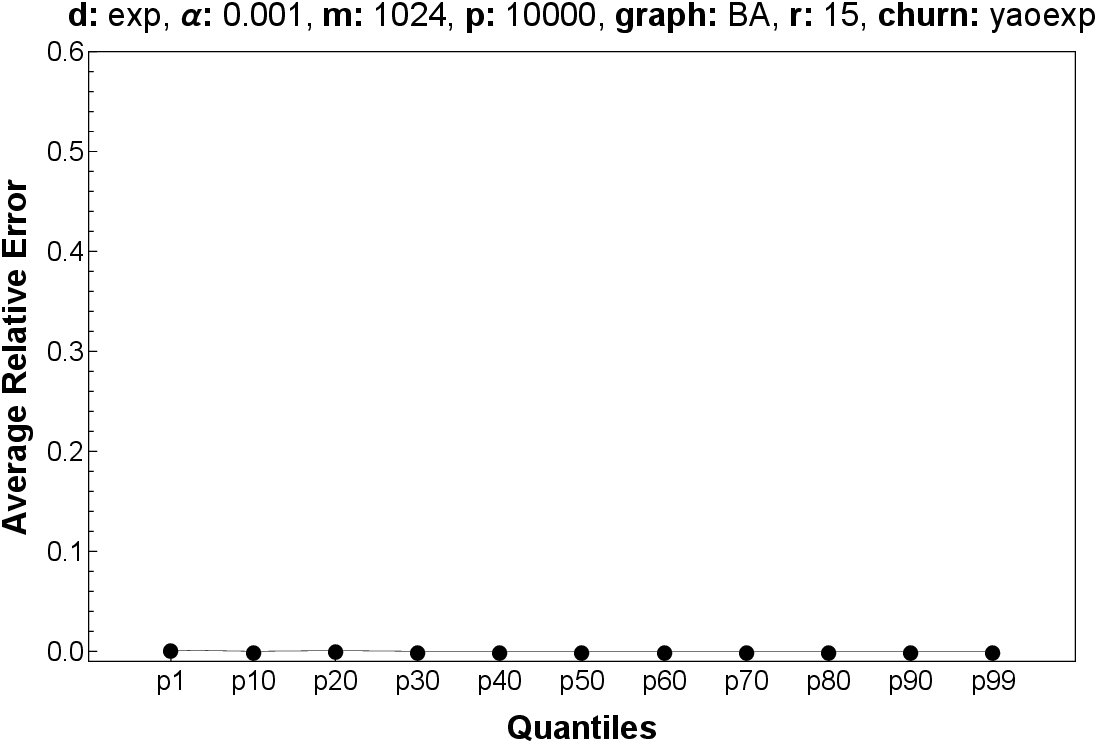}
		} &
		
		\subfloat[]{
		    \includegraphics[width=0.42\textwidth]{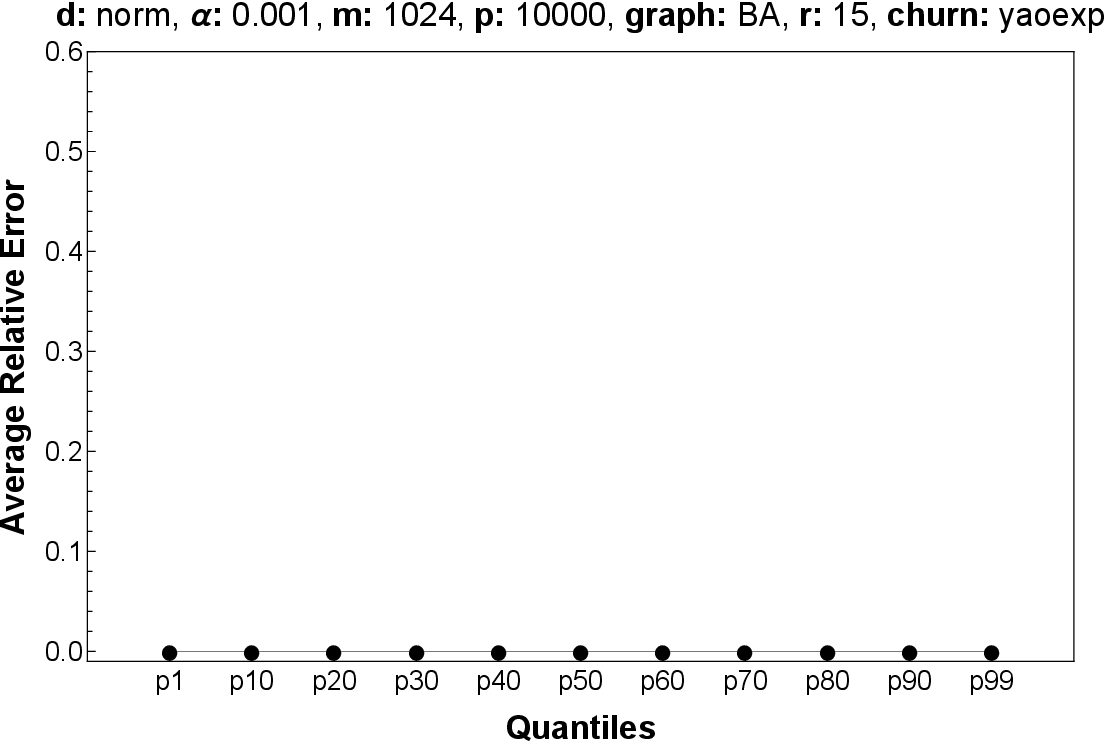}
		} \\

        \subfloat[]{
		    \includegraphics[width=0.42\textwidth]{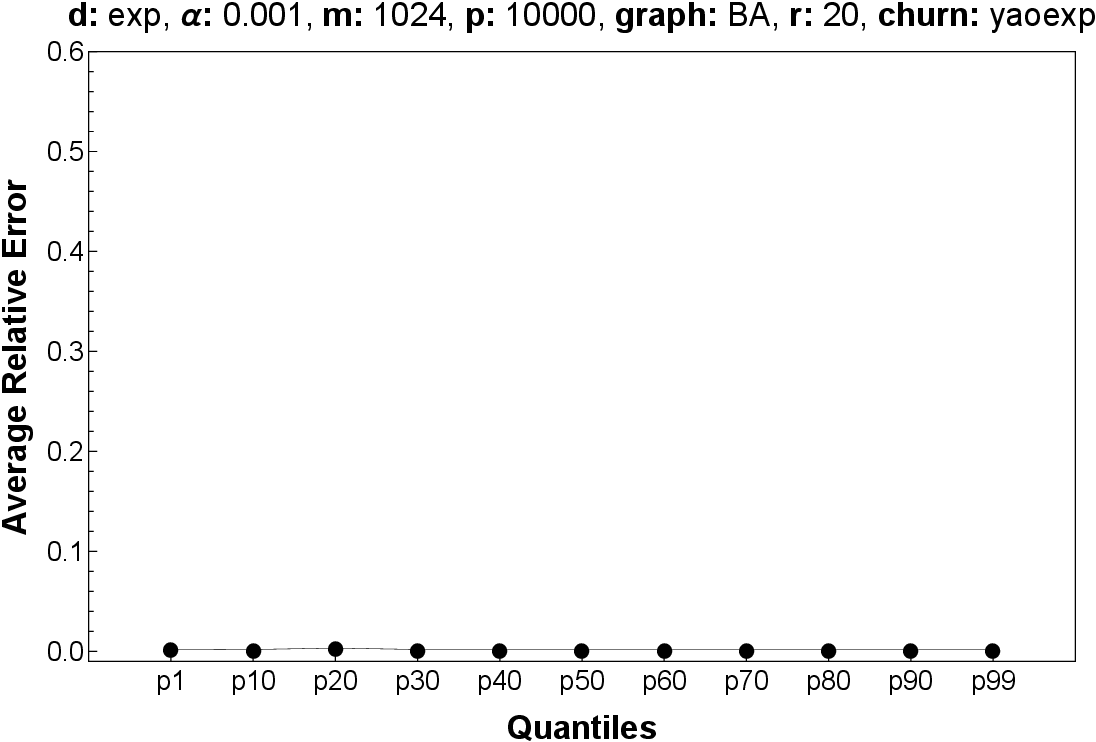}
		} &

		\subfloat[]{
		    \includegraphics[width=0.42\textwidth]{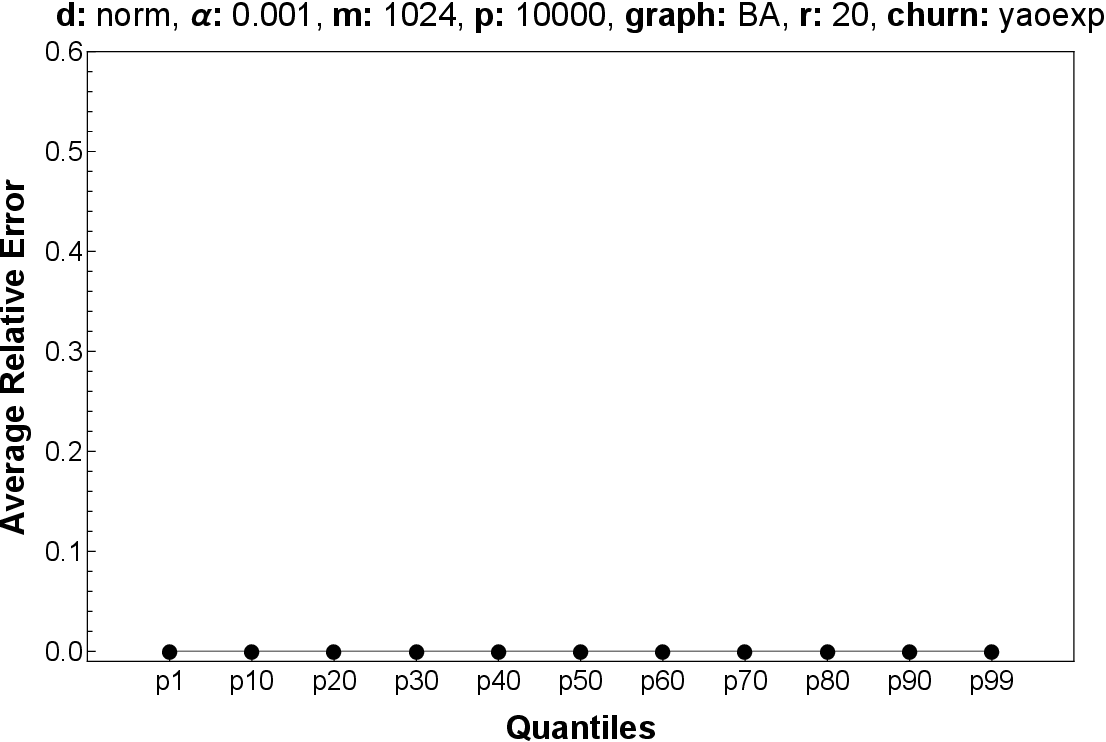}
		} \\
		
        \subfloat[]{
		    \includegraphics[width=0.42\textwidth]{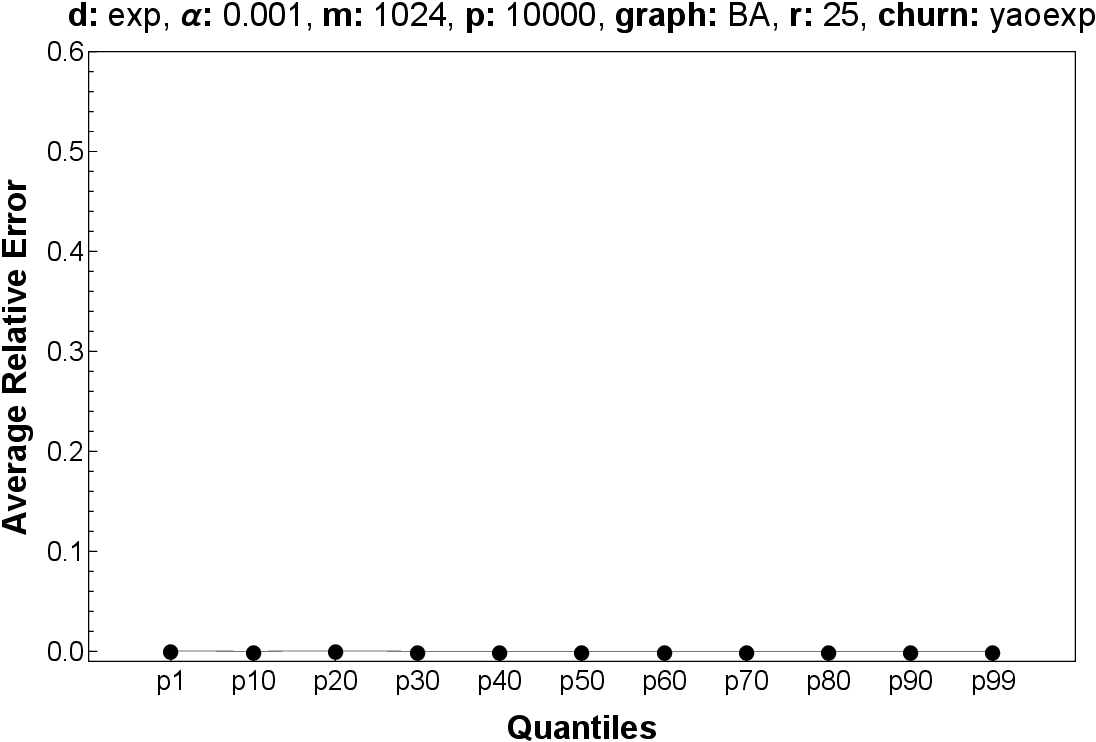}
		} &

		\subfloat[]{
		    \includegraphics[width=0.42\textwidth]{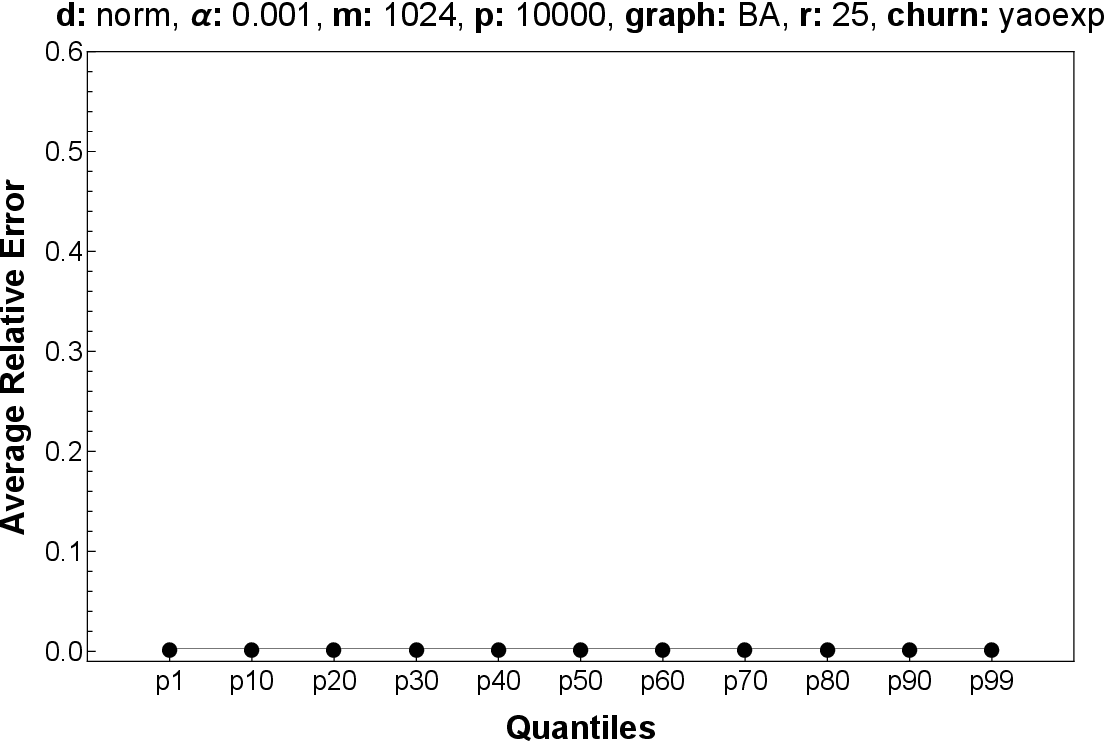}
		} \\
		
	\end{tabular}	
	\caption{Protocol convergence varying the number of rounds, in presence of \emph{Yao exponential} churning, exponential (left column) and normal input (right column), $\alpha=0.001$, $m=1024$, over a network of $10000$ peers on a Bar\'abasi-Albert random graph.} 
	\label{fig.peers.ba.yaoexp2}
\end{figure*}

Under the churning effects, we assume that a peer can detect a neighbour failure, then:

\begin{itemize}
    \item if a peer fails before sending a push message or after receiving a pull message, that is, when no communications are ongoing, then no actions have to be performed;
    \item if a peer $p$ fails before sending a pull message to a peer $r$ in response to its push message, then the peer $r$ detects the failure and simply cancels the push–pull exchange, so that its state does not change;
    \item if a peer $p$ fails after sending a push message to a peer $r$ and before receiving the corresponding pull message, then the peer $r$ detects the failure and restores its own local state as it was before the push–pull exchange.
\end{itemize}

The figures show the results of tests for the set of quantiles in Table \ref{params}. The Relative Error between the sequential algorithm estimation $\hat{x}_q$ and the values estimated by each peer through the distributed algorithm, $\tilde{x}_{q,i}$ have been computed and averaged among the $p$ peers in the network: 

\begin{equation}
\text{ARE}_q = \frac{1}{p}\sum^{p}_{i=1}\frac{\left | \tilde{x}_{q,i} - \hat{x}_q \right |}{\hat{x}_q}.
\label{eq_are}
\end{equation}

Figures \ref{fig.peers.ba.fs1}-\ref{fig.peers.ba.yaoexp2} show how the churning of nodes affects the convergence of the distributed gossip-based protocol with different input distributions (each figure column refer to a different distribution), varying the number of rounds executed and keeping fixed the other parameters. \\
Figures \ref{fig.peers.ba.fs1} and \ref{fig.peers.ba.fs2} show the results obtained using the Fail \& Stop model with probability of failure equal to $0.01$. \\

Figures \ref{fig.peers.ba.yao1} and \ref{fig.peers.ba.yao2} refer to the Yao model variant with probability of an offline peer joining again the network drawn from a shifted Pareto distribution. Whilst Figures \ref{fig.peers.ba.yaoexp1} and \ref{fig.peers.ba.yaoexp2} show the results relative to the Yao model variant with an Exponential distribution.

As expected, in presence of churning the speed of convergence is lower, especially for the \textit{adversarial} input. The most severe scenario is the Fail \& Stop model where peers failing can not re-join the network and the strongly connected property of the graph representing the network is not guaranteed anymore (if the graph is not strongly connected, information can only be exchanged among the peers belonging to the same connected component and the distributed gossip protocol can not work as expected). This is particularly evident from the plots relative to the \textit{adversial} input which show how the errors do not tend to converge to zero anymore. 

In the Yao models the churning effect is less severe, owing to the fact that the peers can join again the network. The convergence can be reached though at a lower pace, as it is shown especially by the plots relative to the \textit{adversial} input, whilst the other input distributions are less affected by the churning.

\subsection{Real dataset}
\begin{figure*}[htb]
    \centering
    \begin{tabular}{cc}

		\subfloat[]{
		    \includegraphics[width=0.42\textwidth]{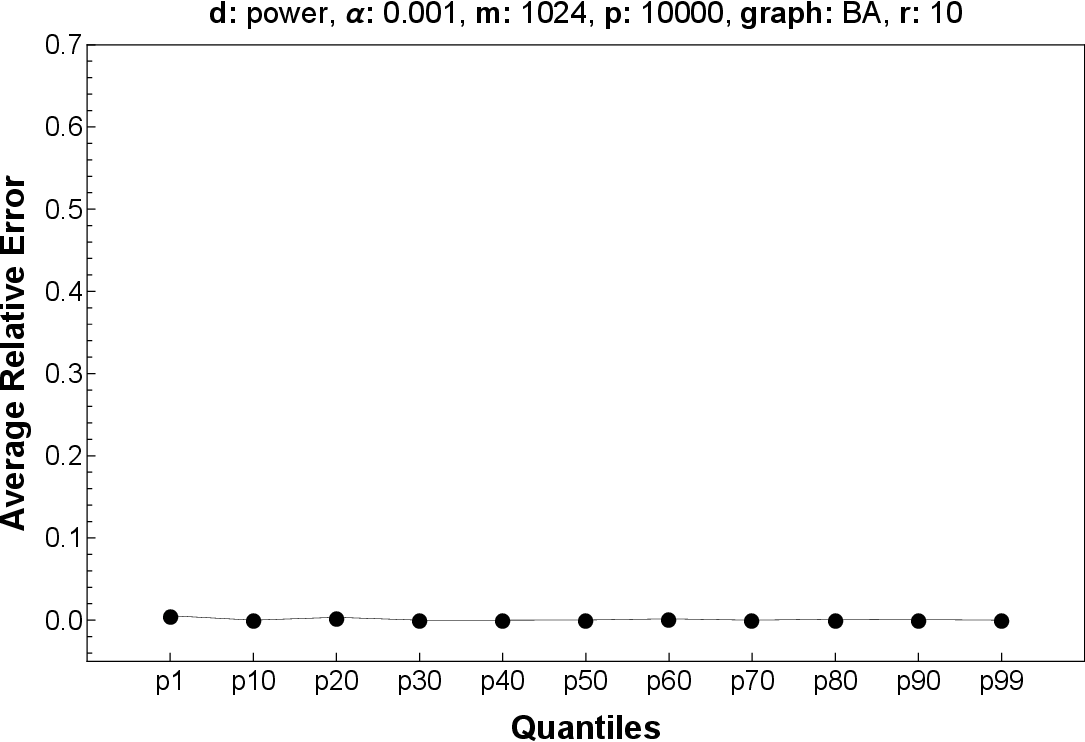}
		} &	
		
		\subfloat[]{
		    \includegraphics[width=0.42\textwidth]{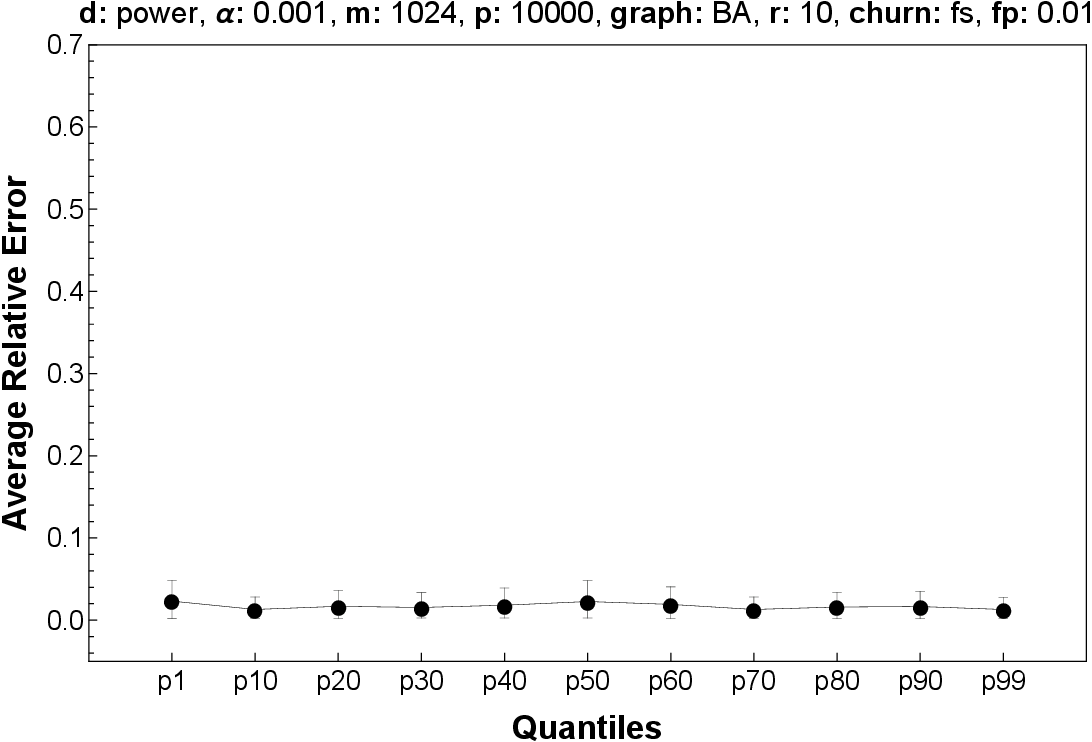}
		} \\	

        \subfloat[]{
		    \includegraphics[width=0.42\textwidth]{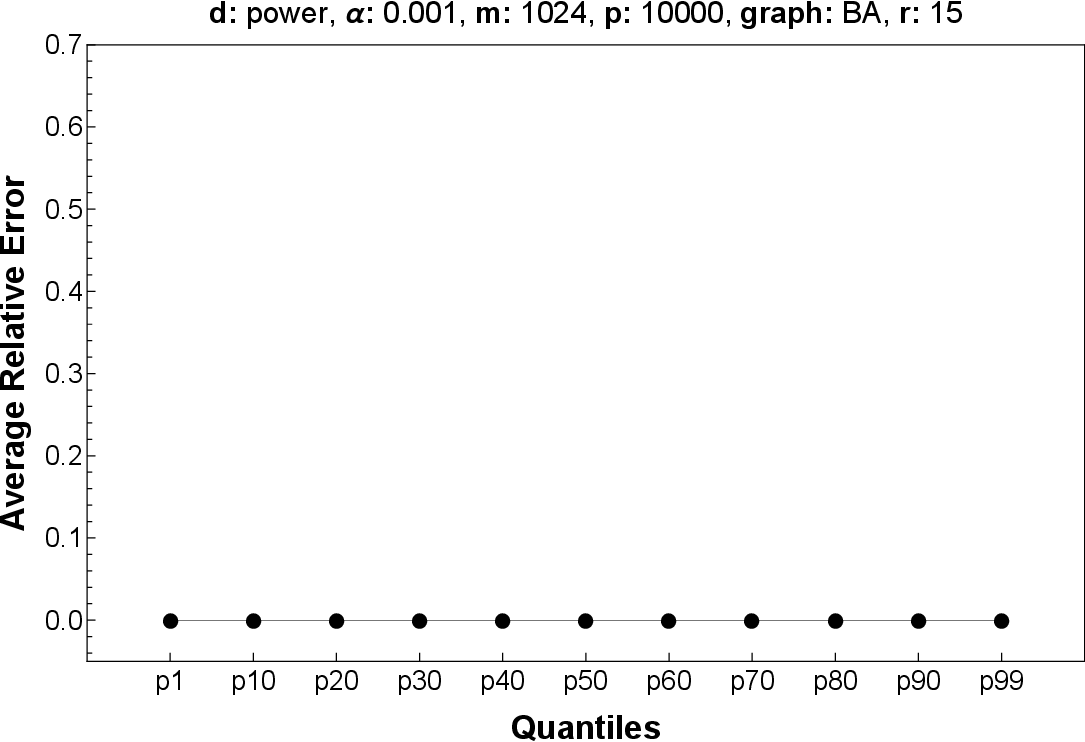}
		} &
		
		\subfloat[]{
		    \includegraphics[width=0.42\textwidth]{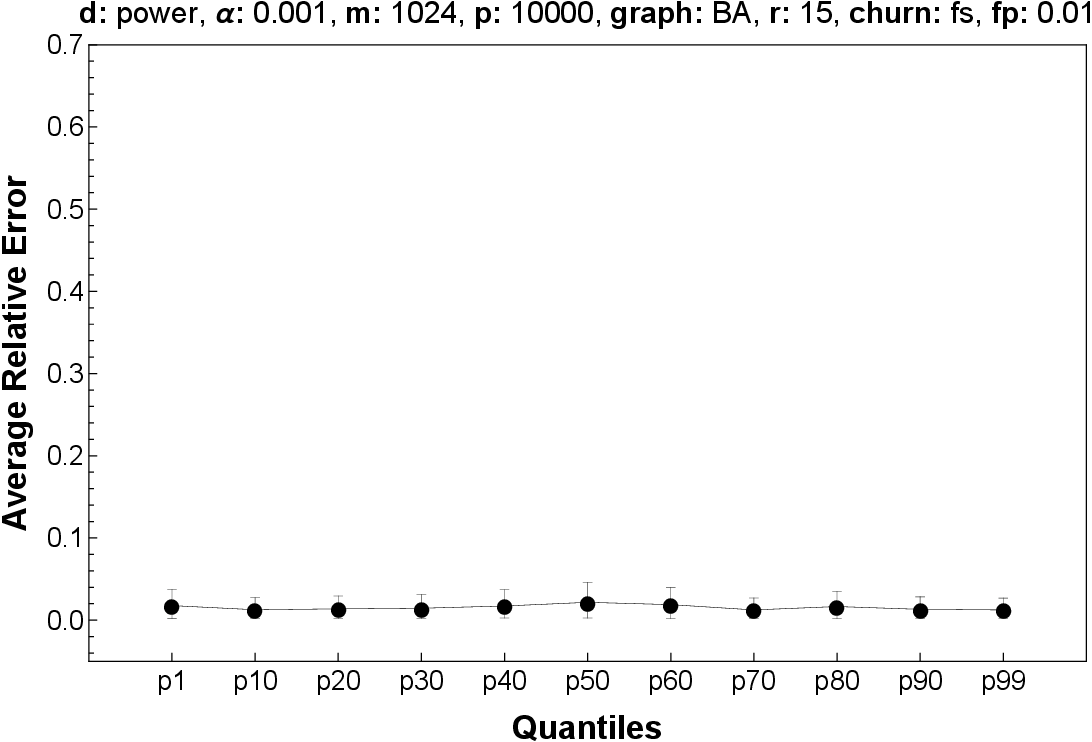}
		} \\

        \subfloat[]{
		    \includegraphics[width=0.42\textwidth]{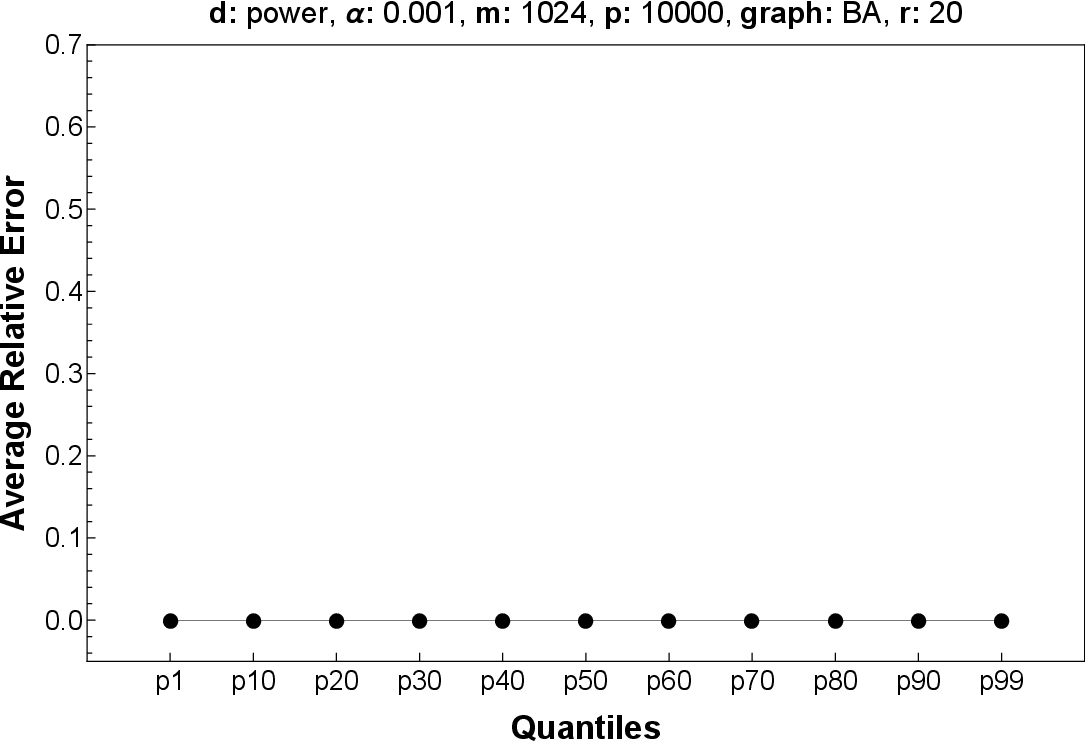}
		} &

		\subfloat[]{
		    \includegraphics[width=0.42\textwidth]{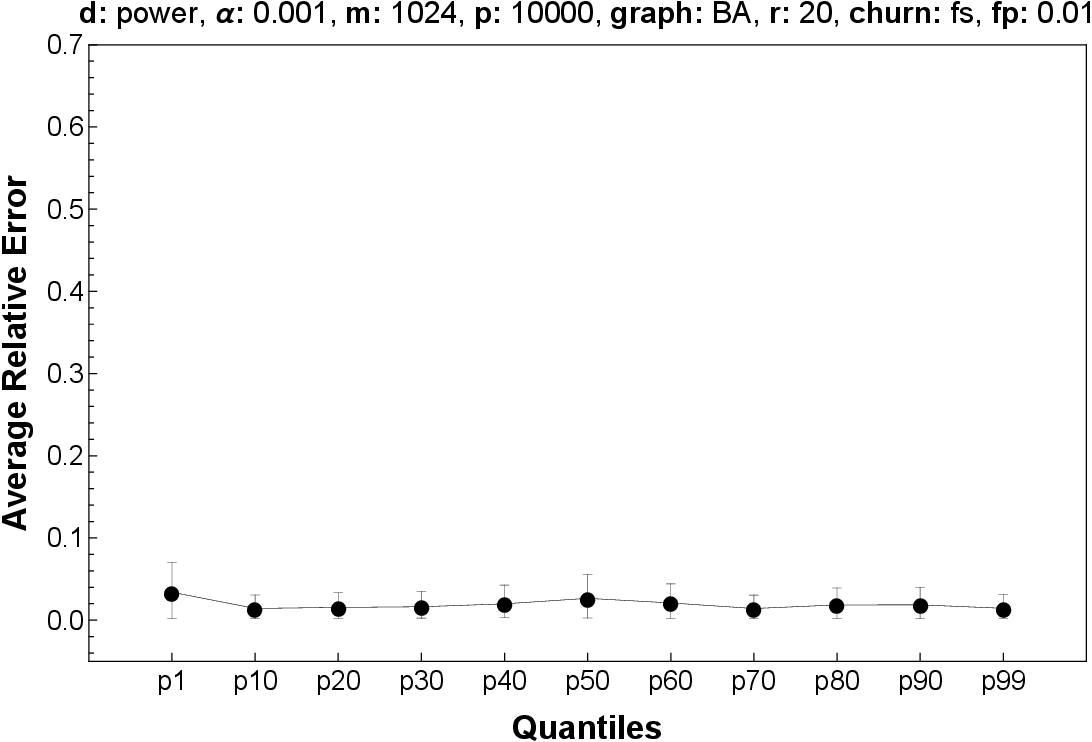}
		} \\
		
        \subfloat[]{
		    \includegraphics[width=0.42\textwidth]{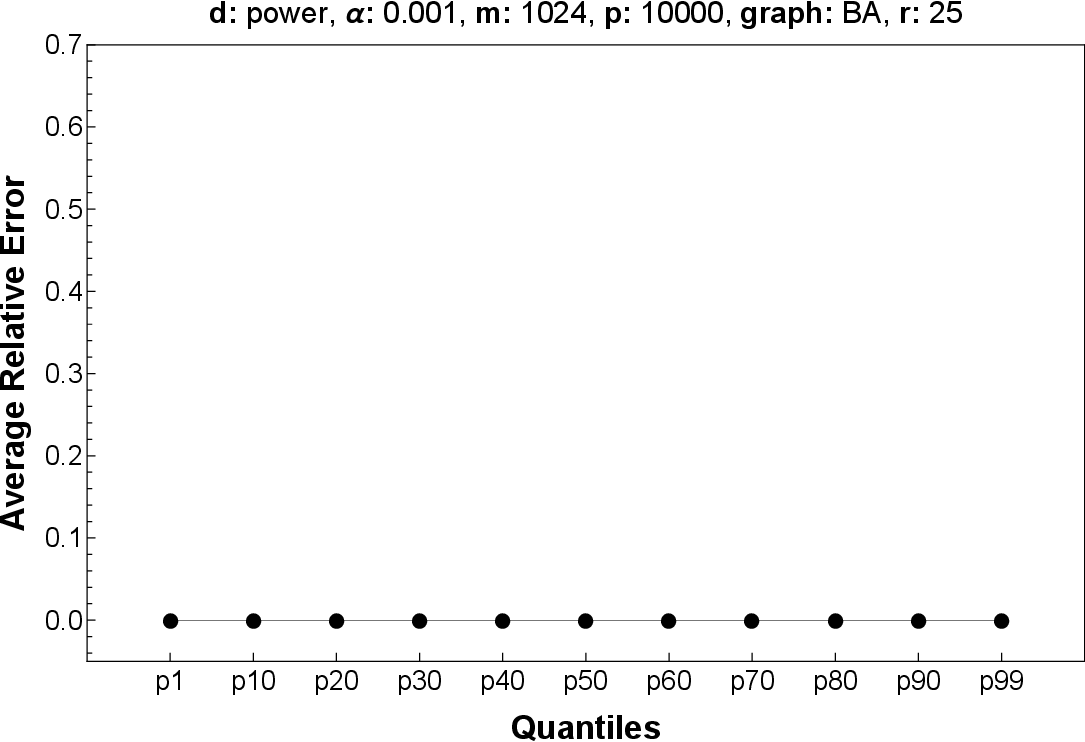}
		} &

		\subfloat[]{
		    \includegraphics[width=0.42\textwidth]{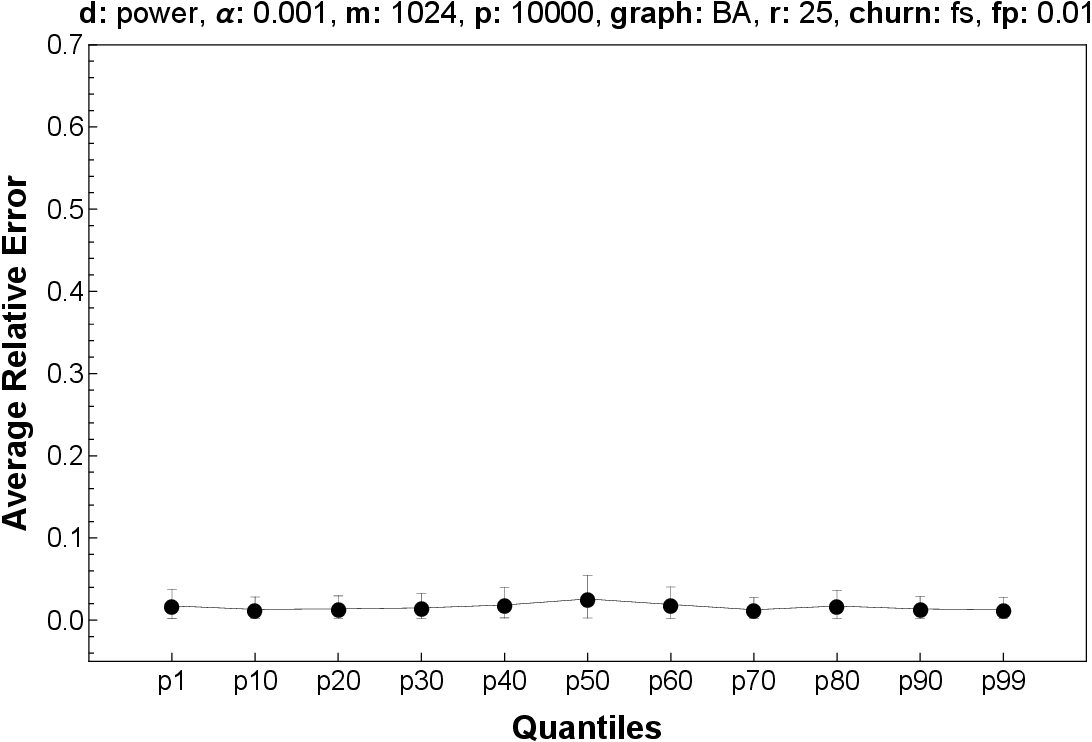}
		} \\
		
	\end{tabular}	
	\caption{Protocol convergence for the Power dataset and varying the number of rounds, in absence of churning (left column) and in presence of \emph{Fail \& Stop} churning (right column), $\alpha=0.001$, $m=1024$, network of $10000$ peers on a Bar\'abasi-Albert random graph.} 
	\label{fig.peers.ba.real1}
\end{figure*}

\begin{figure*}[htb]
    \centering
    \begin{tabular}{cc}

		\subfloat[]{
		    \includegraphics[width=0.42\textwidth]{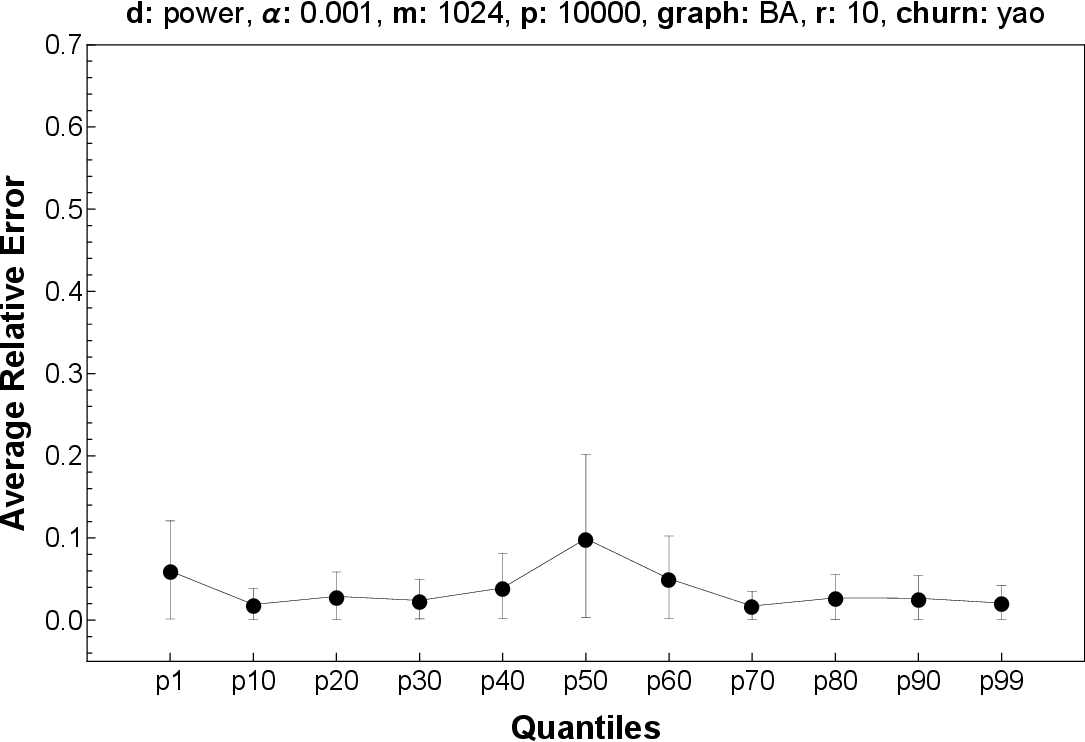}
		} &	
		
		\subfloat[]{
		    \includegraphics[width=0.42\textwidth]{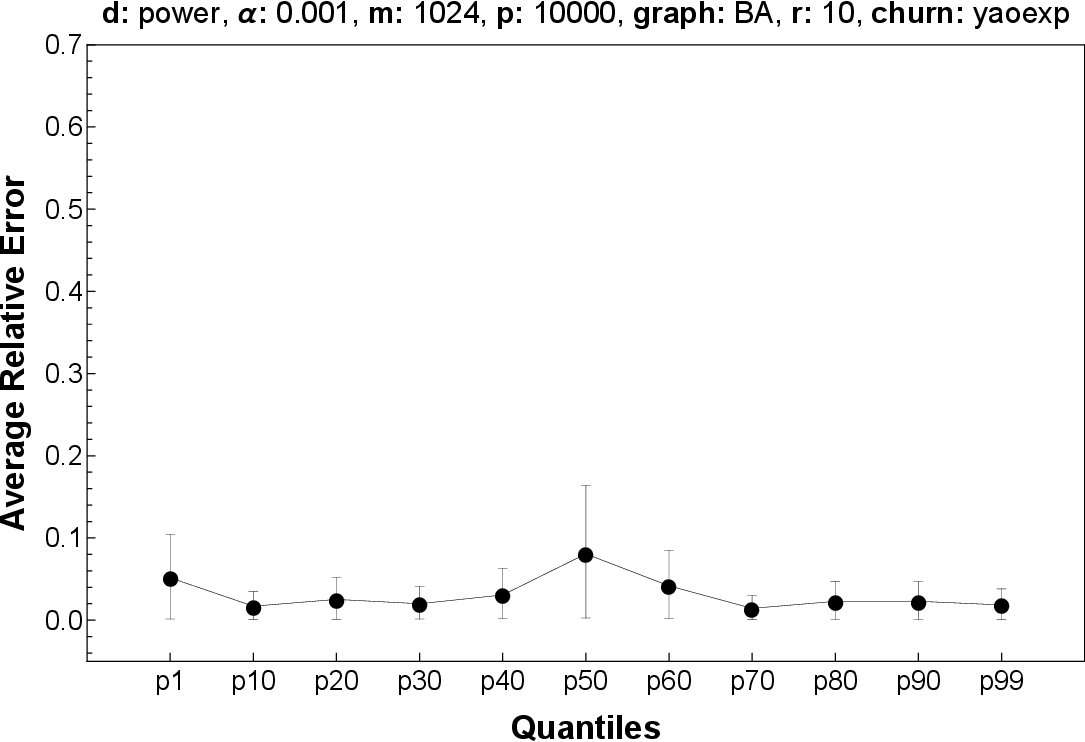}
		} \\	

        \subfloat[]{
		    \includegraphics[width=0.42\textwidth]{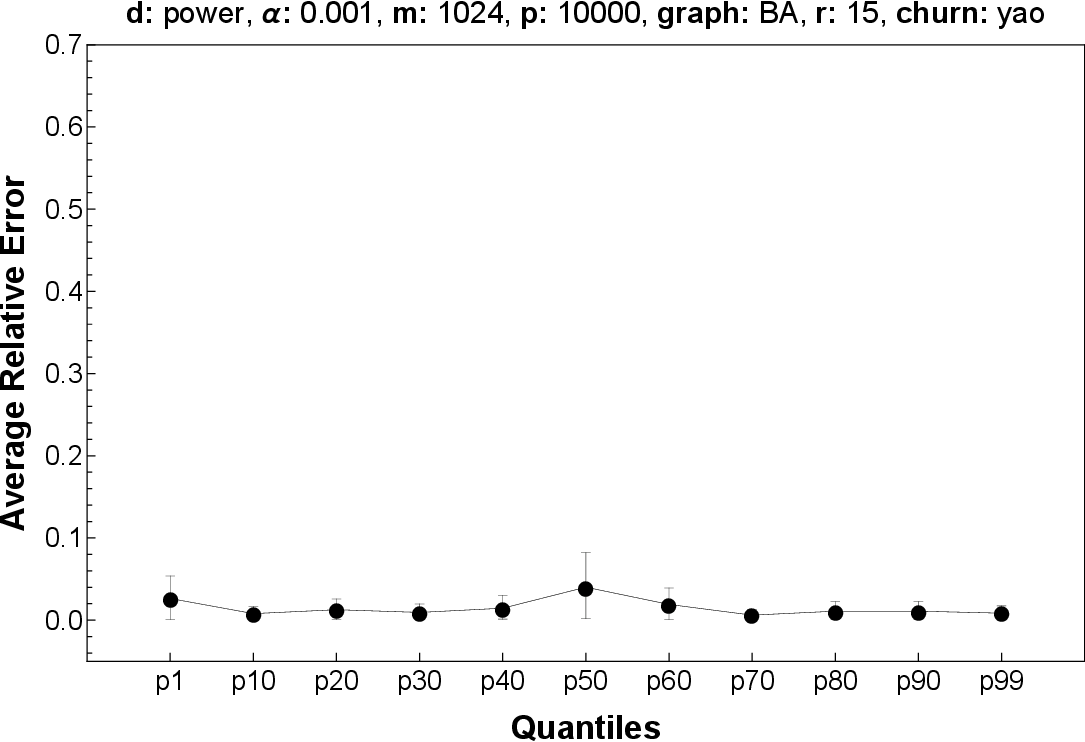}
		} &
		
		\subfloat[]{
		    \includegraphics[width=0.42\textwidth]{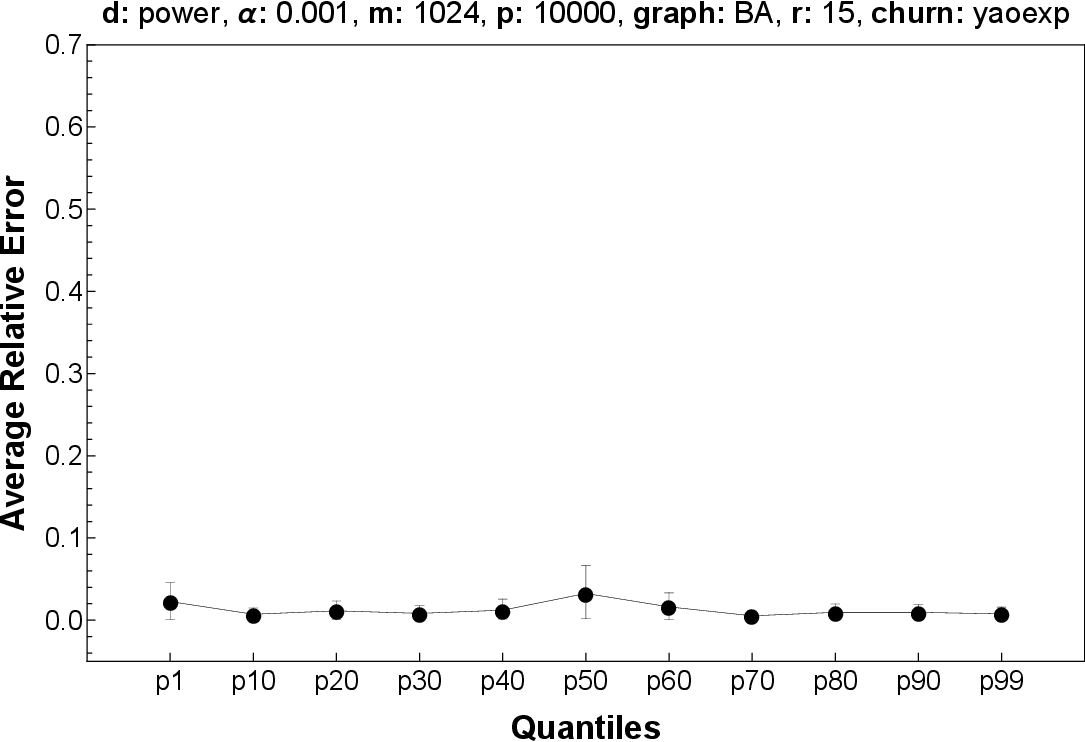}
		} \\

        \subfloat[]{
		    \includegraphics[width=0.42\textwidth]{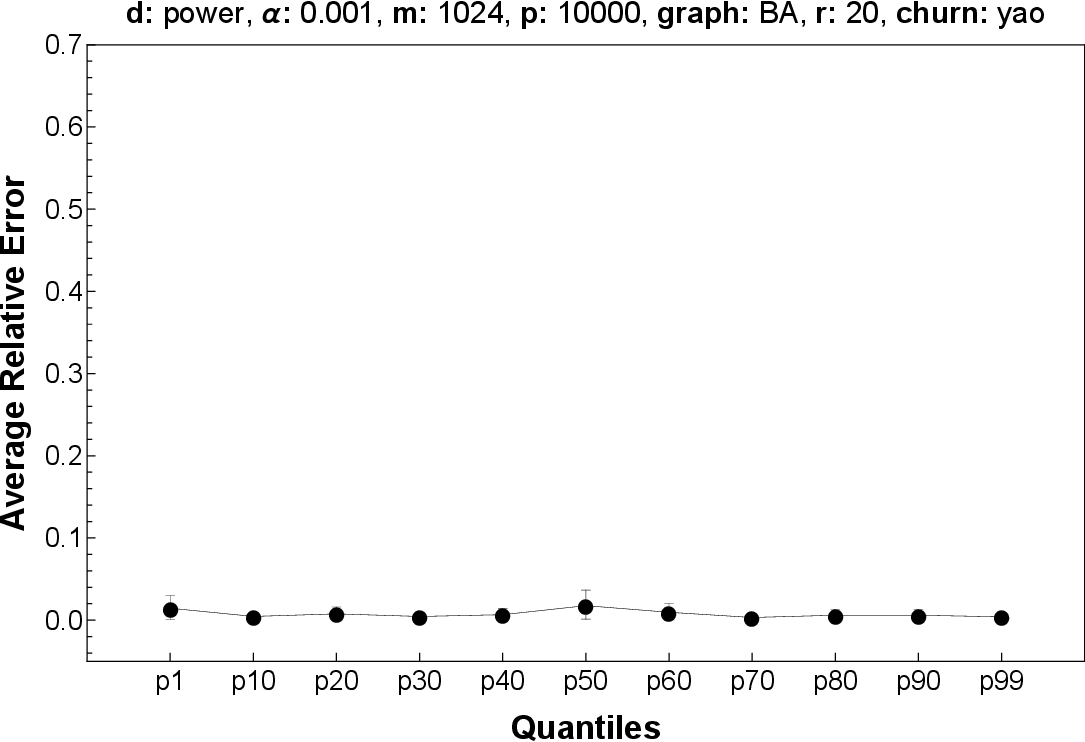}
		} &

		\subfloat[]{
		    \includegraphics[width=0.42\textwidth]{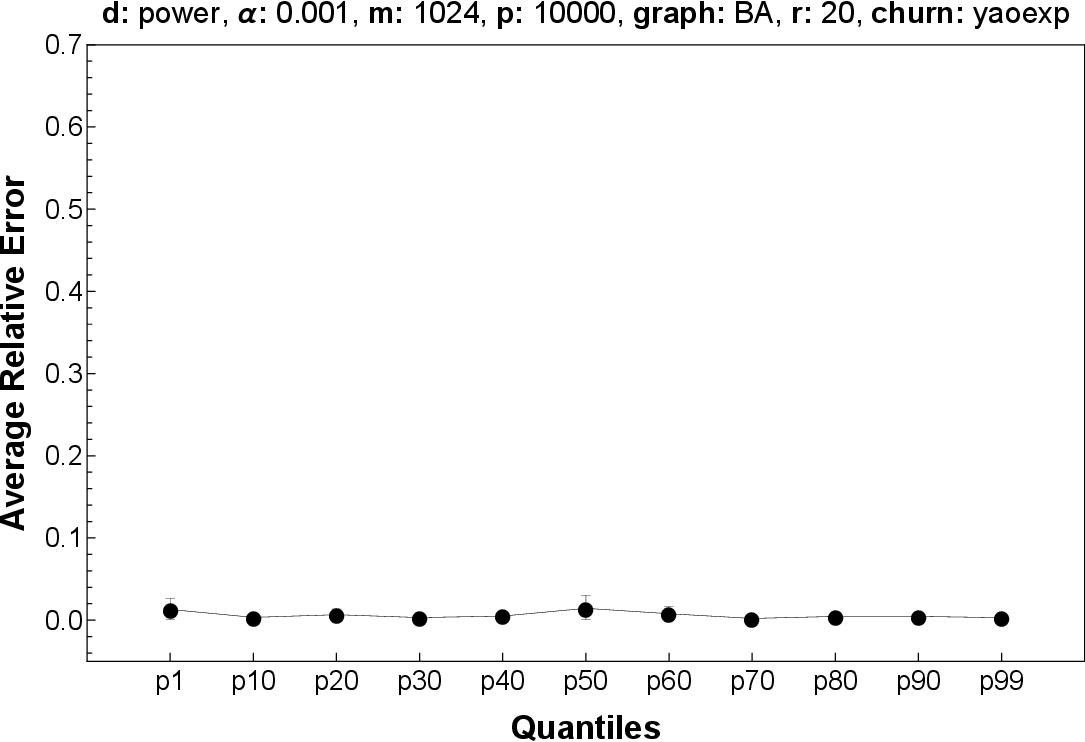}
		} \\
		
        \subfloat[]{
		    \includegraphics[width=0.42\textwidth]{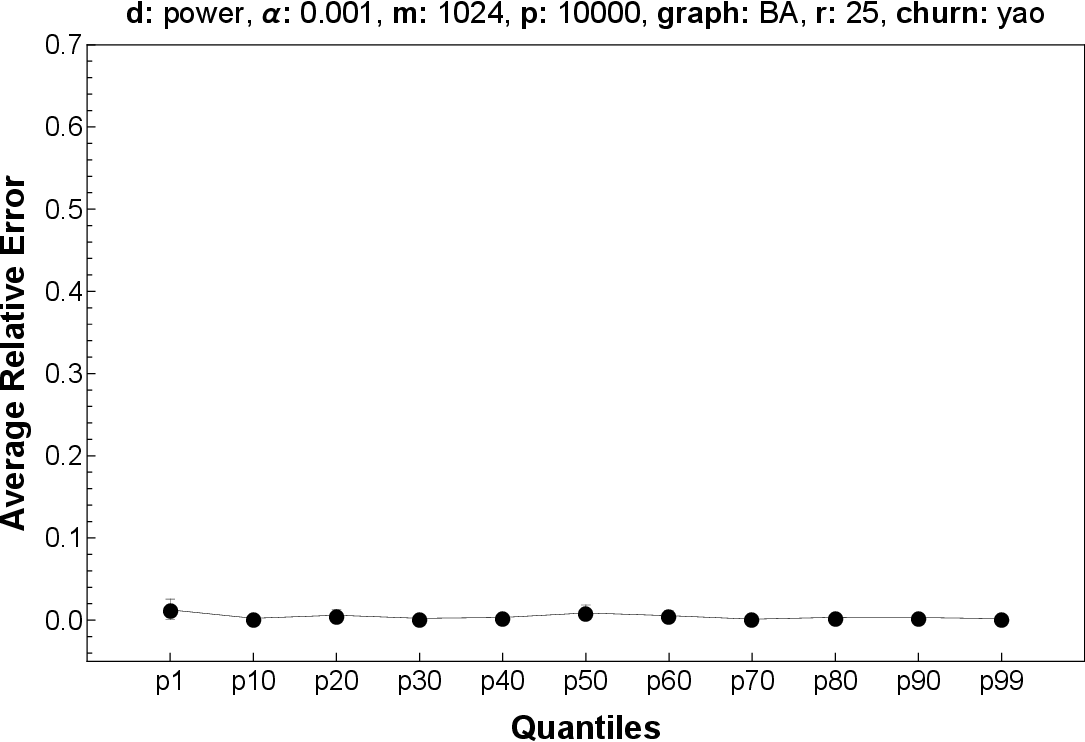}
		} &

		\subfloat[]{
		    \includegraphics[width=0.42\textwidth]{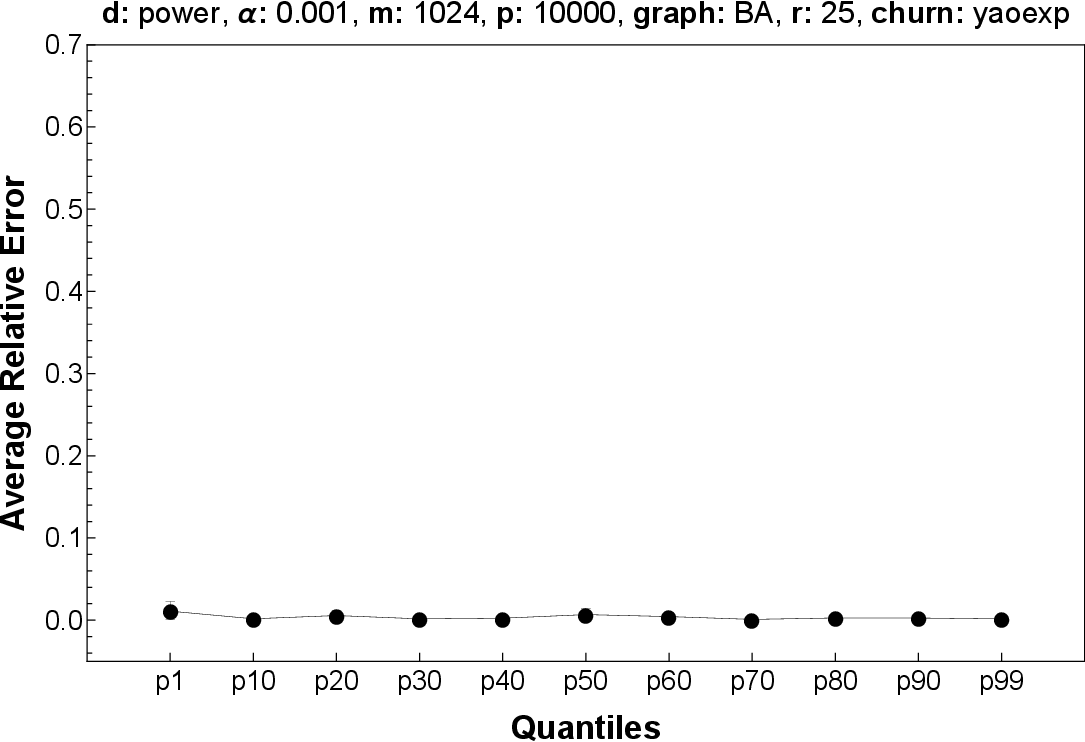}
		} \\
		
	\end{tabular}	
	\caption{Protocol convergence for the Power dataset and varying the number of rounds, in presence of \emph{Yao} churning (left column) and \emph{Yao exponential} churning (right column), $\alpha=0.001$, $m=1024$, network of $10000$ peers on a Bar\'abasi-Albert random graph.} 
	\label{fig.peers.ba.real2}
\end{figure*}

Tests on a real dataset have been performed in order to verify the behaviour of the distributed algorithm also in a real scenario. The dataset used, named \textit{power}, has been drawn from the global active power measurements of the UCI Individual Household Electric Power Consumption dataset \cite{powerconsumption-dataset}. Figures \ref{fig.peers.ba.real1} and \ref{fig.peers.ba.real2} report the results obtained executing our algorithm over the power dataset, with and without churning, for a network of $10000$ peers and leaving all of the parameters at their default values. The plots confirm the good behaviour of the algorithm also in this case, showing that few rounds are enough for convergence, which remains true even in the case of churning.

\section{Conclusions}
\label{conclusions}

In this work, we have dealt with the problem of computing quantiles in unstructured P2P networks. In particular, we focused on the design of a distributed version of the \textsc{UDDSketch} algorithm, and obtained a fully decentralized, gossip-based protocol for the computation of quantiles. 

Through the experimental results presented in Section \ref{results} we showed that the Distributed \textsc{UDDSketch} algorithm provides very good accuracy for datasets characterized by different distributions. In fact, the analysis revealed that the local sketch of a peer engaged in the gossip-based distributed averaging protocol converges to the sketch that would be obtained by running \textsc{UDDSketch} on the union of the local datasets  (or substreams) held by the peers.

\section*{Data availability}
The dataset analysed during the current study - Individual Household Electric Power Consumption - is available in the UC Irvine Machine Learning Repository \cite{powerconsumption-dataset}.

\section*{Code availability}
The source code is freely available at https://github.com/cafaro/DUDDSketch

\clearpage

\bibliographystyle{elsarticle-num}
\bibliography{bibliography}

\end{document}